\documentclass[twocolumn, aps, pra, amsmath, amssymb, nofootnotebib, superscriptaddress]{revtex4-2}

\usepackage{hyperref}
\hypersetup{colorlinks = true, linkcolor = [rgb]{0.19411,0.51882,0.667058}, urlcolor = [rgb]{0.125490,0.29542,0.1647058}, citecolor = [rgb]{0.75882,0.37411,0.14117}}

\usepackage{bm}
\usepackage{amsmath}
\usepackage{amsfonts}
\usepackage{amssymb,upgreek}
\usepackage{graphicx}
\usepackage{mathtools}

\providecommand{\U}[1]{\protect\rule{.1in}{.1in}}

\newcommand{\nn}{\nonumber \\}
\usepackage{xcolor}

\usepackage{braket}

\DeclareMathOperator{\Cov}{Cov}

\renewcommand{\vec}[1]{\bm{\mathrm{#1}}}
\newcommand{\mat}[1]{\bm{\mathrm{#1}}}

\newcommand{\vac}{\text{vac}}

\usepackage{graphicx}
\usepackage{tikz}
\def\tick{\tikz\fill[scale=0.4](0,.35) -- (.25,0) -- (1,.7) -- (.25,.15) -- cycle;} 

\begin{document}

\title{Quantum-enabled optical large-baseline interferometry: applications, protocols and feasibility} 

\author{Zixin Huang}
\affiliation{School of Science, STEM College, RMIT University, Melbourne, VIC 3000, Australia}
\affiliation{Centre for Quantum Software and Information, Faculty of Engineering and
Information Technology, University of Technology Sydney, NSW 2007, Australia}

\author{Oleg Titov}
\affiliation{Phase $\&$ Rate, Canberra, Australia}

\author{Miko\l aj K. Schmidt}
\affiliation{School of Mathematical and Physical Sciences, Macquarie University, NSW 2109, Australia}

\author{Benjamin Pope}
\affiliation{School of Mathematical and Physical Sciences, Macquarie University, NSW 2109, Australia}

\author{Gavin K. Brennen}
\affiliation{School of Mathematical and Physical Sciences, Macquarie University, NSW 2109, Australia}

\author{Daniel Oi}
\affiliation{SUPA Department of Physics, University of Strathclyde, Glasgow G4 0NG, United Kingdom}

\author{Pieter Kok}
\affiliation{School of Mathematical and Physical Sciences, The University of Sheffield, Sheffield, S3 7RH, United Kingdom.}

\date{\today}

\begin{abstract}
Optical Very Long Baseline Interferometry (VLBI) offers the potential for unprecedented angular resolution in both astronomical imaging and geodesy measurements. Classical approaches face limitations due to photon loss, background noise, and their need for dynamical delay lines over large distances.
This review surveys recent developments in quantum-enabled optical VLBI that address these challenges using entanglement-assisted protocols, quantum memory storage, and nonlocal measurement techniques. While its application to astronomy is well known, we also examine how these techniques may be extended to geodesy -- specifically, the monitoring of Earth’s rotation.
 Particular attention is given to quantum-enhanced telescope architectures, including repeater-based long-baseline interferometry and quantum error-corrected encoding schemes, which offer a pathway toward high-fidelity optical VLBI.
To aid the discussion, we also compare specifications for key enabling technologies to current state-of-the-art experimental components.
By integrating quantum technologies, future interferometric networks may achieve diffraction-limited imaging at optical and near-infrared wavelengths, surpassing the constraints of classical techniques and enabling new precision tests of astrophysical and fundamental physics phenomena.

\end{abstract}

\maketitle

\tableofcontents

\section{Introduction}

The performance of an imaging system or interferometer is limited by diffraction: the resolution is proportional to its aperture and inversely proportional to the wavelength $\lambda$ of the light.  Together, these place a fundamental limit on how well one can image the objects of interest. This is especially true for light originating in astronomical objects, where we cannot illuminate the objects of interest, and we are restricted to analysing the light that reaches us. 

2019 saw a dramatic demonstration of the image resolution as limited by the size of the aperture of the imaging system. The Event Horizon Telescope collaboration combined telescopes from all over the world into a single imaging system with an aperture roughly the size of the Earth. This collective effort resulted in the first-ever image of a black hole and its surroundings. 

The underlying physics of the black hole imaging setup is purely classical:
the array operates at radio frequencies, where both the amplitude and phase of the received electromagnetic field can be directly measured. The data is then post-processed to reconstruct an image. 
However, direct measurement is not possible in optics, because even the fastest electronics cannot directly measure the oscillations of the electric field at optical frequencies. Nevertheless, if we can find alternate solutions, we can potentially gain 3-5 orders of magnitude improvement in resolution. 
This unlocks possibilities for imaging astrophysical structures otherwise inaccessible to our current imaging systems.

Several challenges hinder the progress in building large-baseline optical interferometers, including the weight of the instrument itself, photon loss, background noise, and the requirements for long dynamical delay lines. These factors ultimately limit the achievable distance between telescope sites. Quantum technologies can help bypass transmission losses: using quantum memories and entanglement, we can replace the direct optical links, allowing for much larger distances. 
 If deployed on the Moon, we can exploit the natural vacuum as a beam guide for long-baseline operation without atmospheric distortion.

Among the key scientific applications of optical VLBI, two stand out. 
First, VLBI offers a path toward high angular resolution astronomical imaging.
 Second, and less commonly considered, optical VLBI can achieve unprecedented spatial and temporal resolution in geodetic metrology, such as monitoring Earth’s rotation, tectonic motion, and lunar libration.

Optical VLBI refers to interferometric regimes where real-time compensation of optical path lengths is not feasible, necessitating the use of closure phases and post-processed visibilities~\cite{monnier2003}. Instruments like CHARA~\cite{chara} and VLTI~\cite{vlti} operate in this regime, and future concepts such as the Big Fringe Telescope~\cite{bft} aim to extend it. Our work does not propose optical VLBI as a new idea, but rather explores how quantum-enabled protocols might overcome key technical limitations, particularly in photon loss and scalability.

These approaches complement classical strategies such as heterodyne~\cite{townes98} and intensity interferometry~\cite{nunez12}.
However, the quantum approach uniquely targets large-baseline, high-precision measurements and scalable architecture without the need for formation flying or complex delay-line infrastructure.

Today’s most advanced interferometric imaging systems operate at radio and microwave wavelengths. Instruments such as the Event Horizon Telescope (EHT) achieve resolutions of tens of microarcseconds by combining a maximum baseline of $\sim$12,000~km (roughly the Earth's diameter) with a typical wavelength of 3~mm. Shifting to optical wavelengths can potentially increase resolution by several orders of magnitude. An optical system operating at 600 nm could match the EHT with only a 2.4 km baseline; at 1550 nm, 6.2 km would suffice.

To place the present approach in context, it is useful to recall the historical development of astronomical interferometry and its connection to quantum correlation measurements.
The Hanbury Brown–Twiss (HBT) interferometry \cite{brown1956correlation} was among the first quantum optical effects in astronomy. By correlating photon fluxes, HBT showed that spatial information about a thermal source can be recovered through second-order coherence, even without physically combining light beams. Nevertheless, amplitude interferometry—which includes conventional optical telescopes and VLBI—offers a fundamentally more powerful approach, as it measures first-order coherence and thus accesses phase information.
While intensity interferometry can support large effective apertures without an optical link--and under select constraints may be as powerful as, or more than amplitude interferometry \cite{bojer2022quantitative}, it tends to suffer from low signals due to the small average photon number per optical mode. Amplitude interferometry, by contrast, achieves the ultimate resolution permitted by optical coherence.

\color{black}
\section{Astronomical imaging}

The advantage of a quantum approach is not merely improved angular resolution based on extending the baseline, but achieving the best precision in complex visibility measurements. Many science cases—such as detecting exoplanets, imaging stellar surfaces, or tracking relativistic orbital precession—rely on high signal-to-noise measurements rather than resolving power alone. 

This section discusses some science cases for using optical VLBI, including imaging of exoplanets, tracking the stars around black holes, and measuring separations of double-star systems.

\subsubsection{Imaging exoplanets}
\begin{table*}[t]
    \centering
    \begin{tabular}{|l|c|c|c|}
        \hline
        \textbf{Exoplanet} & \textbf{Semi-Major Axis (AU)} & \textbf{Distance (pc)} & \textbf{Separation (arcseconds)} \\
        \hline
        Proxima Centauri b  & 0.0485  & 1.301  & 0.037  \\
        Barnard's Star b    & 0.0406  & 1.834  & 0.022  \\
        Ross 128 b          & 0.0496  & 3.374  & 0.015  \\
        Luyten's Star b     & 0.0911  & 3.785  & 0.024  \\
        Wolf 1061 c         & 0.084   & 4.287  & 0.020  \\
        \hline
    \end{tabular}
    \caption{    \label{tab:exoplanet_angular_separation} Angular separation of nearby exoplanets from their parent stars.}
\end{table*}

Imaging exoplanets is a challenging task: they are both extremely small in angular size and orders of magnitude fainter than their host stars. Direct imaging of these objects requires:
\begin{itemize}\itemsep=0pt
\item high angular resolution to spatially separate the planet from the star.
\item high visibility (contrast) to detect the planet’s faint emission against the much brighter parent star.
\end{itemize}
Optical VLBI offers a potential solution to both challenges: at a wavelength of 600~nm and a baseline of 100~km, 
the achievable angular resolution is approximately 1.24~microarcseconds ($\mu$as)—well beyond the reach of existing instruments, including the Event Horizon Telescope. At this resolution, it becomes conceivable to directly image features in the atmospheres of nearby exoplanets, opening new possibilities for studying their composition, structure, and potentially even weather systems. We show the separation of the star from the exoplanet in Table~\ref{tab:exoplanet_angular_separation} for reference.

\subsubsection{Tracking stars orbiting black holes}

Studying the orbits of stars in the strong gravitational fields near black holes provides a unique opportunity to test Einstein’s General Relativity (GR) in regimes that cannot be accessed in laboratory experiments. In Newtonian gravity, the orbit of a test particle around a massive body is a closed ellipse, as described by Kepler’s first law. However, GR predicts that the curvature of spacetime near massive objects causes the orbit to precess — a phenomenon known as Schwarzschild precession. The precession angle per orbit is given by
\begin{align}
\Delta \Phi = 6\pi\frac{GM}{a (1-e^2) c^2},
\end{align}
where $G$ is the gravitational constant, $M$ is the mass of the central object, $a$ is the semi-major axis, $e$ is the eccentricity of the orbit, and $c$ is the speed of light. This effect, which is first order in $(v/c)^2 \sim 10^{-4}$, has been experimentally confirmed~\cite{genzel2024experimental}. 

\begin{figure}[h]
\includegraphics[trim = 0cm 4cm 0cm 1.5cm, clip, width=0.9\linewidth]{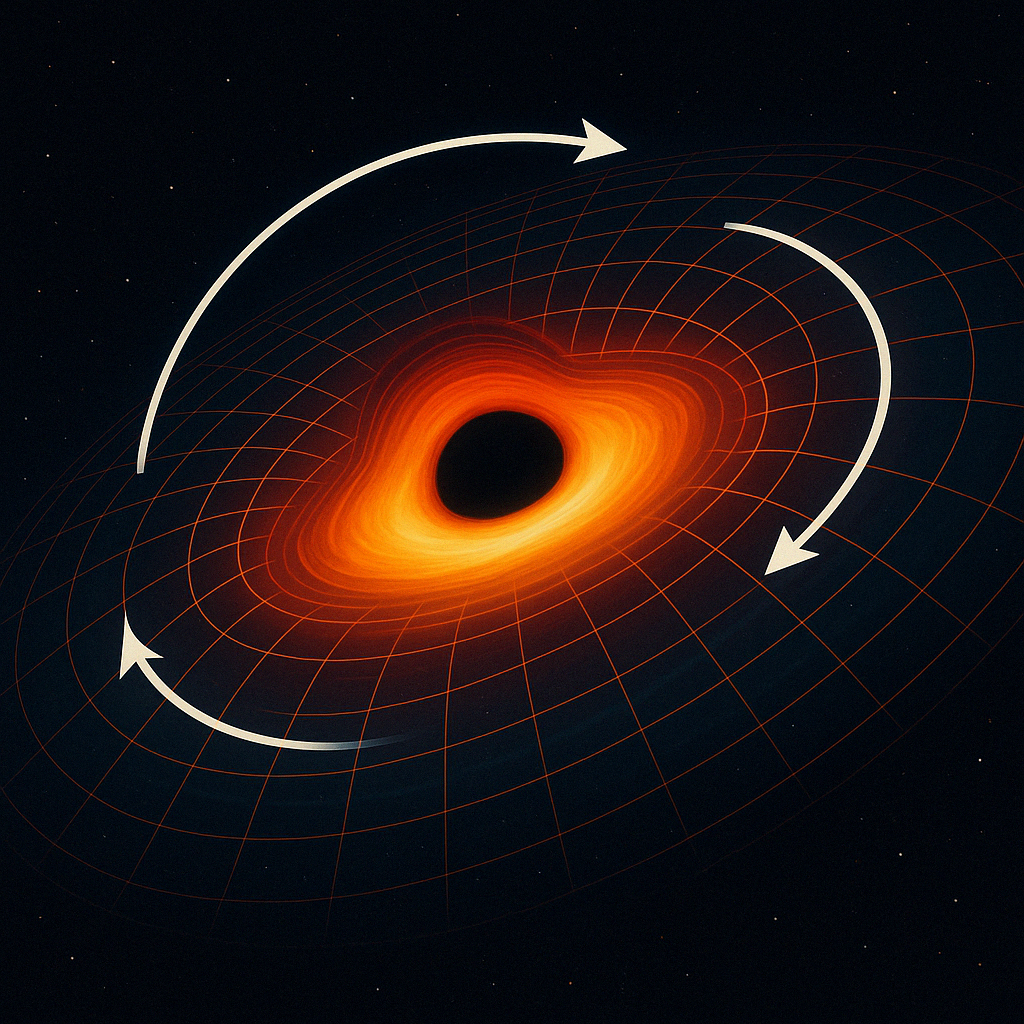} 
\caption{\label{fig:lense_thirring} The Lense-Thirring effect: a star's orbit precesses due to the frame-dragging caused by a rotating massive object, such as a spinning black hole.}
\end{figure}

In cases where the black hole is spinning, there is an additional relativistic effect known as frame dragging, or the Lense–Thirring effect: the rotation of the black hole causes spacetime itself to twist (Fig.~\ref{fig:lense_thirring}), subtly altering the motion of nearby stars. The corresponding precession per orbit is given by
\begin{align}
\Delta \Phi_\text{LT} = 2\chi \left( \frac{R_S}{a(1-e^2)}\right)^{1.5},
\end{align}
where $\chi$ is the dimensionless spin parameter of the black hole (ranging from 0 to 1), and $R_S$ is the Schwarzschild radius. For the star S2, which orbits the supermassive black hole near the Galactic center, the Lense–Thirring precession is of the order of 0.05 arcminutes per orbit — about 240 times smaller than the Schwarzschild precession~\cite{genzel2024experimental}. Resolving such small effects requires extremely high angular resolution. The GRAVITY collaboration (see, e.g., Refs.~\cite{abuter2017first,amorim2019test,abuter2021constraining}) currently achieves a near-infrared resolution of 3 milli-arcseconds and astrometric precision of 10–100~$\upmu$as, and a baseline of 130~m. To detect the Lense–Thirring precession at comparable wavelengths (near infrared), a baseline of at least \mbox{$130 \times 240 = 32$~km} would be required.

\subsubsection{Measuring the stellar separation of A. Centauri.}
Alpha Centauri consists of two Sun-like stars (Alpha Centauri A and B) in a close binary orbit.
The angular separation of the two main stars in the Alpha Centauri binary system — Alpha Centauri A and Alpha Centauri B — varies over their 79.91-year orbit. It ranges from about 2 arcseconds at closest approach (periastron) to about 22 arcseconds at their farthest (apastron).
By studying their motion, we can precisely measure their masses using Kepler’s laws and Newtonian mechanics.  This helps in stellar modelling, as mass is a key determinant of a star’s lifetime and properties. Tracking deviations from expected orbital motion can reveal exoplanets, especially Earth-like planets in the habitable zone.

\section{Geodesy}

\begin{figure}[h!]
\includegraphics[trim = 0cm 0.0cm 0cm 0cm, clip, width=0.9\linewidth]{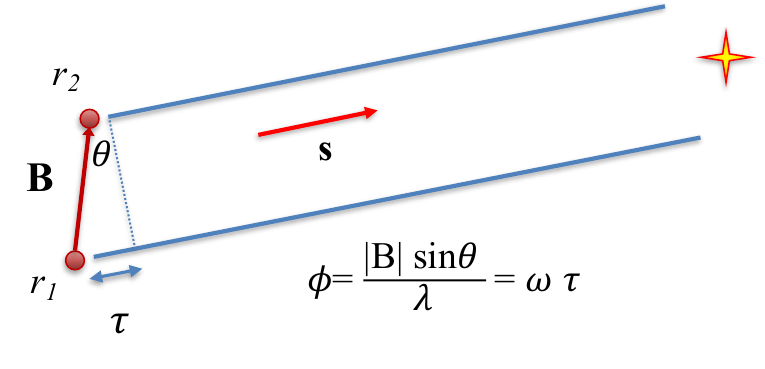} 
 \caption{\label{fig:geometry_geodesy} Geometry for using VLBI for geodesy. The position of the reference star is assumed to be fixed, and the goal is to measure the baseline vector $\bm{B}$.}
\end{figure}

Geodesy uses precise measurements of Earth's shape, orientation, and gravity field—often through VLBI and satellite techniques—to monitor tectonic motion, the Earth's rotation, and maintain global reference frames.
VLBI measures the time difference in the arrival of signals from a distant quasar at two Earth-based antennas. The basic geometric delay $\tau$ measured by two radio telescopes with Earth-based positions $r_{1}$ and $r_{2}$ is given by 
\begin{align}
\tau  = -\frac{\vec B \cdot s}{c}
\end{align}
where $\vec B$ is the baseline vector $\vec B= r_{2}-r_{1}$, $~s$ is the radio source vector, and $c$ is the speed of light, see Fig~\ref{fig:geometry_geodesy}.

Using large numbers of time difference measurements from many quasars observed with a global network of antennas, VLBI determines the inertial reference frame defined by the quasars and simultaneously. Since the antennas are fixed, their locations track the instantaneous orientation of the Earth. Relative changes in the antenna locations from a series of measurements indicate tectonic plate motion, regional deformation, and local uplift or subsidence.

When a quasar emits EM waves, they arrive at different telescopes at slightly different times due to Earth's shape and rotation.
  The exact time delay between telescopes depends on the Earth’s rotation state at that moment.
By comparing the signals and modelling their delay based on Earth's motion, we can infer small variations in rotation speed and axis position.

Geodetic VLBI plays a crucial role in geodesy. Currently, the tectonic motion is measured with an accuracy up to 0.1 mm/year, though the seasonal variations of the individual radio telescopes' positions are also monitored~\cite{Lovell2013}. Some of the radio telescopes were affected by strong earthquakes (Japan, 2010; Chile, 2011)~\cite{Yin2013}. Hence, the co-seismic and post-seismic displacement can be monitored. A global catastrophe, like the Sumatra Boxing Day earthquake on 26-Dec-2004, may cause a global shift in positions of all geodetic sites around the Earth with an amplitude of $\sim$1~mm. 
In addition, geodetic VLBI studies the post-glacial uplift in Fenno-Scandinavia, the northern part of North America, and Antarctica. The global melting of glaciers causes a dramatic change in the immediate positions of radio telescopes, and changes the shape of the planet due to the secular shift of the inertia momentum.

\subsection{Earth rotation}

\subsubsection{Time Scale}

\begin{figure}[h!]
\includegraphics[ width=1.0\linewidth]{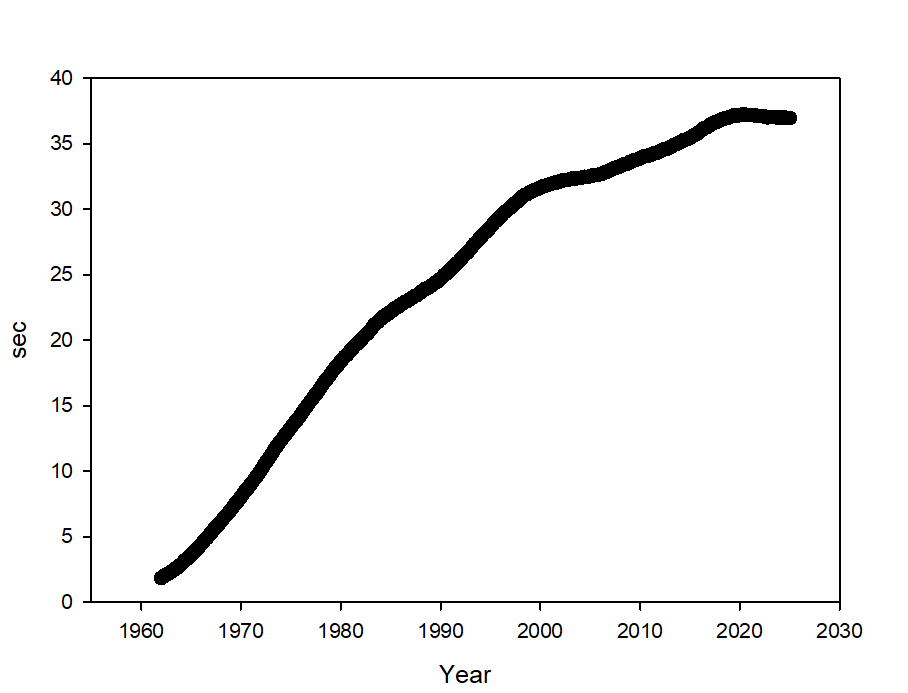} 
\caption{\label{fig:lod}Increase in the length of day since 1962 \cite{malkin2024should}.}
\end{figure}

One of the primary applications of geodetic VLBI is the daily monitoring of UT1 (Universal Time 1), which tracks Earth's rotational time. Earth's rotation is not constant due to interactions with atmospheric winds, ocean currents, and the fluid outer core. These variations necessitate daily UT1 updates to account for fluctuations in the length of the day. The length of day (LOD) has increased by 37 sec since 1962 (Fig~\ref{fig:lod}).

Another key application is long-term drift monitoring, which allows scientists to detect millisecond-scale changes in Earth's rotation over decades. Such long-term observations reveal that Earth's rotation is gradually slowing, with an average deceleration of 1.7 milliseconds per century due to tidal friction from the Moon’s gravitational influence. 

The leap second problem arises from the need to synchronise two fundamentally different timekeeping systems: Atomic Time (TAI/UTC) and Astronomical Time (UT1). Atomic Time, based on hydrogen and caesium atomic clocks, is highly stable and defines the SI second with unparalleled precision. In contrast, Astronomical Time (UT1) is tied to Earth's rotation, which varies unpredictably. As a result, UT1 drifts relative to atomic time, necessitating the periodic introduction of leap seconds to maintain synchronisation.
In the 1990s, the leap second was introduced almost every year. Although the Earth's rotation deceleration nowadays is not so high and the leap second is introduced rarely (the last time was in 2015), the situation may change again and needs to be monitored continually~\cite{Nelson2001}. 

Leap seconds introduce significant challenges for computing, satellite navigation, and global timekeeping systems. 
Since leap seconds occur irregularly and unpredictably, they disrupt time-dependent operations, requiring software updates and manual interventions. To mitigate these issues, discussions are ongoing regarding the potential elimination or redefinition of leap seconds in the future~\cite{Agnew2024}. Optical VLBI offers a substantial improvement in the temporal resolution of UT1 measurements, enabling the detection of changes in Earth's rotation on shorter and more precise timescales.

\subsubsection{Polar motion}

Polar motion is the motion of the instantaneous Earth rotation axis around the geographic pole. The regular motion consists of two major components, namely the annual rotation with a period of approximately 365 days and the Chandler wobble with a period of approximately 430 days. The Chandler wobble is a free rotation effect with variable period and amplitude. Currently, it has disappeared from the total Earth's pole motion (presumably temporarily) for unknown reasons. VLBI helps in tracking polar motion, as the Earth's rotational axis drifts slightly due to mass redistributions, such as the melting of ice sheets and shifts in the planet’s mantle~\cite{bizouard2013},~\cite{Karbon2017}. Optical VLBI can enhance our ability to track polar motion with finer spatial resolution, potentially improving sensitivity to small-scale mass redistributions.

\subsubsection{Precession and nutation}

The Earth's axis draws a pattern in the sky. This pattern splits into two effects, precession and nutation. Precession has a period of 26,000 years and traces out a cone with an opening angle of $23.5^\circ$. The slow motion of the axis leads to a displacement of all apparent positions of the celestial objects of 50 arcsec every year. Nutation is a smaller signal with a major amplitude of about 9 arcsec and a period of 18.6 years, caused by the gravitational attraction of the Moon. Analysis of the numerous subtle effects in the nutation, e.g. free core nutation (FCN) and the free inner core nutation (FICN), helps to gain insights into the Earth's deep interior dynamics and composition~\cite{Shirai2001,Mathews2002,Lambert2007}.

\subsection{Astrometry}

Geodetic VLBI produces the International Celestial Reference Frame (ICRF) by measuring the positions of several thousand strong radio sources. The current International Celestial Reference Frame (ICRF3) adopted by the International Astronomical Union is based on the positions of 4536 radio sources observed between 1979 and 2018, and 303 radio sources define the fixed position of the Earth's fundamental rotation axis with an uncertainty of 30~$\upmu$as~\cite{charlot2020third}. Detection errors of the individual positions of radio sources vary considerably due to an uneven number of observations, from 6~$\upmu$as to a few mas.
Recent observations after 2018 increased the number of monitored radio sources to approximately 5600~\cite{Krasna2023}.

The main ICRF3 reference catalogue is based on observations at a frequency of 8.4 GHz (the X band). In parallel, there are additional catalogues for frequencies of 24~GHz and 32~GHz with fewer objects. These catalogues are important for tracking deep space missions~\cite{Jacobs2009}.

\vspace{5mm}
The preceding sections outlined the observational motivations for optical interferometry in astronomy and geodesy. We now introduce the quantum layer—the set of quantum resources and protocols that enable, in effect, direct interferometry on the stellar signal. The goal of this layer is to preserve  the optical coherence of the incoming light. In conventional direct interferometry, beam splitters act as non-local measurements that combine optical fields from separate apertures; however, in large-baseline configurations, bringing the light physically together is impractical. Pre-distributed entanglement effectively allows this non-local interference to be implemented virtually, reproducing the same measurement outcomes without optical co-location. In this sense, the quantum layer operationally extends classical interferometry, which is known to be optimal for parameter estimation in the linear-optical regime \cite{PhysRevLett.124.080503}.

\section{The model and parameter estimation}

High-resolution astronomical imaging and high-precision geodesy are fundamentally linked by their reliance on the same core technology: interferometry. In astronomical imaging, interferometers are used to resolve fine spatial structures by coherently combining light collected from a distant source. In geodesy, the principle is inverted: the celestial source serves as a fixed reference, and the interferometric measurement is used to track the motion of the Earth itself. 
Given the shared technology, we adopt a common physical model for the stellar source  and focus on the well-understood two-mode case, where quantum Fisher information serves as a natural metric for comparison.

Consider a single frequency band whose two-mode continuous variable quantum description is given by a mean vector and covariance matrix of the form~\cite{mandel1995optical,Pearce2017optimal,PhysRevLett.129.210502},

\begin{align} 
\label{eq:star}
\vec r_\star & := (
\begin{array}{cccc}
0 & 0 & 0 &0
\end{array})^T,
\\
\mat \sigma_\star & := 
\left(
\begin{array}{cccc}
 \epsilon +1 & 0 & \gamma  \epsilon  \cos \phi   & -\gamma  \epsilon  \sin \phi  \\
          0  & \epsilon +1                        & \gamma  \epsilon  \sin \phi  &  \gamma  \epsilon  \cos \phi \\
  \gamma  \epsilon  \cos \phi & \gamma  \epsilon  \sin \phi  & \epsilon +1 & 0  \\
 -\gamma  \epsilon  \sin \phi  & \gamma  \epsilon  \cos \phi & 0  & \epsilon +1
\end{array}
\right),
\label{eq:star2}
\end{align}
where we have used the quadrature ordering $(q_A, p_A, q_B, p_B)$, with subscripts referring to Alice~($A$) and Bob ($B$).
In the limit that $\epsilon \ll1$, which is typically true for astronomy, we can describe the state by the density matrix,
\begin{align}\label{eq:coupled}
\rho_{\star} \approx &(1-\epsilon)\ket{\text{vac},\text{vac}}\bra{\text{vac},\text{vac}}_{A B} +  \nn
&\epsilon \left(\frac{1+\gamma}{2}\right) \ket{\psi_+^\phi}\bra{\psi_+^\phi} + 
\epsilon \left(\frac{1-\gamma}{2}\right)\ket{\psi_-^\phi}\bra{\psi_-^\phi} 
\end{align}
\noindent where $\ket{\psi_\pm^\phi} = (\ket{1}_{A} \ket{\text{vac}}_{B} \pm e^{i\phi} \ket{\text{vac}}_{A} \ket{1}_{B})/\sqrt{2} $. 
Here, the subscript $\ket{1}$ denotes a single photon Fock state and $\ket{\text{vac}}$ denotes the vacuum.

For imaging, the parameters of interest are $\phi$ and $\gamma$, where $\phi \in [0 ,2\pi )$ is related to the location of the sources, and $\gamma \in [0,1]$ is proportional to the Fourier transform of the intensity distribution via the van Cittert-Zernike theorem~\cite{mandel1995optical}.

The ultimate precision in parameter estimation 
is specified by the quantum Cram\'er--Rao bound~\cite{caves,caves1} (see also~\cite{giovannetti2011advances,giovannetti2006quantum}).
For estimation of a parameter $\theta$ encoded into a quantum state~$ \hat \rho_\theta$, the Cram\'er--Rao bound sets a lower bound on the variance $(\Delta \theta)^2 = \langle \theta^2 \rangle - \langle \theta \rangle^2$ of any unbiased estimator~$\theta$.
For unbiased estimators, the quantum Cramér--Rao bound establishes that 
\begin{align} \label{eq:var}
 (\Delta  \theta) ^2 \geqslant  \frac{1}{N  J_\theta( \hat \rho_\theta)} \, ,
\end{align}
$N$ is the number of copies of $\hat \rho_\theta$ used and $J_\theta$ is the quantum Fisher information (QFI) associated with the state $\hat\rho_\theta$.

If there are multiple parameters we want to estimate, where $\vec\theta = (\theta_1, \theta_2, \dots)$, we can define a \emph{QFI matrix} $\mat{J}$.
The matrix elements are given by
\begin{align}
J_{jk}
:= \frac{1}{2}\text{Tr}[\hat\rho_{\vec{\theta}}(\hat L_j \hat L_k + \hat L_k \hat L_j)],
\end{align}
 where $\hat L_j$ is the symmetric logarithmic derivative with respect to $\theta_j$~\cite{paris2009quantum}. 
The inverse of the QFI matrix provides a lower bound on the covariance matrix
 $\big[\Cov(\vec\theta) \big]_{jk} = \braket{\theta_j \theta_k } - \braket{\theta_j}\braket{\theta_k}$,
 \begin{align}
 \Cov(\vec\theta) \geq \frac{1}{N } \vec J^{-1}.
\end{align}
The QFI matrix elements for the incoming stellar state are \cite{huang2024limited}
\begin{subequations}
\begin{align} 
\label{eq:ideal_qfi_matrix}
J_{\phi } 
&=\frac{2 \gamma ^2 \epsilon }{2+ \epsilon(1-\gamma ^2) } ,
\\
J_{\gamma } &= \frac{2 \epsilon  \left(2 + \epsilon + \epsilon \gamma ^2 \right)}{\left(1-\gamma ^2\right) \left(4+ 4\epsilon + \epsilon ^2 \left(1-\gamma ^2\right) \right)} ,
\\
J_{\phi\gamma} &= 0.
\end{align}
\end{subequations}

If we take the trace norm of the QFI matrix which is equal to $J_\phi+J_\gamma$
this can be significantly larger than $\epsilon$, and that care should be taken to discuss the sum $J_\phi$ of $J_\gamma $, because $J_\phi/\epsilon \leq 1$, whereas 
$J_\gamma/\epsilon$ can, in principle, approach infinity.

For geodesy, the task is to measure the distance between two telescope sites, given that a stable astronomical source is at a known location (Fig.~\ref{fig:geometry_geodesy}). The most accurate way of measuring this distance is by estimating the optical path length difference between the two sites, which manifests as a phase shift  $\phi$.

\begin{table*}[t]
\begin{center}
\renewcommand{\arraystretch}{1.5}
\setlength{\tabcolsep}{18pt}
\begin{tabular}{ |l |c |c| }\hline
  Requirements  & Direct interferometry   & quantum-enabled \\ 
  \hline
  Phase reference and phase stabilisation     &  \tick       &  \tick     \\
 Variable delay lines                        &  \tick    &           \\
 (Fast) Time-bin multiplexing                   &            &   \tick  \\ 
 High cooperativity optical cavities              &            &   \tick  \\ 
 High-fidelity light-matter coupling/gates &        &    \tick  \\ 
Entanglement distribution/ distillation &             &   \tick   \\
Long-lived quantum memories                            &             &   \tick       \\
Multi-system quantum gates                 &               & \tick    \\ 
\hline
\end{tabular}
\end{center}
\caption{\label{tab:qtech} The technology required for direct interferometry vs quantum VLBI's. }
\end{table*}

Now, since the source is a point source, the quantum state shared between the two stations can be represented as
\begin{align}
\label{eq:geodesy_point}
\ket{\psi} &= \frac{1}{\sqrt2} (\ket{01}_{AB} + e^{i\phi}\ket{10}_{AB}), \nn
\phi       &= \omega \tau = \frac{B \sin\theta}{\lambda}.
\end{align}
%
Note that, although Eq.~\eqref{eq:geodesy_point} (and Eq.~\eqref{eq:ground_photon}) represents a single photon distributed between two spatial modes, such a state is commonly regarded as an entangled state of the electromagnetic field—entangled between the optical modes and photon-number degrees of freedom. In this sense, it simultaneously describes a single-photon superposition and an entangled state. For the original extensive discussion about this point, see Refs.~\cite{hardy1994nonlocality,hardy1995hardy,PhysRevLett.76.2205}. 
\color{black}

From quantum metrology theory, we know that the QFI of $\phi $ is equal to 1; i.e. $\Delta \phi \approx 1$. Assume $\theta$ and $\lambda$ to be fixed, this means
\begin{align}
\Delta B \geq \frac{\sin\theta}{\lambda} \frac{\Delta\phi }{\sqrt n},
\label{eq:uncertainty_B}
\end{align}
where $n$ represents the total number of detected photons (rather than the instantaneous photon number per detection event). Photon-number-resolving detectors are not strictly required here; this bound can be obtained statistically from $n$ repeated single-photon detection events.
\color{black}

Eq.~\eqref{eq:uncertainty_B} highlights a key advantage of optical wavelengths: because baseline uncertainty scales inversely with $\lambda$, shorter wavelengths significantly improve geodetic precision for the same photon count and geometry.
Having established the scaling advantages at optical wavelengths, we next examine the technological layers that make it possible to implement these quantum-enabled measurements across large baselines.
\color{black}

\section{Technological requirements of Optical VLBI}
The technological components described in this section collectively realise the quantum layer introduced earlier, whose purpose is to enable, in principle, the non-local measurement corresponding to a beam-splitter operation between distant telescopes. This allows direct interferometry on the stellar signal without physically bringing the light together. 
Before developing the formal model, we note that the “quantum gate” layer discussed below refers not to general-purpose computation, but to part of the measurement process that allow us to extract the parameters of interest. Pre-distributed entanglement reproduces the non-local interference central to classical interferometry.
\color{black}
Realising quantum-enabled optical VLBI involves a multi-layered technological framework that integrates several key components (see Table~\ref{tab:qtech}):

\begin{enumerate}
    \item The entire setup needs to be phase-stabilised to within a small fraction of the wavelength. A shared phase reference is needed.

    \item Starlight is collected and time-multiplexed into different bins (see Fig.~\ref{f:time_bins}).

    \item Photons in each time bin are coupled to quantum memories, typically via an optical cavity, to enhance the otherwise weak light-atom interaction.
    
    \item Non-local measurement is performed between quantum memories of the same time bin between the telescope sites. Per time bin, we need at least:
        \begin{enumerate}
            \item One single-qubit gate to absorb the photon into the memory,
            \item One set of (logical) multi-qubit gates to encode the memory state onto a quantum error correcting code,
            \item One set of (logical) gates to perform parity checks,
            \item One set of logical single-qubit gates to perform the measurements. This one is non-Clifford in general.
        \end{enumerate}
    \end{enumerate}
Given that each shared time-bin contains a state of mean photon number $\epsilon$, the logical gate error rate needs to be much smaller than $\epsilon$. Below, we outline these core requirements and the current state of the art.

\subsection{Phase stabilisation}

Maintaining phase coherence across distant telescope sites is critical for interferometry, as the entire system must be stabilised to within a small fraction of the optical wavelength. A shared optical phase reference is therefore necessary. 

Recent twin-field QKD experiments reported 98\%  interference visibility over 200 km at 1550 nm~\cite{PhysRevX.9.021046}, suggesting that phase stabilisation over VLBI-scale baselines is indeed technically feasible. A quantitative analysis on the shared phase reference is given in Ref.~\cite{zhang2025criteria}.

\subsection{Time-bin multiplexing}

\begin{figure}[t!]
\includegraphics[trim = 0cm 0.0cm 0cm 0cm, clip, width=0.7\linewidth]{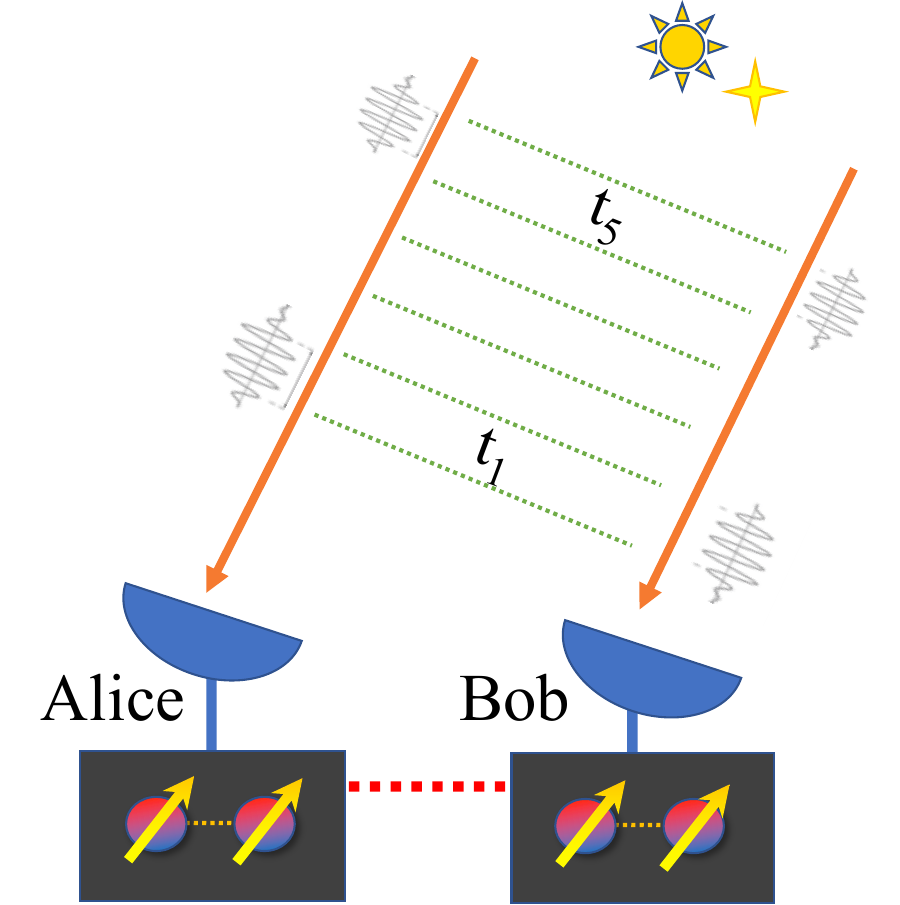} 
 \caption{\label{f:time_bins} The coherence time of the received signal depends inversely on the bandwidth. Signals with a path length difference larger than the coherence time will not interfere with one another. In this illustration, there is one photon shared non-locally between the two stations in bin $t_1$, and in $t_5$: the photon in $t_1$ will not interfere with the one in $t_5$. }
\end{figure}

Starlight must be parsed into discrete time bins, with the temporal width of each bin defined by the coherence time of the incoming thermal light (see Fig~\ref{f:time_bins}), because signal separated by longer than the coherence time will not interfere. 
The coherence time of the light is inversely proportional to its bandwidth $\Delta \nu$ ~\cite{mandel1995optical},
\begin{align}\label{eq:coherence_length}
\Delta t \sim \frac{1}{\Delta \nu}.
\end{align}
Therefore, as the bandwidth increases, the coherence time—and therefore the required time-bin size—decreases proportionally.
As an example, for light generated by thermal sources with a narrow spectral width ($\Delta \nu \sim 100$ MHz), the corresponding coherence time is of order $10$ ns, and coherence length is of order 3~m (we note that the bandwidth of the stellar sources is typically quoted in either frequency or wavelength, and provide examples of conversion in Appendix~\ref{app:a}).

To realise time-binning, we would require fast optical switches. State-of-the-art switches operate at rates of up to 38~MHz with approximately 80\% transmission efficiency~\cite{munzberg2022fast}, corresponding to a minimum time-bin size of roughly \( 26 \) ns. At a wavelength of 1550~nm, a spectral bandwidth of 1~nm corresponds to a temporal coherence time of about 8~ps---nearly 3000 times shorter than the achievable switching interval.  
A recent hybrid silicon–lithium-niobate devices have demonstrated sub-nanosecond ($\approx$ 1 GHz) switching speeds \cite{Gao2019}, showing excellent potential for fast electro-optic control, though further reduction of the present $\sim$6 dB insertion losses ($25\%$ transmission) will be necessary for large-scale quantum-enabled VLBI implementations.\color{black}

One possible strategy to alleviate this constraint is to implement frequency multiplexing, dividing the incoming bandwidth into, for example, 100 discrete spectral channels. This effectively increases the usable time-bin size per channel, easing the demands on switching speed. Combined with recent advances in fast optical switches---particularly those based on barium titanate (BTO) platforms developed by groups such as PsiQuantum \cite{psiquantum2025manufacturable}---this suggests that an {order-of-magnitude improvement} in switching speed is realistic in the near term. Such improvements would significantly narrow the gap between current switching capabilities and the time-resolution requirements of optical VLBI.

For frequency multiplexing, optical demultiplexers—such as dense wavelength division multiplexers
se diffraction Bragg gratings, etalons, typically provide passbands of tens to 100's of MHz \cite{fs_dwdm_2023}, and high-resolution astronomical spectrographs, such as HARPS \cite{eso_harps_overview}, have frequency resolution of order GHz. If we would like demultiplexing schemes with ~1 MHz bandwidth in practice, significant technological advances needed for such filters.\color{black}

\subsection{Implementing quantum nodes with optical cavities}

\begin{figure}[h]
\centering
    \includegraphics[width=.5\textwidth]{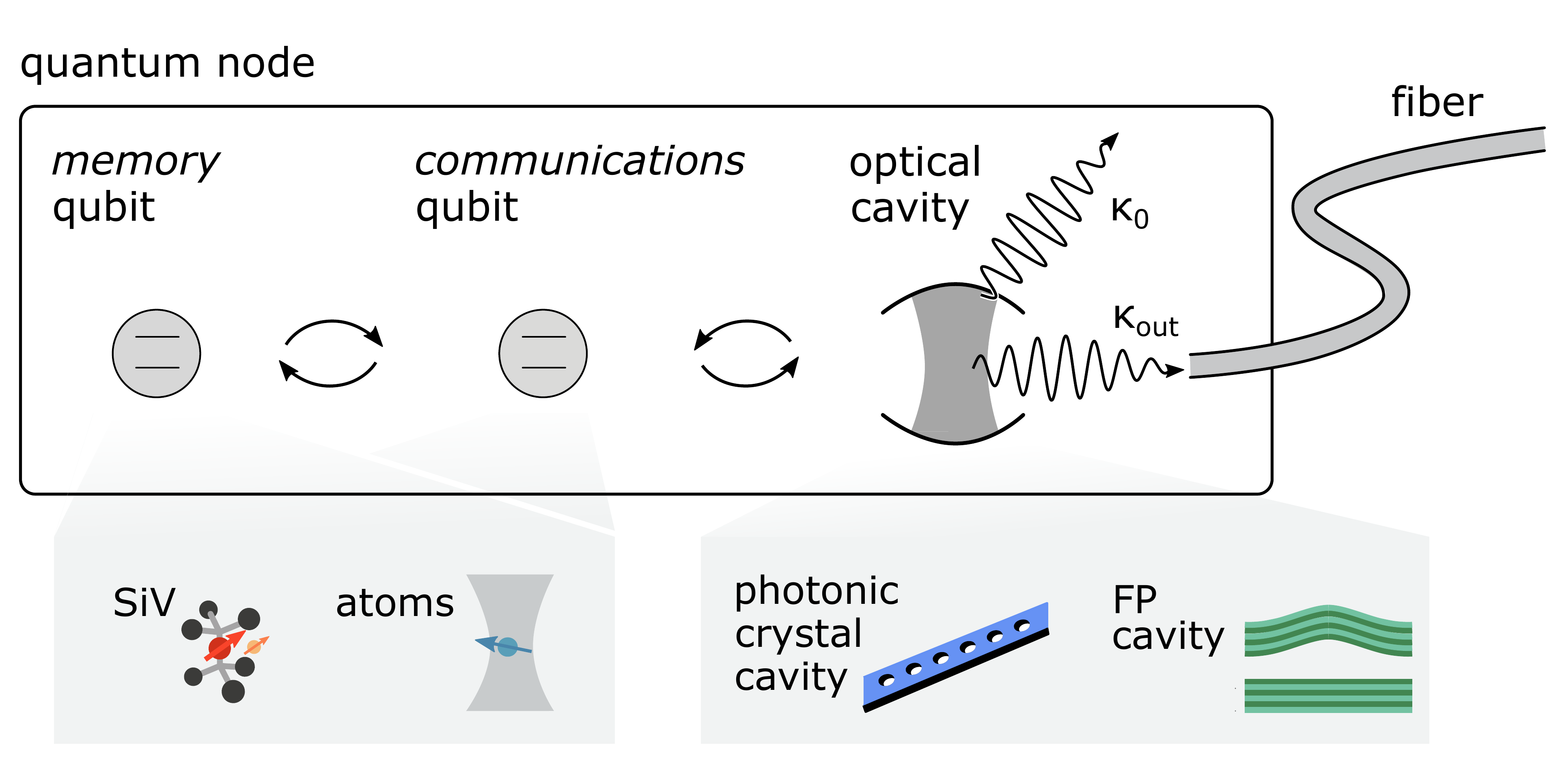}
\caption{\label{fig:qnodes} Schematic of a quantum node architecture with memory and communication qubits overcoupled to the communication bus -- a fibre -- enhanced by an optical cavity.}
\end{figure}

To coherently transfer information from an external optical field into an atom, we can make use of the strong light-matter interaction enabled by optical cavities (Fig.~\ref{fig:qnodes}). In order to do this efficiently, it is necessary to implement a unitary transformation of all of the incident modes to the target cavity mode, before the light is coupled into \textit{memory} qubits in a quantum node. 
Toward this task, it has been shown~\cite{Stone:21} that by exploiting the spatial dependence of impedance mismatch in a pair of misaligned Fabry-Perot resonators, high efficiency spatial mode conversion of optical photons is possible. It was demonstrated that light from a Hermite-Gauss HG$_{00}$ mode can be converted into an arbitrary target HG$_{m0}$ mode with conversion/transmission efficiency $>75\%$, by varying the length of a Fabry-Perot resonator over a few nanometers.


The many platforms for implementing quantum nodes in physical systems are built around {memory} qubits, which store the information in long-lived nuclear spin, dark or metastable states, perfectly isolated from the environment. To control and read out that memory, quantum nodes would typically use a direct, controlled coupling to a cavity, or mediate it via a short-lived \textit{communication} qubit~\cite{RevModPhys.94.041003}. The cavity is used to boost the intrinsically low efficiency of these processes and enhance the in and out-coupling from the memory to the fibre.

 To this end, the cavity operates in the over coupled regime, meaning that its coupling to the fibre mode $\kappa_\text{out}$ should dominate over other losses $\kappa_0$, and support high \textit{cooperativity} $C = g^2/\kappa\gamma \gg 1$, defined by the cavity-qubit coupling $g$, and decoherence rate of the qubit $\gamma$. 

High cooperativity serves to accelerate a particular, desired emission pathway for the qubit (i.e. into the fibre-coupled optical mode, thus reducing the role of other emission and dephasing channels), and boost the efficiency of in-coupling the incident photon into the qubit toward values comparable to $C/(1+C)$. Since the cooperativity can be conveniently approximated by the ratio of the quality factor and the effective volume of the mode $C\propto Q/(V/\lambda^3)$ (with the latter normalised by the diffraction-limited volume $\lambda^3$), we can maximise it by embracing macroscopic high-$Q$ cavities (like the Fabry-Perot resonators), microscopic systems operating near $V/\lambda^3 \sim 1$ (fiber cavities), or deeply sub-wavelength nanophotonic devices (photonic crystal cavities)~\cite{RevModPhys.94.041003, RevModPhys.90.031002}.

The particular choice of the platform for a quantum node is dictated by the realisation of the memory qubits, for example:

\subsubsection{Atomic defects, incorporating nitrogen, silicon, and germanium atoms}
{Atomic defects, incorporating nitrogen, silicon, or germanium atoms next to vacancies in diamond (NV, SiV, and GeV)} can implement both the communication and memory qubits, encoding information in the electronic transitions of the defect, nuclear spins of the nearby atoms---$^{13}$C for NVs, and $^{29}$Si for SiVs---respectively~\cite{pompili21,PhysRevX.6.021040}. These two types of defects also offer very different challenges and opportunities: NV exhibits far longer coherence times at room temperature ($\sim 1$ s, several orders of magnitude over SiV), but also is far more sensitive to electrical noise and coupling to phonons, which degrades its optical response. Both defects can couple to moderate-$Q$ subwavelength nanoscale photonic crystal cavities, fabricated in diamond itself. 

An example quantum network architecture was recently realised by Knaut \textit{et al.}~\cite{knaut2024entanglement}, with two nodes built around SiV in diamond photonic crystal cavities, operating below 200 mK. Memory qubits were realised using the nuclear spin of a nearby $^{29}$Si (with 10 ms decoherence times, controlled by RF electronics), coupled to the fiber modes via a communications qubit (electronic spin states of SiV) and a cavity with moderate quality factor $Q\sim 10^3$ and cooperativity $C\sim 10$. Even accounting for the additional inefficiencies, including cavity-fibre coupling, correcting the frequency mismatching between the SiVs in the two nodes, downconverting optical emission from the SiV at 737 nm closer to the telecom wavelength at 1350 nm, and losses suffered by the photonic qubits over 45 km propagation between the nodes, this simple network realised entanglement between the two memories with fidelity of 0.7 at 1 Hz rate. 

With each mode built around micron-scale cavities, moderate requirements for cooling, and rapid improvements in techniques for diamond nanofabrications, these compact designs constitute a promising platform for implementing large-scale quantum nodes.
    
\subsubsection{Neutral atoms and ions}
{Neutral atoms and ions, trapped (optically or electrically) in ultra-high vacuum, and coupled to high-$Q$ optical cavities}~\cite{RevModPhys.87.1379}, are characteristed with efficient isolation from the environment, resulting in memory qubits exhibiting coherence times approaching 1~s (for neutral atoms~\cite{korber_decoherence-protected_2018}) over 1~h (for ions~\cite{wang_single_2021}). Entanglement between remote nodes was demonstrated with 0.85 fidelity has been demonstrated with neutral atoms over a decade ago~\cite{ritter_elementary_2012}, and --- more recently --- over 0.9 with trapped ions~\cite{PhysRevLett.124.110501}. 

A realisation of a quantum repeater, using two $^{87}$Rb atoms implementing memory qubits in a high-$Q$ cavity was reported by Langenfeld \textit{et al.} ~\cite{Langenfeld}. The decoherence rate of the qubits exceeded 20 ms, and the interfacing with optical photons was implemented via a Raman process, enhanced by a mm-length cavity with $Q\sim 10^8$, with the resulting cooperativity $C\approx 4$.

While trapped atoms are arguably the most mature technology for realising quantum nodes, the enormous footprint of ultra-high vacuum and optical setups significantly limits their scalability, and opportunities for deployment in large-scale quantum networks.

\subsection{Quantum gates}

This light matter described in the previous section must preserve the photon's temporal and phase information, as these encode the interferometric signal of interest. Moreover, because each photon may only occupy a narrow time bin and arrive probabilistically, the entire absorption-and-storage operation must be repeatable and low-latency, with errors significantly below the mean photon number $\epsilon$ per bin. Any imperfection at this stage reduces the signal-to-noise ratio of the subsequent quantum measurement, and hence the overall sensitivity of the VLBI array.

Recent experimental advances have demonstrated high gate fidelity and speeds for various platforms, including
\begin{itemize}
    \item {Ion traps}: Single-qubit gates with error \( 1.5 \times 10^{-6} \) and 600\,ns gate time~\cite{PhysRevLett.131.120601}; two-qubit gates in 270\,$\mu$s~\cite{postler2022demonstration}.
    \item {Neutral atoms}: Rydberg-blockade gates with \( \sim5 \times 10^{-3} \) error in 250\,ns~\cite{evered2023high}; F\"orster resonance gates achieving 1\% error in 6.5\,ns~\cite{chew2022ultrafast}; single-site readout errors \( \sim7 \times 10^{-3} \) over 36\,ms~\cite{Phuttitarn:24}.
    \item {Superconducting qubits}: Single- and two-qubit gate times of 20\,ns and 40\,ns, respectively, as demonstrated by Google~\cite{arute2019quantum}.
\end{itemize}

Importantly, quantum sensing in this context does not require universal, fault-tolerant quantum computation. Maintaining gate errors below standard quantum error correction thresholds is sufficient for reliable operation.

\subsection{Entanglement distribution and repeaters}
In quantum-enabled VLBI, pre-shared entanglement substitutes for direct optical links between telescope sites, enabling nonlocal measurements without transmitting fragile stellar photons across long baselines. The performance of such a system is fundamentally limited by the rate and fidelity of Bell pair distribution.

Recent progress in entangled photon sources includes:
polarization-entangled photons at telecom wavelengths have enabled key rates exceeding 1 Gbit/s \cite{neumann2022experimental};
     in SiC systems, pair generation rates of $\sim 10^4$  pairs/s have been demonstrated\cite{rahmouni2024entangled};
    commercial sources: off-the-shelf systems, such as those from Quantum Computing Inc., now reach 5M pairs/$\mu$W/s;
    on-chip generation of time-energy entangled pairs has achieved 22 MHz using only 36 $\mu$W of pump power ~\cite{chopin2023ultra}.

In terms of repeaters, the latest progress include:
     nanophotonic diamond resonators, where solid-state spin memories coupled to high-cooperativity cavities (737 nm, $C= 105\pm11$) have achieved Bell measurements at 0.1 Hz with 85\% detection efficiency, surpassing the repeaterless bound over 350 km of fiber \cite{bhaskar2020experimental};
    Heralded memory-to-memory entanglement: for a separation over 3.5 meters, using GHz-bandwidth absorptive memories and polarization-entangled photons, with 80.4$\%$ fidelity  \cite{liu2021heralded}; memory–memory entanglement among three fiber-connected nodes spread over 7.9 km – 12.5 km at a rate of $\sim 1$ Hz \cite{liu2024creation}.
    In multiplexed architectures: recent efforts demonstrate photon-pair sources with up to 200 spectral modes at telecom wavelengths, compatible with quantum memories and offering GHz bandwidths \cite{chakraborty2025towards}; 
    teleportation and entanglement distribution over a 14.4 km urban dark-fiber link with a single trapped-ion memory and frequency-converted photons at a rate of $\sim 1$ Hz \cite{knaut2024entanglement}. A demonstration by USTC performed non-local photonic gates between memories over 7 km, \cite{liu2024nonlocal}, pushing beyond entanglement generation to quantum processing between nodes.

\subsection{Scalability and architectures}
These developments mark a substantial progress from earlier linear-optics-only approaches, thanks to improvements in source quality, detection efficiency, and the emergence of quantum memories and spectral/temporal multiplexing techniques.
However, significant challenges remain before practical, high-rate repeater networks are feasible:
rates remain low when memory is involved;
multiplexing is not fully integrated yet—many setups demonstrate multiplexed sources or filters, but do not combine this with active feed-forward control and working quantum memories;
most systems often require frequency conversion, which adds loss and complexity;
scalability is an open challenge—interfacing multiple quantum memories, handling timing jitter, and maintaining fidelity across many links will require engineering solutions beyond current lab-scale demonstrations.

Recent advances in photonic quantum computing (PQC) naturally raise the question of whether such technologies could directly implement the gate operations envisioned here. PQC typically refers to linear-optical or measurement-based photonic circuits implementing interferometric gates such as beam splitters, phase shifters, and conditional operations. These systems operate deterministically, with on-chip control over photon generation and timing. In contrast, quantum-enabled optical VLBI functions as a distributed network spanning large distances, where stellar photons arrive stochastically and must be correlated across potentially thousands of kilometres of time delays. In this regime, it is doubtful whether PQC alone can maintain coherence or compensate for asynchronous arrivals---any failed fusion operation would lose the signal photon. Quantum memories therefore remain essential, providing the temporal buffering, and entanglement storage required for non-local measurements between remote telescopes.

Implementing full entanglement distillation or quantum error correction on satellites is unlikely to be practical in the near term, given the severe constraints on mass, power, and thermal control. However, satellites could, in principle, perform direct optical interferometry without requiring these quantum resources. Recent proposals for space-based quantum communication have suggested constellations of satellites equipped with focusing optics to bypass diffraction and achieve extremely long effective baselines \cite{PhysRevApplied.20.024048}. Such configurations highlight the potential of space-based interferometry, though maintaining path-length matching and phase stabilization across inter-satellite distances would remain an engineering challenge. For a comparison of 
free-flying interferometer to that of a lunar array, see Ref.~\cite{carpenter2025nasa}, where similar aspects would hold for the quantum versions.

\section{Literature survey}

Many protocols have made advances to quantum-enabled VLBI, we review a selection of 
Refs.~\cite{tsang2011quantum,
gottesman2012longer,
brown2023interferometric,
khabiboulline2019optical,
khabiboulline2019quantum,bland2021quantum,
PhysRevLett.129.210502,
czupryniak2023optimal,
huang2024vacuum,
huang2024limited,wang2023astronomical,wang2025limitations,bojer2022quantitative,
tang2025phase,liu2024super,zhang2025criteria}.

\subsection{Quantum Nonlocality in Weak-Thermal-Light Interferometry~\cite{tsang2011quantum}.}
In Ref.~\cite{tsang2011quantum}, Tsang showed that, for any local measurement strategies, the Fisher information scales as $\epsilon^2$. Non-local measurements are required if we wish to achieve an FI linear in $\epsilon$. In particular,
$\ket{\psi^\delta_\pm}\bra{\psi^\delta_\pm}$, then the FI matrix for direct detection is
\begin{align}
\mat{I} = \frac{\epsilon}{1-\text{Re}(\gamma e^{-i\delta})^2}
\left(
\begin{array}{cc}
 \cos^2 \delta & \sin \delta \cos \delta\\
 \sin \delta \cos \delta & \sin^2\delta \\
\end{array}
\right)
\end{align}
with eigenvalues
\begin{align}
\lambda_1  = 0,\qquad
\lambda_2 = \frac{\epsilon}{1-\text{Re}(\gamma e^{-i\delta})^2}
\end{align}
The trace norm of the FI matrix is then 
\begin{align}
||\mat{I}|| =\lambda_1 + \lambda_2 \geq \epsilon
\end{align}
Then for $M$ measurements,
\begin{align}
||\mat{I}^M|| = M ||\mat{I}|| \geq M \epsilon
\end{align}
Note that $||\mat{I}^M||$ can be significantly larger than $\epsilon$ for $\gamma\rightarrow 1$.

\subsection{Longer-Baseline Telescopes Using Quantum Repeaters~\cite{gottesman2012longer}}

The problem with direct interferometry is that it is difficult to transport the single photon state over long
distances without incurring loss. Ref.~\cite{gottesman2012longer} proposes a protocol where, instead of sending a valuable quantum state directly
over a noisy quantum channel, entanglement is distributed between the two telescopes instead.

Suppose we can distribute the following entangled state (``ground" photon),
\begin{align}
\label{eq:ground_photon}
\ket{\psi_\mathrm{shared}} = \frac{1}{\sqrt{2}}(\ket{0}_A\ket{1}_B + e^{i\delta}\ket{1}_A\ket{0}_B),    
\end{align}
where $\delta$ is determined by a controllable phase, allowing completion of the protocol to determine $\phi$.

Now, each half of $\ket{\psi_\mathrm{shared}} $ is then interfered with the stellar state in Eq.~\eqref{eq:coupled} at the respective telescope station with a 50:50 beam splitter.
Then, we post-select and consider the events where a two-photon coincidence is observed at stations $A$ and $B$ simultaneously. 
 The total probabilities of seeing photons in correlated, and anticorrelated outputs are respectively
\begin{align}
    &\frac{1}{2} (1 + \mathrm{Re}[\gamma e^{-i(\phi-\delta)}]),\nn
    &\frac{1}{2} (1 - \mathrm{Re}[\gamma e^{-i(\phi+\delta)}]).
\end{align}

This protocol requires the ``ground" photon to to interfere with a stellar photon, which means that the two need to be matched in their respective temporal and frequency modes. Per stellar photon measured, the entanglement consumption scales as $1/\epsilon $: see unary encoding below.

Ref.~\cite{PhysRevA.106.032424} discusses a multi-photon extension of Ref.~\cite{gottesman2012longer}. This is complementary to
Ref.~\cite{marchese2023large}, which extended the analysis to distributing more ground photons, where parallel-distributed ground photons interfere with the astronomical photon in a balanced multimode beam splitter. Ref.~\cite{modak2024large} investigated the detrimental effect of having distinguishable ground photons.

\subsection{Unary encoding}
Consider the simplest unary encoding where there is one memory qubit for each time bin. Suppose there are $M$ time bins in total. After a single photon is stored, in the 5th time bin, for example, the  memories will be in the state
\begin{align}
\frac{1}{\sqrt2}(&\ket{000010...}_A\ket{000000...}_B \nn 
&+e^{i\phi}\ket{000000...}_A\ket{000010...}_B),
\end{align}
where the entangled state is the 5th qubits of systems $A$ and $B$.

Since the stellar state is mostly vacuum, we must figure out which memory qubits contain a stellar photon. This can be
accomplished by using pre-shared Bell pairs and 2 CZ gates to perform a parity check, projecting out the vacuum~\cite{khabiboulline2019optical}. 
Denote the preshared Bell pair as $\ket{\Phi}$,

\begin{align}
\rho_{A B} \otimes \ket{\Phi^+} &\stackrel{2 \times \text{CZ}}{\rightarrow}\nn
& 
\ket{0,0}\bra{0,0}_{AB} \otimes\ket{\Phi^+}\bra{\Phi^+} \nn
&+\epsilon \left(\frac{1+\gamma}{2}\right) \ket{\psi_{+}^\phi}\bra{\psi_{+}^\phi}\otimes \ket{\Phi^-}\bra{\Phi^-}  \nn
 &+\epsilon \left(\frac{1-\gamma}{2}\right)\ket{\psi_{-}^\phi}\bra{\psi_{-}^\phi} \otimes\ket{\Phi^-}\bra{\Phi^-}.
\end{align}
That is, if a photon is present in the memory, a parity check leads to $\ket{\Phi}$ acquiring a minus phase.

In a naive encoding, the unary encoding requires one entangled pair for each time bin, thus representing a large consumption of entanglement.
 Performing the entanglement-enabled parity checks will result in the following 
\begin{align}
\ket{\Phi^+}^{\otimes M-1} \otimes \ket{\Phi^-},
\end{align}
consuming $M$ Bell pairs.

\subsection{Unary encoding with binary search}

\begin{figure}[h]
\centering
    \includegraphics[width=0.3\textwidth]{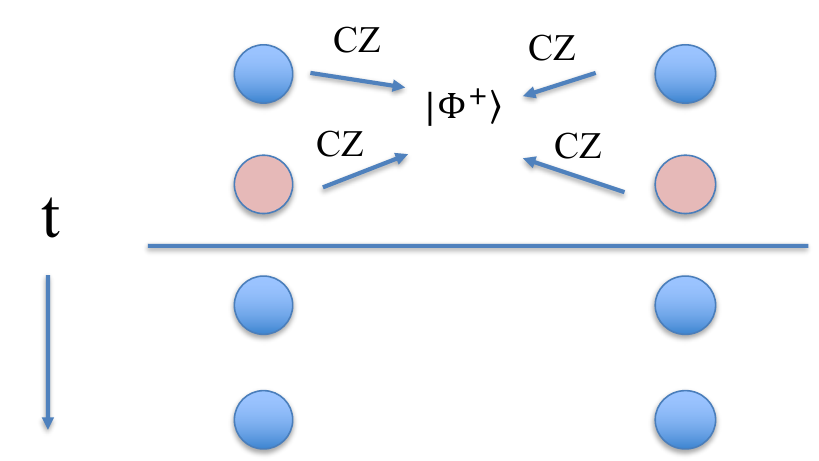}

\caption{\label{fig:b_search} Schematic for binary search for the time bin into which the photon arrived.
 }
\end{figure}

We can use an encoding that is efficient in the entanglement consumption, where the encoding is akin to a binary search. We can choose a block size $M$ such that $M \epsilon \sim 1$.
For each run of the parity check, we divide the memory block into two (Fig. \ref{fig:b_search}).
 To search for where the photon is within the time bin,
     we apply CZ gates between all of Alice's or Bob's memory qubits.

In our example with 4 bins, a photon arrives in the second bin and is shared non-locally between Alice and Bob. The state is
\begin{align}
\rho = &\ket{0,0}\bra{0,0}_{A_1B_1}\otimes \rho^{\star'}_{A_2B_2} \nn
       & \otimes\ket{0,0}\bra{0,0}_{A_3B_3}\otimes \ket{0,0}\bra{0,0}_{A_4B_4},
\end{align}
where
\begin{align}
\rho^{\star'} &=  \left(\frac{1+\gamma}{2}\right) \ket{\psi_+^\phi}\bra{\psi_+^\phi} + 
 \left(\frac{1-\gamma}{2}\right)\ket{\psi_-^\phi}\bra{\psi_-^\phi}.
\end{align}
We divide the block of 4 time bins into two halves. The first half is
\begin{align}
 \rho_{1,2}=&\ket{0,0}\bra{0,0}_{A_1B_1} \otimes \rho^{\star'}_{A_2B_2}.
\end{align}
To perform the check, for all qubits in the half-block, CZ gates are applied between the memory qubit and the shared Bell pair.
\begin{align}
\rho_{ 1,2} \otimes \ket{\Phi^+}\bra{\Phi^+} &\stackrel{4 \times \text{CZ}}{\rightarrow} \rho_{ 1,2} \otimes \ket{\Phi^-}\bra{\Phi^-}.
\end{align}
If there is no photon in the block, such as bins 3 and 4, then the Bell state remains the same
\begin{align}
\rho_{ 3,4} \otimes \ket{\Phi^+}\bra{\Phi^+} &\stackrel{4 \times \text{CZ}}{\rightarrow} \rho_{ 3,4} \otimes \ket{\Phi^+}\bra{\Phi^+}.
\end{align}

The procedure continues by iteratively dividing the block containing the photon into halves, until a single time bin remains.
This encoding is linear in the memory requirements, where the number of memory qubits scales as $1/\epsilon$. However, the entanglement consumption has been reduced to $\sim \log_2(1/\epsilon)$.

\subsection{Binary encoding: optical interferometry with quantum Networks~\cite{khabiboulline2019optical}}

The binary encoding proposed in~\cite{khabiboulline2019optical} is efficient in memory and entanglement consumption. In this protocol, we
\begin{itemize}
    \item assume approximately one photon arrives over $M$ time bins, which $M\sim1/\epsilon$.
    \item label each time bin $m \in \mathbb{Z}_+$, with its binary representation $m_2$.
    \item define logical qubits $\ket{\bar 0} \equiv \ket{0...0}$ and $ \ket{\bar 1_m} \equiv \ket{m_2}$. E.g. the fifth time bin
    $$\ket{\bar 1_5} \equiv \ket{1010...0}$$
 
\end{itemize}
A encoded logical CNOT (CX$_m$) is performed between the photonic degrees of freedom and the appropriate memory qubits at each time bin
    \begin{align}
    \ket{0}(\ket{\bar 0},\ket{\bar 1_j} )&\stackrel{\bar{ \text{CX}_m} }{\rightarrow} \ket{0}(\ket{\bar 0},\ket{\bar 1_j} ), \\
    \ket{1}(\ket{\bar 0},\ket{\bar 1_j} )&\stackrel{\bar{ \text{CX}_m} }{\rightarrow} \ket{1}(\ket{\bar 1_m},\ket{\bar 1_j+\bar 1_m} ).
    \end{align}
This encoding keeps track of the time-of-arrival information of the photon: empty time bins leave the memories unchanged. Defining
$\bar \rho_0  \equiv \ket{\bar 0, \bar 0}\bra{\bar 0,\bar 0}$
%
, we perform the CX$_m$ operation to find
\begin{align}
\rho_{AB} \otimes \bar \rho_0 
             &\stackrel{ \bar{ \text{CX}_m} }{\rightarrow} (1-\epsilon) \rho_\text{vac} \otimes \bar \rho_0  \nn
             &+\frac{\epsilon(1+|g|)}{2} \ket{\Psi^+_{\phi,m}}\bra{\Psi^+_{\phi,m}} \nn 
             &+\frac{\epsilon(1-|g|)}{2} \ket{\Psi^-_{\phi,m}}\bra{\Psi^-_{\phi,m}},
\end{align}
where
\begin{align}
\ket{\Psi^\pm_{\phi,m}} = (\ket{0,1}_{A,B} \ket{\bar 0,\bar 1_m} \pm e^{i\phi} \ket{1,0}_{A,B}  \ket{\bar 1_m, \bar 0}).
\end{align}

Then the photonic degree of freedom is decoupled from the memory qubit by measuring it in the X basis. This step is likely challenging, which can be solved in the next protocol. Then, applying the appropriate correction, the memory qubits end up in the state
\begin{align}
\rho_{AB} \rightarrow ~&(1-\epsilon) \bar \rho_0~+
                \frac{\epsilon(1+|g|)}{2} \ket{\psi^+_{\phi,m}}\bra{\psi^+_{\phi,m}} \nn 
              &+\frac{\epsilon(1-|g|)}{2} \ket{\psi^-_{\phi,m}}\bra{\psi^-_{\phi,m}},
\end{align}
where $\ket{\psi^\pm_{\phi,m}} =\frac{1}{\sqrt2}(\ket{\bar 0,\bar 1_m} \pm e^{i\phi} \ket{\bar 1_m,\bar 0})$.

In summary, if a photon arrives in the $(m-1)$-th time bin, then it is stored in such a way that the time bin is prepared according to $\ket{\bar 1_m}$.

This protocol uses the same set of memory qubits for all time bins, which require gates to be applied in between each subsequent time bin encoding: the implication is that the gate speed needs to catch up with the time bin size.

\noindent \begin{center}\textit{Errors due to multiphoton contribution}\end{center}

Let us examine errors due to multiphoton contribution in this protocol and how to alleviate them. 

Starlight is thermal, if we choose the time bin block size such that $M\epsilon \sim 1$, then the mean photon number in the block is 1; if so, the 
probability of having more than 1 photon in the block is also non-negligible. 
Examine the state post-encoding: 
\begin{align}
\rho_{AB} = p_0\rho_\vac + p_1\rho_1 + (1-p_0 - p_1)\rho_\text{multi},
\end{align}
where $p_0$ is the probability of receiving the vacuum, $p_1$ is the probability of having a single photon,
and $\rho_\text{multi}$ is the state of the memory where more than one photon has landed. 
In the case of $\rho_\text{multi}$, the memory qubit is considered completely depolarised.

For $\gamma =1$, we have
$p_0 = {1}/(\epsilon+1)$, $p_1 = {\epsilon}/{(1+\epsilon)^2}$ \cite{khabiboulline2019optical}.
%
If we have $\rho_{AB}^{\otimes M}$ in a block of $M$, then final decoded state in the
readout qubit has coeﬃcients given by a trinomial distribution of the input coeﬃcients:
\begin{align}
\rho_{AB}^{\otimes M} 
\rightarrow & \left(\frac{1}{1+\epsilon} \right)^M\rho_\vac +
              M\left(\frac{1}{1+\epsilon}\right)^{M-1} \left(\frac{\epsilon}{(1+\epsilon)^2}\right)\rho_1  \nn
             &+ \left(1-\frac{1}{(1+\epsilon)^M}  -\frac{M\epsilon}{ (1+\epsilon)^{M+1} }\right) \rho_\text{multi}.
\end{align}                 
%
%
\begin{figure}[t]
\centering
    \includegraphics[width=0.46\textwidth]{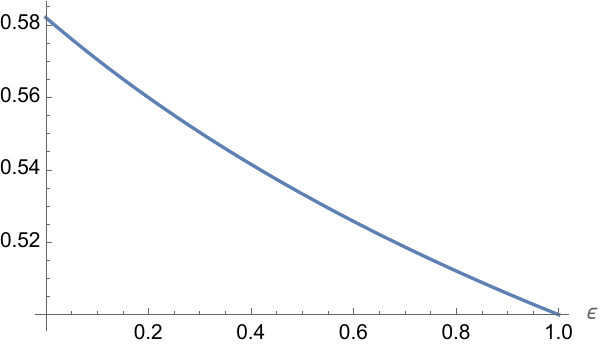}
     \includegraphics[width=0.46\textwidth]{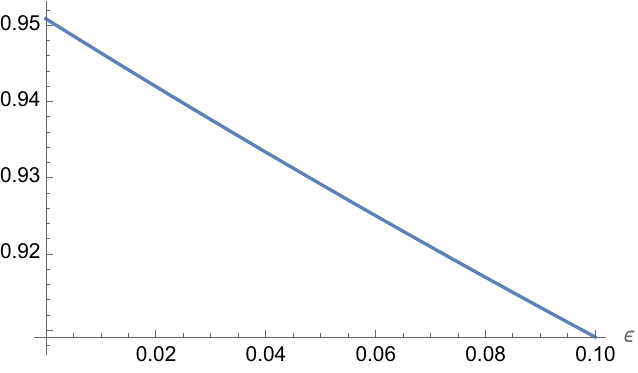}
\caption{\label{fig:c_factor} 
The parameter $c$ in Eq.~\eqref{eq:c_factor} as a function of $\epsilon$. 
The value on the y-axis is $c$, for (top) $M\epsilon = 1$, 
(bottom) $M\epsilon = 0.1$. 
The expense is that the entanglement consumption rate went up by a factor of 10.}
\end{figure}

Assuming the multiphoton state is completely mixed, the post-selected `single photon' state reads as
\begin{align}
\label{eq:c_factor}
\rho = c \rho_1 + (1-c)\rho_\text{mix}, \qquad c = \frac{M\epsilon}{[(1+\epsilon)^M - 1](1+\epsilon)}.
\end{align}
This presents a problem, because for high-contrast $c$ needs to be as close to 1 as possible, and this is fundamental; $c$ depends on how we choose $M \epsilon$ (Fig.~\ref{fig:c_factor}). If we frequency multiplex the signal such that it is spread over more memory qubits, the probability of infidelity decreases.

\noindent \begin{center}\textit{Entanglement consumption rate}\end{center}

For this protocol, the number of required memory qubits is approximately $\log_2(1/\epsilon)$~\cite{khabiboulline2019quantum}.
We can calculate the entanglement consumption rate:
\begin{enumerate}
    \item The time bin size is given by the inverse bandwidth, Eq.~\eqref{eq:coherence_length}, $\tau = 1/\Delta \nu$. 
     \item The average number of photons per second is  \mbox {$\Delta \nu  \epsilon$}.
     \item Per stellar photon, the parity check requires $\log_2(1/\epsilon)$; we arrive at the entanglement consumption rate being
     approximately $\Delta \nu \times \epsilon \log_2(1/\epsilon)$.
     \item We likely need to increase this by a factor of approximately 10 to circumvent multiphoton events (see Fig.~\ref{fig:c_factor}). The multi-photon problem may be lifted if we expand to broadband operation (section below).
\end{enumerate}
For  $\Delta \nu$ = 10 GHz, $\epsilon = 7 \times 10^{-7}$, entanglement distribution rate is $\Delta \nu \epsilon \log_2(1/\epsilon) \sim$ {200 kHz}.
 See Table~\ref{tab:e_consumption} for further examples.
%
%
\begin{table}[h]
\begin{center}
\begin{tabular}{ |c |c|c|c| c|}  
 \hline
$\lambda$   &  $\Delta \nu$ (Hz) &   $\Delta \lambda$  & $\epsilon$    & $\Delta \nu \epsilon \log_2(1/\epsilon)$ per sec \\ \hline
555 nm      &   1 THz            &   1 nm              &   $10^{-7}$   &   $2\times 10^6$       \\ 
555 nm      &   1 THz            &   1 nm            &   $10^{-10}$   &   $3\times 10^3$        \\ 
555 nm      &   1 THz            &   1 nm            &   $10^{-11}$   &   40                       \\\hline
760 nm      &   500 GHz           &     1 nm          &   $10^{-7}$   &   $1.2\times 10^6$       \\ 
760 nm      &   500 GHz           &    1 nm           &   $10^{-10}$   &   $1.2\times 10^3$       \\
760 nm      &   500 GHz           &    1 nm           &   $10^{-12}$   &   20 \\ \hline
%
%
%
1.65 $\mu$m  &   110 GHz          &    1 nm          &   $10^{-7}$   &   $2.5\times 10^5$       \\ 
1.65 $\mu$m  &   110 GHz          &    1 nm           &   $10^{-10}$   &   $3.6\times 10^2$       \\
1.65 $\mu$m  &   110 GHz          &    1 nm           &   $10^{-12}$   &   4.4       \\ \hline
\end{tabular}
\end{center}
\caption{\label{tab:e_consumption} Entanglement consumption rate for various wavelengths, bandwidths, and mean photon numbers.}
\end{table}

\subsection{Quantum-assisted telescope arrays~\cite{khabiboulline2019quantum}}

The above protocol is generalised to broadband operation in Ref~\cite{khabiboulline2019quantum}. 
The incoming light is split into R frequency bands. As an example, consider that a photon arrives in the 5th time bin, and is a mixture of two frequency bands
\begin{align}
\rho \approx (1-\epsilon) \rho_\vac + \frac{\epsilon}{2}(\rho_{\nu1} + \rho_{\nu2} ).
\end{align}

The notation here for the 2nd frequency band is:
\begin{align}
&\underbrace{\ket{0_{A_5},0_{B_5}}}_\text{freq 1} \underbrace{\ket{1_{A_5},0_{B_5} } }_\text{freq 2}
\underbrace{\ket{1011_A,0000_B}}_\text{memories}, \nn
& \underbrace{\ket{0_{A_5},0_{B_5}}}_\text{freq 1} \underbrace{\ket{0_{A_5},1_{B_5} } }_\text{freq 2}
\underbrace{\ket{0000_A,1011_B}}_\text{memories}.
\end{align}
The first 3 memory qubits encode the time bin, and the 4th encodes the frequency band:
\begin{align}
&\ket{ \underbrace{101}_\text{time} \underbrace{1}_\text{freq} {}_A, 0000_B  }
\end{align}

This variation has parallel operation over frequencies, at the expense of memory scaling as $R\log_2M$, but the entanglement consumption scales logarithmically with $R$ and $M$.

\subsection{Incorporating quantum error correction~\cite{PhysRevLett.129.210502}}
\begin{figure}[t]
\includegraphics[trim = 0cm 0.0cm 0cm 0cm, clip, width=1.0\linewidth]{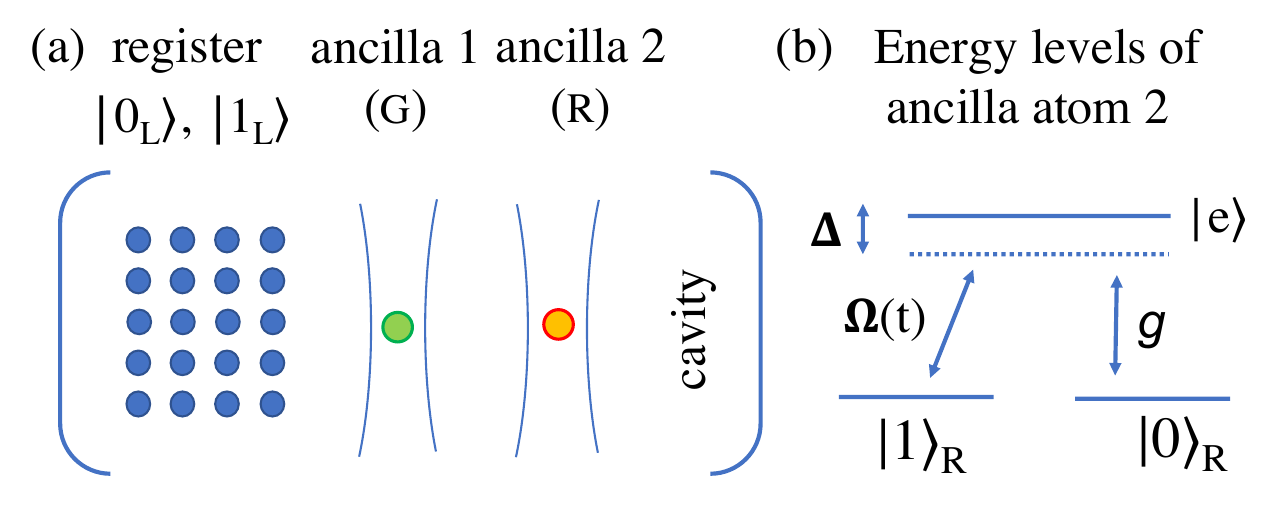} 
 \caption{\label{f:stirap} \textbf{Cavity-assisted coherent single-photon transfer.} (a) A system of qubits has logical states $\ket{0_L},\ket{1_L}$; ancillary qubit 1 is initially prepared into a Bell state with the register, $1/\sqrt2 (\ket{0_L 0_G} + \ket{1_L 1_G})$ ; ancilla 2 is used in the STIRAP interaction to interact with the star photon. (b) Energy levels of ancilla atom 2 used for the STIRAP interaction.}
\end{figure}

We can incorporate quantum error correction (QEC) as follows.
We assume the stellar photon is described by the state in Eq.~\eqref{eq:coupled}. 

In this protocol, the set-up is depicted in Fig.~\ref{f:stirap}: (a) inside a cavity, we use three different sets of systems. We denote the blue array as the register.
The blue array is initialised in a codespace of a QEC code encoding a single logical qubit, spanned by its logical codewords $\ket{0_L}$ and $\ket{1_L}$. We also need ancilla qubit 1 (green), and ancilla atom 2 (red). Note that the three types of matter qubits could consist of different electronic sublevels of the same species of atom if desired. 

Suppose we now prepare the register and the green ancilla (here the subscript $G$ denotes green) in the Bell state 
\begin{align}
\ket{\Phi_0}=\frac{1}{\sqrt2}(\ket{0_L}\ket{0_G} + \ket{1_L}\ket{1_G}).
\end{align}
Now, the red ancilla is initially prepared in state $\ket{0}_R$, so our set-up is in state
%
$\ket{\Psi_0} =\ket{\Phi_0} \otimes \ket{0}_R.$
%
Suppose Alice and Bob each have a copy of $\ket{\Psi_0}$, and they perform STIRAP individually 
(Fig.~\ref{f:stirap}.
They share a single photon from the star
\mbox{$\frac{1}{\sqrt2}(\ket{1_p}_A\ket{\text{vac}}_B \pm e^{i \phi}\ket{\text{vac}}_A\ket{1_p}_B).$}
%
%
%
In the presence of the photon, the STIRAP interaction transforms $\ket{0}_R \rightarrow \ket{1}_R$, and the phase relationship in the photon is preserved.
This means that the state of the red ancillae (on Alice and Bob's sites) is now
\begin{align}
\frac{1}{\sqrt 2} \left(\ket{1_R, 0_R}_{AB} \pm e^{i \phi}\ket{0_R, 1_R}_{AB} \right).
\end{align}
Performing a Bell measurement on the red and green ancillae teleports the state onto the registers. 
After the Pauli operator correction dependent on the measurement outcome, the state of the registers between Alice and Bob becomes an entangled state, and the entanglement arises entirely from the starlight photon.

\begin{figure}[h!]
\includegraphics[trim = 0cm 0.0cm 0cm 0cm, clip, width=1.0\linewidth]{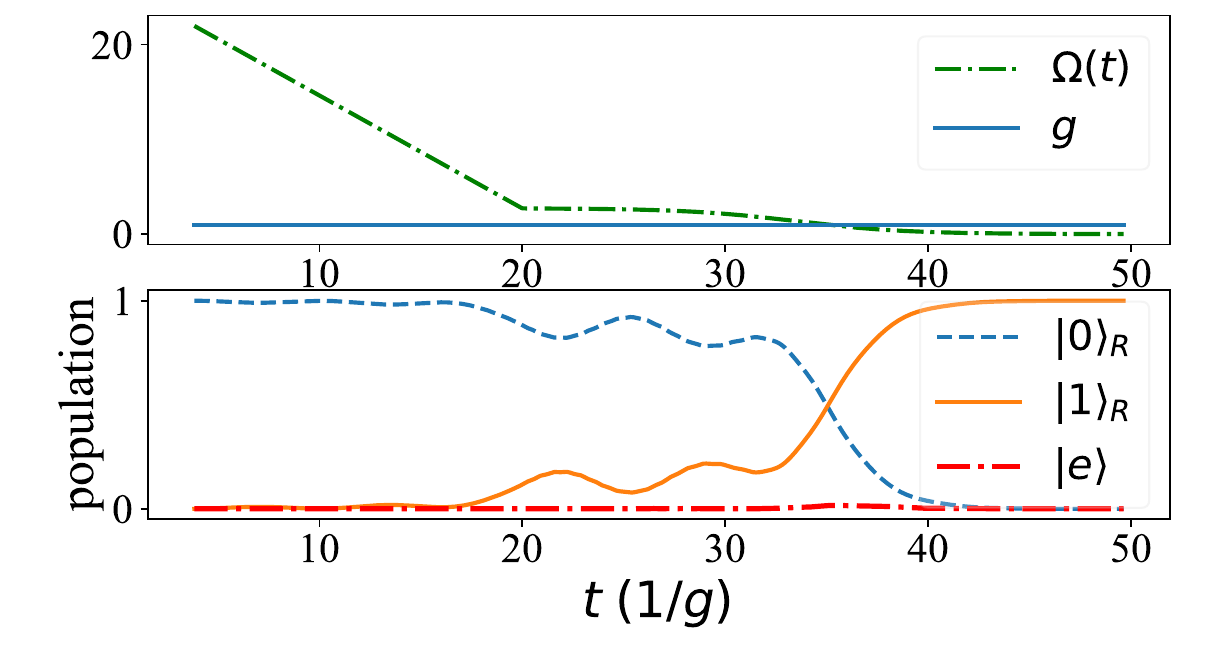} 
 \caption{\label{f:population_transfer} \textbf{Population transfer for of a three-state adiabatic passage}. Top: the interaction strengths of $g$ and $\Omega$ as a function of time $t$ in units of $1/g$. Bottom: occupancy of in $\ket{e},\ket{0}_R$ and $\ket{1}_R$ in ancillary atom 2 (r). The detuning parameter is set to $\Delta(t) = g^2 +\Omega^2(t) $ to satisfy the adiabatic condition~\cite{PhysRevA.80.013417}. Note that the excited state $\ket{e}$ is not populated.}
\end{figure}
The technically difficult part of this protocol may include:
    1) The STIRAP step needs to be extremely low noise: the stellar state is unencoded at this point, and the error rate must be $\ll \epsilon$. 2) For a complete population transfer, here $t=50/g$, which is slow. The cavity quality needs to be sufficiently high to retain the stellar photon until the transfer is complete. 3) STIRAP schemes tend to have very narrow bandwidth, in rare Earth ions the typical bandwidth is around 10k Hz

As an example of a suitable STIRAP scheme, consider a $^{87}$Rb atom trapped in an optical tweezer and coupled to a fibre Fabry-Perot cavity as recently demonstrated~\cite{Lukin:2025}. Adapted to the goal here of coherent single-photon transfer, we follow the prescription in~\cite{srivastava2024entanglementenhancedquantumsensingoptimal}. Choose qubit states $\ket{0}= \ket{5 ^{2}S_{1/2}, F= 1, m_{F}= 0}$, $\ket{1}= \ket{5 ^{2}S_{1/2}, F= 2, m_{F}= 0}$, and $\ket{e}= \ket{5 ^{2}P_{3/2}, F= 3, m_{F}= 0}$, where the linewidth of the $\ket{1} \leftrightarrow \ket{e}$ transition ($\lambda = 780\,$nm) is $\gamma =2\pi \times 6\,$MHz (FWHM). We assume a cavity finesse $F \approx 2 \times 10^5$, a waist radius ${\rm w}_r \approx 2\,\mu$m and a length $L \approx 40\,\mu$m resulting in a cooperativity of $C = 3\lambda^2F/(2\pi^3 {\rm w}_r^2) \approx 1500$ (as demonstrated in~\cite{Hunger_2010}) with a coupling strength of $g = \sqrt{3\lambda^2 C \gamma}/(2\pi^2 {\rm w}_r^2L) \approx 2\pi \times  400\,$MHz and $\kappa =\pi C/L F \approx 2\pi \times 20\,$MHz (FWHM), so that $\gamma/\kappa \approx 0.3$. Tweezer-induced dephasing rates on state $\ket{1}$ can be treated as negligible $\gamma_{\phi}/g = 0.03 \times 10^{-6}$~\cite{manetsch2024tweezer}, so the dominant source of error would be cavity decay. Over a population transfer period of $T$, with on average $1/2$ photon occupation in the cavity, one estimates a fidelity of $F=e^{-T\kappa/2}$ which for $T=50g^{-1}$ would imply $F\approx 0.29$. Better cavities providing a factor of $5$ improvement in $g/\kappa$ would enable $F>0.77$, which could be further improved by post-selecting on null results from single photon detectors monitoring leaked photons \cite{Niemietz2021-nf}.

\subsection{Continuous-variable investigations}

States received from astronomical sources are inherently thermal~\cite{mandel1995optical}. Therefore, it is natural to consider this problem in the framework of continuous-variable (CV) quantum information~\cite{serafini2017quantum,RevModPhys.84.621}.

In Ref.~\cite{huang2024limited} three schemes are scrutinised: (1) direct interferometry (DI), where the two modes of the stellar state are physically brought together for interference. Each mode suffers transmission loss parameterised by $\eta \in [0,1]$. 
 (2) A local strategy, where heterodyne detection is performed separately on the modes held by Alice and Bob, and no loss is incurred.
 (3) A CV teleportation-based~\cite{PhysRevLett.80.869} strategy, where a two-mode squeezed vacuum (TMSV) is distributed to Alice and Bob. During distribution, each mode of the TMSV suffers transmission loss parameterised by $\eta$. Bob performs joint homodyne measurements as prescribed by standard CV teleportation and sends his measurement outcomes to Alice.

In the lossless regime, it is shown that a squeezing parameter of $r\approx 2$ (18 dB) is required to reach $\approx$ 95\% of the QFI achievable with DI; such a squeezing level is beyond what has been achieved experimentally. In the low-loss regime, the CV teleportation strategy becomes inferior to DI, and the performance gap widens as loss increases. Curiously, in the high-loss regime, a small region of loss exists where the CV teleportation strategy slightly outperforms both DI and local heterodyne, representing a transition in the optimal strategy.

However, the advantage is very limited because it occurs for a small region of loss, and the magnitude of the advantage is also small. Practical difficulties further impede achieving any quantum advantage, limiting the merits of a CV teleportation-based strategy for stellar interferometry. Consistent with Ref.~\cite{wang2023astronomical}, we conclude that TMSV teleportation-based strategies are practically infeasible if there is no quantum repeater available.

\subsection{Vacccum beam guide for quantum networks~\cite{huang2024vacuum}}

This paper introduces a simple and elegant solution for long-distance distribution of quantum states: a ground-based optical channel composed of a vacuum tube with precisely aligned lenses.  The authors demonstrate that a VBG with realistic parameters can achieve an attenuation rate as low as $3\times 10^{-5} $ dB/km. The system avoids the complexity of quantum repeaters by requiring only passive optical elements. Key sources of loss—lens imperfections, residual air absorption, and beam misalignment—are analysed and shown to be manageable with current technology. 

 This architecture is not only promising for quantum communication but also enables advances in quantum-enabled O-VLBI. In particular, we see two distinct, yet complementary, directions in which the VBG infrastructure could be adapted:

\subsubsection{High-Efficiency Entanglement Distributor}

In a repeater-based VLBI architecture, VBGs could serve as the physical backbone for entanglement distribution between telescope nodes. This offers several advantages.
 Low-loss transmission over thousands of kilometres enables high-rate distribution of entangled photons with reduced need for quantum repeaters.
 The VBG channel can support a variety of quantum encodings (e.g., time-bin, polarisation, frequency-bin) suitable for distributed quantum metrology and interferometry.

\subsubsection{Using the VBG as an interferometer arm}
Alternatively, the vacuum beam guide (VBG) itself could be repurposed as an interferometric transmission channel, effectively forming one arm of a large optical interferometer.

In a configuration without quantum memories, starlight collected at separate telescopes would be coupled directly into the VBG and transported to a central detection station for interference-based measurement. Because the VBG avoids the material losses and chromatic dispersion associated with optical fibres, it provides an excellent medium for maintaining coherence over long baselines. However, in this arrangement active path-length compensation must be reintroduced to preserve fringe stability within a fraction of the optical wavelength. This could be accomplished through adjustable delay lines and phase-locking mechanisms, potentially guided by a bright reference source or classical beacon.

In a configuration with quantum memories, the incoming stellar photons are stored at the central station, where the memories act as tuneable  optical delay lines that compensate for geometric path-length differences between telescopes. Interference can then be performed locally using a range of possible recombination schemes. This approach eliminates the need for distributing entanglement to each telescope and requires only a single bank of quantum memories at the combiner. It thus simplifies the overall architecture while still leveraging quantum storage to replace the extremely long physical delay lines that would otherwise be needed for baselines of hundreds to thousands of kilometres.

\color{black}
In addition, if implemented on the Moon, the vacuum beam guide would benefit from the natural hard vacuum of the lunar environment, eliminating the need for complex evacuated tubes. 
\color{black}

\section{Open questions}

Implementing quantum-enabled VLBI is extremely technologically demanding: we need to resolve (the many) individual time modes corresponding to sparsely arriving stellar photons. This requires highly efficient coupling into quantum memories with mode-selective control. The large amount of time and frequency modes make this a difficult task.

There are open questions that remain.  First, what is the best way to transfer the information in a single photon into a quantum memory? What are the trade-offs between adiabatic loading schemes (e.g. STIRAP) and faster dynamical techniques? Adiabatic loading schemes offer robustness but are slow, while dynamical schemes allow faster coupling at the cost of having high noise to timing and control. The optimal choice depends on memory coherence time and system timing constraints.

How many temporal modes do we actually need? Does this depend solely on $\epsilon$, or do detector and memory limitations impose stricter bounds? What should the mode shapes look like in practice?  For binary encoding, the number of memory qubits scales as $\log(1/\epsilon)$, but practical constraints such as detector dead time and memory reset times also play a role.
How can we most efficiently implement multiplexing without overwhelming the memory and feedforward control? 

The atmosphere fluctuates on the timescale of milliseconds. What is the best way of correcting for the phase variations? What requirements are needed for a potential artificial guiding star? Can we collect enough photons by using broadband light to avoid using a guiding star?

How might established techniques such as closure-phase observables be adapted within quantum-enhanced architectures? More broadly, can quantum estimation theory enable new imaging schemes that go beyond pairwise visibility measurements and the standard van Cittert–Zernike framework?

 In the absence of full quantum repeaters, what level of VLBI enhancement is realistically achievable using repeater-less quantum links? Can the distribution of additional ground photons, particularly with more robust temporal or mode entanglement, meaningfully improve signal-to-noise ratio? Furthermore, how might such protocols connect with or inform developments in all-optical quantum repeater architectures?

Beyond classical Fisher information and maximum likelihood methods, could quantum-enhanced or learning-assisted inference techniques play a role in astronomical imaging? Recent work at the interface of quantum estimation and learning theory (e.g.~\cite{wang2025advancing}) suggests a potentially fruitful direction for developing robust, data-efficient reconstruction techniques tailored to interferometric observations.

\section{Concluding remarks}
This report has outlined the feasibility and potential of quantum VLBI. 
By integrating quantum information protocols, many long-standing challenges in optical VLBI can be mitigated.

From a technological perspective, we examined a range of architectures. 
Encouragingly, most of the individual hardware components—such as high-cooperativity optical cavities, long-lived quantum memories, and entangled photon sources—are now approaching the threshold of practical feasibility. Experimental progress in photon collection, phase stabilisation, and entanglement distribution suggests that several essential subsystems could be prototyped with near-term technologies.
The primary bottlenecks that remain are the speed and fidelity of quantum gate operations. In particular, time-bin multiplexed architectures impose stringent requirements on gate execution times and error rates that exceed current capabilities in most physical platforms. 

Nevertheless, the development of quantum technologies toward these benchmarks is rapid. Continued improvements in quantum control and photonic integration may soon enable the deployment of quantum-enhanced VLBI systems capable of achieving diffraction-limited resolution at optical wavelengths. Such a capability would unlock new frontiers in both astronomical imaging and precision Earth science.

\begin{acknowledgements}
We thank Max Charles, John Bartholomew, Michael Steel, Christian Schwab, John Jeffers, Lorenzo Maccone, and Chiara Macchiavello, Mark Wilde, Yujie Zhang, Yunkai Wang and Virginia Lorenz for insightful discussions.
ZH is supported by an ARC DECRA Fellowship (DE230100144) ``Quantum-enabled super-resolution imaging'', 
a RMIT Vice Chancellor’s Senior Research Fellowship, and the EPSRC International Network in Space Quantum Technologies (EP/W027011/1). MKS is supported by an Australian Research Council Discovery Early Career Researcher Award (DE220101272). 
DO is supported by the EPSRC Quantum Technology Hub in Quantum Communication (EP/T001011/1), EPSRC Integrated Quantum Networks Research Hub (EP/Z533208/1), and ESA (4000147561/25/NL/FGL/ss). 
 PK is supported by EPSRC's Large Baseline Quantum-Enhanced Imaging Networks grant (EP/V021303/1) and Mathematical Tools for Practical Quantum Imaging Protocols grant (UKRI069).

\end{acknowledgements}



\begin{thebibliography}{98}%
\makeatletter
\providecommand \@ifxundefined [1]{%
 \@ifx{#1\undefined}
}%
\providecommand \@ifnum [1]{%
 \ifnum #1\expandafter \@firstoftwo
 \else \expandafter \@secondoftwo
 \fi
}%
\providecommand \@ifx [1]{%
 \ifx #1\expandafter \@firstoftwo
 \else \expandafter \@secondoftwo
 \fi
}%
\providecommand \natexlab [1]{#1}%
\providecommand \enquote  [1]{``#1''}%
\providecommand \bibnamefont  [1]{#1}%
\providecommand \bibfnamefont [1]{#1}%
\providecommand \citenamefont [1]{#1}%
\providecommand \href@noop [0]{\@secondoftwo}%
\providecommand \href [0]{\begingroup \@sanitize@url \@href}%
\providecommand \@href[1]{\@@startlink{#1}\@@href}%
\providecommand \@@href[1]{\endgroup#1\@@endlink}%
\providecommand \@sanitize@url [0]{\catcode `\\12\catcode `\$12\catcode
  `\&12\catcode `\#12\catcode `\^12\catcode `\_12\catcode `\%12\relax}%
\providecommand \@@startlink[1]{}%
\providecommand \@@endlink[0]{}%
\providecommand \url  [0]{\begingroup\@sanitize@url \@url }%
\providecommand \@url [1]{\endgroup\@href {#1}{\urlprefix }}%
\providecommand \urlprefix  [0]{URL }%
\providecommand \Eprint [0]{\href }%
\providecommand \doibase [0]{https://doi.org/}%
\providecommand \selectlanguage [0]{\@gobble}%
\providecommand \bibinfo  [0]{\@secondoftwo}%
\providecommand \bibfield  [0]{\@secondoftwo}%
\providecommand \translation [1]{[#1]}%
\providecommand \BibitemOpen [0]{}%
\providecommand \bibitemStop [0]{}%
\providecommand \bibitemNoStop [0]{.\EOS\space}%
\providecommand \EOS [0]{\spacefactor3000\relax}%
\providecommand \BibitemShut  [1]{\csname bibitem#1\endcsname}%
\let\auto@bib@innerbib\@empty
\bibitem [{\citenamefont {Tsang}(2011)}]{tsang2011quantum}%
  \BibitemOpen
  \bibfield  {author} {\bibinfo {author} {\bibfnamefont {M.}~\bibnamefont
  {Tsang}},\ }\bibfield  {title} {\bibinfo {title} {Quantum nonlocality in
  weak-thermal-light interferometry},\ }\href@noop {} {\bibfield  {journal}
  {\bibinfo  {journal} {Physical review letters}\ }\textbf {\bibinfo {volume}
  {107}},\ \bibinfo {pages} {270402} (\bibinfo {year} {2011})}\BibitemShut
  {NoStop}%
\bibitem [{\citenamefont {Gottesman}\ \emph {et~al.}(2012)\citenamefont
  {Gottesman}, \citenamefont {Jennewein},\ and\ \citenamefont
  {Croke}}]{gottesman2012longer}%
  \BibitemOpen
  \bibfield  {author} {\bibinfo {author} {\bibfnamefont {D.}~\bibnamefont
  {Gottesman}}, \bibinfo {author} {\bibfnamefont {T.}~\bibnamefont
  {Jennewein}},\ and\ \bibinfo {author} {\bibfnamefont {S.}~\bibnamefont
  {Croke}},\ }\bibfield  {title} {\bibinfo {title} {Longer-baseline telescopes
  using quantum repeaters},\ }\href@noop {} {\bibfield  {journal} {\bibinfo
  {journal} {Physical review letters}\ }\textbf {\bibinfo {volume} {109}},\
  \bibinfo {pages} {070503} (\bibinfo {year} {2012})}\BibitemShut {NoStop}%
\bibitem [{\citenamefont {Khabiboulline}\ \emph
  {et~al.}(2019{\natexlab{a}})\citenamefont {Khabiboulline}, \citenamefont
  {Borregaard}, \citenamefont {De~Greve},\ and\ \citenamefont
  {Lukin}}]{khabiboulline2019optical}%
  \BibitemOpen
  \bibfield  {author} {\bibinfo {author} {\bibfnamefont {E.~T.}\ \bibnamefont
  {Khabiboulline}}, \bibinfo {author} {\bibfnamefont {J.}~\bibnamefont
  {Borregaard}}, \bibinfo {author} {\bibfnamefont {K.}~\bibnamefont
  {De~Greve}},\ and\ \bibinfo {author} {\bibfnamefont {M.~D.}\ \bibnamefont
  {Lukin}},\ }\bibfield  {title} {\bibinfo {title} {Optical interferometry with
  quantum networks},\ }\href@noop {} {\bibfield  {journal} {\bibinfo  {journal}
  {Physical review letters}\ }\textbf {\bibinfo {volume} {123}},\ \bibinfo
  {pages} {070504} (\bibinfo {year} {2019}{\natexlab{a}})}\BibitemShut
  {NoStop}%
\bibitem [{\citenamefont {Khabiboulline}\ \emph
  {et~al.}(2019{\natexlab{b}})\citenamefont {Khabiboulline}, \citenamefont
  {Borregaard}, \citenamefont {De~Greve},\ and\ \citenamefont
  {Lukin}}]{khabiboulline2019quantum}%
  \BibitemOpen
  \bibfield  {author} {\bibinfo {author} {\bibfnamefont {E.~T.}\ \bibnamefont
  {Khabiboulline}}, \bibinfo {author} {\bibfnamefont {J.}~\bibnamefont
  {Borregaard}}, \bibinfo {author} {\bibfnamefont {K.}~\bibnamefont
  {De~Greve}},\ and\ \bibinfo {author} {\bibfnamefont {M.~D.}\ \bibnamefont
  {Lukin}},\ }\bibfield  {title} {\bibinfo {title} {Quantum-assisted telescope
  arrays},\ }\href@noop {} {\bibfield  {journal} {\bibinfo  {journal} {Physical
  review A}\ }\textbf {\bibinfo {volume} {100}},\ \bibinfo {pages} {022316}
  (\bibinfo {year} {2019}{\natexlab{b}})}\BibitemShut {NoStop}%
\bibitem [{\citenamefont {Huang}\ \emph {et~al.}(2022)\citenamefont {Huang},
  \citenamefont {Brennen},\ and\ \citenamefont
  {Ouyang}}]{PhysRevLett.129.210502}%
  \BibitemOpen
  \bibfield  {author} {\bibinfo {author} {\bibfnamefont {Z.}~\bibnamefont
  {Huang}}, \bibinfo {author} {\bibfnamefont {G.~K.}\ \bibnamefont {Brennen}},\
  and\ \bibinfo {author} {\bibfnamefont {Y.}~\bibnamefont {Ouyang}},\
  }\bibfield  {title} {\bibinfo {title} {Imaging stars with quantum error
  correction},\ }\href {https://doi.org/10.1103/PhysRevLett.129.210502}
  {\bibfield  {journal} {\bibinfo  {journal} {Physical Review Letters}\
  }\textbf {\bibinfo {volume} {129}},\ \bibinfo {pages} {210502} (\bibinfo
  {year} {2022})}\BibitemShut {NoStop}%
\bibitem [{\citenamefont {Huang}\ \emph
  {et~al.}(2024{\natexlab{a}})\citenamefont {Huang}, \citenamefont
  {Salces-Carcoba}, \citenamefont {Adhikari}, \citenamefont {Safavi-Naeini},\
  and\ \citenamefont {Jiang}}]{huang2024vacuum}%
  \BibitemOpen
  \bibfield  {author} {\bibinfo {author} {\bibfnamefont {Y.}~\bibnamefont
  {Huang}}, \bibinfo {author} {\bibfnamefont {F.}~\bibnamefont
  {Salces-Carcoba}}, \bibinfo {author} {\bibfnamefont {R.~X.}\ \bibnamefont
  {Adhikari}}, \bibinfo {author} {\bibfnamefont {A.~H.}\ \bibnamefont
  {Safavi-Naeini}},\ and\ \bibinfo {author} {\bibfnamefont {L.}~\bibnamefont
  {Jiang}},\ }\bibfield  {title} {\bibinfo {title} {Vacuum beam guide for large
  scale quantum networks},\ }\href@noop {} {\bibfield  {journal} {\bibinfo
  {journal} {Physical Review Letters}\ }\textbf {\bibinfo {volume} {133}},\
  \bibinfo {pages} {020801} (\bibinfo {year} {2024}{\natexlab{a}})}\BibitemShut
  {NoStop}%
\bibitem [{\citenamefont {{Monnier}}(2003)}]{monnier2003}%
  \BibitemOpen
  \bibfield  {author} {\bibinfo {author} {\bibfnamefont {J.~D.}\ \bibnamefont
  {{Monnier}}},\ }\bibfield  {title} {\bibinfo {title} {{Optical interferometry
  in astronomy}},\ }\href {https://doi.org/10.1088/0034-4885/66/5/203}
  {\bibfield  {journal} {\bibinfo  {journal} {Reports on Progress in Physics}\
  }\textbf {\bibinfo {volume} {66}},\ \bibinfo {pages} {789} (\bibinfo {year}
  {2003})},\ \Eprint {https://arxiv.org/abs/astro-ph/0307036}
  {arXiv:astro-ph/0307036 [astro-ph]} \BibitemShut {NoStop}%
\bibitem [{\citenamefont {McAlister}\ \emph {et~al.}(2005)\citenamefont
  {McAlister}, \citenamefont {Ten~Brummelaar}, \citenamefont {Gies},
  \citenamefont {Huang}, \citenamefont {Bagnuolo~Jr}, \citenamefont {Shure},
  \citenamefont {Sturmann}, \citenamefont {Sturmann}, \citenamefont {Turner},
  \citenamefont {Taylor} \emph {et~al.}}]{chara}%
  \BibitemOpen
  \bibfield  {author} {\bibinfo {author} {\bibfnamefont {H.~A.}\ \bibnamefont
  {McAlister}}, \bibinfo {author} {\bibfnamefont {T.}~\bibnamefont
  {Ten~Brummelaar}}, \bibinfo {author} {\bibfnamefont {D.}~\bibnamefont
  {Gies}}, \bibinfo {author} {\bibfnamefont {W.}~\bibnamefont {Huang}},
  \bibinfo {author} {\bibfnamefont {W.}~\bibnamefont {Bagnuolo~Jr}}, \bibinfo
  {author} {\bibfnamefont {M.}~\bibnamefont {Shure}}, \bibinfo {author}
  {\bibfnamefont {J.}~\bibnamefont {Sturmann}}, \bibinfo {author}
  {\bibfnamefont {L.}~\bibnamefont {Sturmann}}, \bibinfo {author}
  {\bibfnamefont {N.}~\bibnamefont {Turner}}, \bibinfo {author} {\bibfnamefont
  {S.}~\bibnamefont {Taylor}}, \emph {et~al.},\ }\bibfield  {title} {\bibinfo
  {title} {First results from the chara array. i. an interferometric and
  spectroscopic study of the fast rotator $\alpha$ leonis (regulus)},\
  }\href@noop {} {\bibfield  {journal} {\bibinfo  {journal} {The Astrophysical
  Journal}\ }\textbf {\bibinfo {volume} {628}},\ \bibinfo {pages} {439}
  (\bibinfo {year} {2005})}\BibitemShut {NoStop}%
\bibitem [{\citenamefont {{M{\'e}rand}}\ \emph {et~al.}(2014)\citenamefont
  {{M{\'e}rand}}, \citenamefont {{Abuter}}, \citenamefont {{Aller-Carpentier}},
  \citenamefont {{Andolfato}}, \citenamefont {{Alonso}}, \citenamefont
  {{Berger}}, \citenamefont {{Blanchard}}, \citenamefont {{Boffin}},
  \citenamefont {{Bourget}}, \citenamefont {{Bristow}}, \citenamefont {{Cid}},
  \citenamefont {{de Wit}}, \citenamefont {{del Valle}}, \citenamefont
  {{Delplancke-Str{\"o}bele}}, \citenamefont {{Derie}}, \citenamefont
  {{Faundez}}, \citenamefont {{Ertel}}, \citenamefont {{Grellmann}},
  \citenamefont {{Gitton}}, \citenamefont {{Glindemann}}, \citenamefont
  {{Guajardo}}, \citenamefont {{Guieu}}, \citenamefont {{Guisard}},
  \citenamefont {{Guniat}}, \citenamefont {{Haguenauer}}, \citenamefont
  {{Herrera}}, \citenamefont {{Hummel}}, \citenamefont {{La Fuente}},
  \citenamefont {{Lopez}}, \citenamefont {{Mardones}}, \citenamefont {{Morel}},
  \citenamefont {{M{\"u}ller}}, \citenamefont {{Percheron}}, \citenamefont
  {{Duc}}, \citenamefont {{Pino}}, \citenamefont {{Poupar}}, \citenamefont
  {{Pozna}}, \citenamefont {{Ramirez}}, \citenamefont {{Rengaswamy}},
  \citenamefont {{Rivas}}, \citenamefont {{Rivinius}}, \citenamefont
  {{Segovia}}, \citenamefont {{Schmid}}, \citenamefont {{Sch{\"o}ller}},
  \citenamefont {{Schuhler}}, \citenamefont {{Woillez}},\ and\ \citenamefont
  {{Wittkowski}}}]{vlti}%
  \BibitemOpen
  \bibfield  {author} {\bibinfo {author} {\bibfnamefont {A.}~\bibnamefont
  {{M{\'e}rand}}}, \bibinfo {author} {\bibfnamefont {R.}~\bibnamefont
  {{Abuter}}}, \bibinfo {author} {\bibfnamefont {E.}~\bibnamefont
  {{Aller-Carpentier}}}, \bibinfo {author} {\bibfnamefont {L.}~\bibnamefont
  {{Andolfato}}}, \bibinfo {author} {\bibfnamefont {J.}~\bibnamefont
  {{Alonso}}}, \bibinfo {author} {\bibfnamefont {J.-P.}\ \bibnamefont
  {{Berger}}}, \bibinfo {author} {\bibfnamefont {G.}~\bibnamefont
  {{Blanchard}}}, \bibinfo {author} {\bibfnamefont {H.}~\bibnamefont
  {{Boffin}}}, \bibinfo {author} {\bibfnamefont {P.}~\bibnamefont {{Bourget}}},
  \bibinfo {author} {\bibfnamefont {P.}~\bibnamefont {{Bristow}}}, \bibinfo
  {author} {\bibfnamefont {C.}~\bibnamefont {{Cid}}}, \bibinfo {author}
  {\bibfnamefont {W.-J.}\ \bibnamefont {{de Wit}}}, \bibinfo {author}
  {\bibfnamefont {D.}~\bibnamefont {{del Valle}}}, \bibinfo {author}
  {\bibfnamefont {F.}~\bibnamefont {{Delplancke-Str{\"o}bele}}}, \bibinfo
  {author} {\bibfnamefont {F.}~\bibnamefont {{Derie}}}, \bibinfo {author}
  {\bibfnamefont {L.}~\bibnamefont {{Faundez}}}, \bibinfo {author}
  {\bibfnamefont {S.}~\bibnamefont {{Ertel}}}, \bibinfo {author} {\bibfnamefont
  {R.}~\bibnamefont {{Grellmann}}}, \bibinfo {author} {\bibfnamefont
  {P.}~\bibnamefont {{Gitton}}}, \bibinfo {author} {\bibfnamefont
  {A.}~\bibnamefont {{Glindemann}}}, \bibinfo {author} {\bibfnamefont
  {P.}~\bibnamefont {{Guajardo}}}, \bibinfo {author} {\bibfnamefont
  {S.}~\bibnamefont {{Guieu}}}, \bibinfo {author} {\bibfnamefont
  {S.}~\bibnamefont {{Guisard}}}, \bibinfo {author} {\bibfnamefont
  {S.}~\bibnamefont {{Guniat}}}, \bibinfo {author} {\bibfnamefont
  {P.}~\bibnamefont {{Haguenauer}}}, \bibinfo {author} {\bibfnamefont
  {C.}~\bibnamefont {{Herrera}}}, \bibinfo {author} {\bibfnamefont
  {C.}~\bibnamefont {{Hummel}}}, \bibinfo {author} {\bibfnamefont
  {C.}~\bibnamefont {{La Fuente}}}, \bibinfo {author} {\bibfnamefont
  {M.}~\bibnamefont {{Lopez}}}, \bibinfo {author} {\bibfnamefont
  {P.}~\bibnamefont {{Mardones}}}, \bibinfo {author} {\bibfnamefont
  {S.}~\bibnamefont {{Morel}}}, \bibinfo {author} {\bibfnamefont
  {A.}~\bibnamefont {{M{\"u}ller}}}, \bibinfo {author} {\bibfnamefont
  {I.}~\bibnamefont {{Percheron}}}, \bibinfo {author} {\bibfnamefont {T.~P.}\
  \bibnamefont {{Duc}}}, \bibinfo {author} {\bibfnamefont {A.}~\bibnamefont
  {{Pino}}}, \bibinfo {author} {\bibfnamefont {S.}~\bibnamefont {{Poupar}}},
  \bibinfo {author} {\bibfnamefont {E.}~\bibnamefont {{Pozna}}}, \bibinfo
  {author} {\bibfnamefont {A.}~\bibnamefont {{Ramirez}}}, \bibinfo {author}
  {\bibfnamefont {S.}~\bibnamefont {{Rengaswamy}}}, \bibinfo {author}
  {\bibfnamefont {L.}~\bibnamefont {{Rivas}}}, \bibinfo {author} {\bibfnamefont
  {T.}~\bibnamefont {{Rivinius}}}, \bibinfo {author} {\bibfnamefont
  {A.}~\bibnamefont {{Segovia}}}, \bibinfo {author} {\bibfnamefont
  {C.}~\bibnamefont {{Schmid}}}, \bibinfo {author} {\bibfnamefont
  {M.}~\bibnamefont {{Sch{\"o}ller}}}, \bibinfo {author} {\bibfnamefont
  {N.}~\bibnamefont {{Schuhler}}}, \bibinfo {author} {\bibfnamefont
  {J.}~\bibnamefont {{Woillez}}},\ and\ \bibinfo {author} {\bibfnamefont
  {M.}~\bibnamefont {{Wittkowski}}},\ }\bibfield  {title} {\bibinfo {title}
  {{VLTI status update: a decade of operations and beyond}},\ }in\ \href
  {https://doi.org/10.1117/12.2057150} {\emph {\bibinfo {booktitle} {Optical
  and Infrared Interferometry IV}}},\ \bibinfo {series} {Society of
  Photo-Optical Instrumentation Engineers (SPIE) Conference Series}, Vol.\
  \bibinfo {volume} {9146},\ \bibinfo {editor} {edited by\ \bibinfo {editor}
  {\bibfnamefont {J.~K.}\ \bibnamefont {{Rajagopal}}}, \bibinfo {editor}
  {\bibfnamefont {M.~J.}\ \bibnamefont {{Creech-Eakman}}},\ and\ \bibinfo
  {editor} {\bibfnamefont {F.}~\bibnamefont {{Malbet}}}}\ (\bibinfo {year}
  {2014})\ p.\ \bibinfo {pages} {91460J},\ \Eprint
  {https://arxiv.org/abs/1407.2785} {arXiv:1407.2785 [astro-ph.IM]}
  \BibitemShut {NoStop}%
\bibitem [{\citenamefont {{van Belle}}\ and\ \citenamefont
  {{Jorgensen}}(2024)}]{bft}%
  \BibitemOpen
  \bibfield  {author} {\bibinfo {author} {\bibfnamefont {G.~T.}\ \bibnamefont
  {{van Belle}}}\ and\ \bibinfo {author} {\bibfnamefont {A.~M.}\ \bibnamefont
  {{Jorgensen}}},\ }\bibfield  {title} {\bibinfo {title} {{The Big Fringe
  Telescope}},\ }in\ \href {https://doi.org/10.1117/12.3020515} {\emph
  {\bibinfo {booktitle} {Optical and Infrared Interferometry and Imaging
  IX}}},\ \bibinfo {series} {Society of Photo-Optical Instrumentation Engineers
  (SPIE) Conference Series}, Vol.\ \bibinfo {volume} {13095},\ \bibinfo
  {editor} {edited by\ \bibinfo {editor} {\bibfnamefont {J.}~\bibnamefont
  {{Kammerer}}}, \bibinfo {editor} {\bibfnamefont {S.}~\bibnamefont
  {{Sallum}}},\ and\ \bibinfo {editor} {\bibfnamefont {J.}~\bibnamefont
  {{Sanchez-Bermudez}}}}\ (\bibinfo {year} {2024})\ p.\ \bibinfo {pages}
  {130951R},\ \Eprint {https://arxiv.org/abs/2408.01386} {arXiv:2408.01386
  [astro-ph.IM]} \BibitemShut {NoStop}%
\bibitem [{\citenamefont {{Townes}}\ \emph {et~al.}(1998)\citenamefont
  {{Townes}}, \citenamefont {{Bester}}, \citenamefont {{Danchi}}, \citenamefont
  {{Hale}}, \citenamefont {{Monnier}}, \citenamefont {{Lipman}}, \citenamefont
  {{Tuthill}}, \citenamefont {{Johnson}},\ and\ \citenamefont
  {{Walters}}}]{townes98}%
  \BibitemOpen
  \bibfield  {author} {\bibinfo {author} {\bibfnamefont {C.~H.}\ \bibnamefont
  {{Townes}}}, \bibinfo {author} {\bibfnamefont {M.}~\bibnamefont {{Bester}}},
  \bibinfo {author} {\bibfnamefont {W.~C.}\ \bibnamefont {{Danchi}}}, \bibinfo
  {author} {\bibfnamefont {D.~D.}\ \bibnamefont {{Hale}}}, \bibinfo {author}
  {\bibfnamefont {J.~D.}\ \bibnamefont {{Monnier}}}, \bibinfo {author}
  {\bibfnamefont {E.~A.}\ \bibnamefont {{Lipman}}}, \bibinfo {author}
  {\bibfnamefont {P.~G.}\ \bibnamefont {{Tuthill}}}, \bibinfo {author}
  {\bibfnamefont {M.~A.}\ \bibnamefont {{Johnson}}},\ and\ \bibinfo {author}
  {\bibfnamefont {D.~L.}\ \bibnamefont {{Walters}}},\ }\bibfield  {title}
  {\bibinfo {title} {{Infrared Spatial Interferometer}},\ }in\ \href
  {https://doi.org/10.1117/12.317159} {\emph {\bibinfo {booktitle}
  {Astronomical Interferometry}}},\ \bibinfo {series} {Society of Photo-Optical
  Instrumentation Engineers (SPIE) Conference Series}, Vol.\ \bibinfo {volume}
  {3350},\ \bibinfo {editor} {edited by\ \bibinfo {editor} {\bibfnamefont
  {R.~D.}\ \bibnamefont {{Reasenberg}}}}\ (\bibinfo {year} {1998})\ pp.\
  \bibinfo {pages} {908--932}\BibitemShut {NoStop}%
\bibitem [{\citenamefont {Nunez}\ \emph {et~al.}(2012)\citenamefont {Nunez},
  \citenamefont {Holmes}, \citenamefont {Kieda},\ and\ \citenamefont
  {LeBohec}}]{nunez12}%
  \BibitemOpen
  \bibfield  {author} {\bibinfo {author} {\bibfnamefont {P.~D.}\ \bibnamefont
  {Nunez}}, \bibinfo {author} {\bibfnamefont {R.}~\bibnamefont {Holmes}},
  \bibinfo {author} {\bibfnamefont {D.}~\bibnamefont {Kieda}},\ and\ \bibinfo
  {author} {\bibfnamefont {S.}~\bibnamefont {LeBohec}},\ }\bibfield  {title}
  {\bibinfo {title} {High angular resolution imaging with stellar intensity
  interferometry using air cherenkov telescope arrays},\ }\href@noop {}
  {\bibfield  {journal} {\bibinfo  {journal} {Monthly Notices of the Royal
  Astronomical Society}\ }\textbf {\bibinfo {volume} {419}},\ \bibinfo {pages}
  {172} (\bibinfo {year} {2012})}\BibitemShut {NoStop}%
\bibitem [{\citenamefont {Brown}\ and\ \citenamefont
  {Twiss}(1956)}]{brown1956correlation}%
  \BibitemOpen
  \bibfield  {author} {\bibinfo {author} {\bibfnamefont {R.~H.}\ \bibnamefont
  {Brown}}\ and\ \bibinfo {author} {\bibfnamefont {R.~Q.}\ \bibnamefont
  {Twiss}},\ }\bibfield  {title} {\bibinfo {title} {Correlation between photons
  in two coherent beams of light},\ }\href@noop {} {\bibfield  {journal}
  {\bibinfo  {journal} {Nature}\ }\textbf {\bibinfo {volume} {177}},\ \bibinfo
  {pages} {27} (\bibinfo {year} {1956})}\BibitemShut {NoStop}%
\bibitem [{\citenamefont {Bojer}\ \emph {et~al.}(2022)\citenamefont {Bojer},
  \citenamefont {Huang}, \citenamefont {Karl}, \citenamefont {Richter},
  \citenamefont {Kok},\ and\ \citenamefont {von
  Zanthier}}]{bojer2022quantitative}%
  \BibitemOpen
  \bibfield  {author} {\bibinfo {author} {\bibfnamefont {M.}~\bibnamefont
  {Bojer}}, \bibinfo {author} {\bibfnamefont {Z.}~\bibnamefont {Huang}},
  \bibinfo {author} {\bibfnamefont {S.}~\bibnamefont {Karl}}, \bibinfo {author}
  {\bibfnamefont {S.}~\bibnamefont {Richter}}, \bibinfo {author} {\bibfnamefont
  {P.}~\bibnamefont {Kok}},\ and\ \bibinfo {author} {\bibfnamefont
  {J.}~\bibnamefont {von Zanthier}},\ }\bibfield  {title} {\bibinfo {title} {A
  quantitative comparison of amplitude versus intensity interferometry for
  astronomy},\ }\href@noop {} {\bibfield  {journal} {\bibinfo  {journal} {New
  Journal of Physics}\ }\textbf {\bibinfo {volume} {24}},\ \bibinfo {pages}
  {043026} (\bibinfo {year} {2022})}\BibitemShut {NoStop}%
\bibitem [{\citenamefont {Genzel}\ \emph {et~al.}(2024)\citenamefont {Genzel},
  \citenamefont {Eisenhauer},\ and\ \citenamefont
  {Gillessen}}]{genzel2024experimental}%
  \BibitemOpen
  \bibfield  {author} {\bibinfo {author} {\bibfnamefont {R.}~\bibnamefont
  {Genzel}}, \bibinfo {author} {\bibfnamefont {F.}~\bibnamefont {Eisenhauer}},\
  and\ \bibinfo {author} {\bibfnamefont {S.}~\bibnamefont {Gillessen}},\
  }\bibfield  {title} {\bibinfo {title} {Experimental studies of black holes:
  status and future prospects},\ }\href@noop {} {\bibfield  {journal} {\bibinfo
   {journal} {The Astronomy and Astrophysics Review}\ }\textbf {\bibinfo
  {volume} {32}},\ \bibinfo {pages} {3} (\bibinfo {year} {2024})}\BibitemShut
  {NoStop}%
\bibitem [{\citenamefont {Abuter}\ \emph {et~al.}(2017)\citenamefont {Abuter},
  \citenamefont {Accardo}, \citenamefont {Amorim}, \citenamefont {Anugu},
  \citenamefont {Avila}, \citenamefont {Azouaoui}, \citenamefont {Benisty},
  \citenamefont {Berger}, \citenamefont {Blind}, \citenamefont {Bonnet} \emph
  {et~al.}}]{abuter2017first}%
  \BibitemOpen
  \bibfield  {author} {\bibinfo {author} {\bibfnamefont {R.}~\bibnamefont
  {Abuter}}, \bibinfo {author} {\bibfnamefont {M.}~\bibnamefont {Accardo}},
  \bibinfo {author} {\bibfnamefont {A.}~\bibnamefont {Amorim}}, \bibinfo
  {author} {\bibfnamefont {N.}~\bibnamefont {Anugu}}, \bibinfo {author}
  {\bibfnamefont {G.}~\bibnamefont {Avila}}, \bibinfo {author} {\bibfnamefont
  {N.}~\bibnamefont {Azouaoui}}, \bibinfo {author} {\bibfnamefont
  {M.}~\bibnamefont {Benisty}}, \bibinfo {author} {\bibfnamefont {J.-P.}\
  \bibnamefont {Berger}}, \bibinfo {author} {\bibfnamefont {N.}~\bibnamefont
  {Blind}}, \bibinfo {author} {\bibfnamefont {H.}~\bibnamefont {Bonnet}}, \emph
  {et~al.},\ }\bibfield  {title} {\bibinfo {title} {First light for gravity:
  Phase referencing optical interferometry for the very large telescope
  interferometer},\ }\href@noop {} {\bibfield  {journal} {\bibinfo  {journal}
  {Astronomy \& Astrophysics}\ }\textbf {\bibinfo {volume} {602}},\ \bibinfo
  {pages} {A94} (\bibinfo {year} {2017})}\BibitemShut {NoStop}%
\bibitem [{\citenamefont {Amorim}\ \emph {et~al.}(2019)\citenamefont {Amorim},
  \citenamefont {Baub{\"o}ck}, \citenamefont {Berger}, \citenamefont
  {Brandner}, \citenamefont {Cl{\'e}net}, \citenamefont {Coud{\'e}~du Foresto},
  \citenamefont {de~Zeeuw}, \citenamefont {Dexter}, \citenamefont {Duvert},
  \citenamefont {Ebert} \emph {et~al.}}]{amorim2019test}%
  \BibitemOpen
  \bibfield  {author} {\bibinfo {author} {\bibfnamefont {A.}~\bibnamefont
  {Amorim}}, \bibinfo {author} {\bibfnamefont {M.}~\bibnamefont {Baub{\"o}ck}},
  \bibinfo {author} {\bibfnamefont {J.}~\bibnamefont {Berger}}, \bibinfo
  {author} {\bibfnamefont {W.}~\bibnamefont {Brandner}}, \bibinfo {author}
  {\bibfnamefont {Y.}~\bibnamefont {Cl{\'e}net}}, \bibinfo {author}
  {\bibfnamefont {V.}~\bibnamefont {Coud{\'e}~du Foresto}}, \bibinfo {author}
  {\bibfnamefont {P.}~\bibnamefont {de~Zeeuw}}, \bibinfo {author}
  {\bibfnamefont {J.}~\bibnamefont {Dexter}}, \bibinfo {author} {\bibfnamefont
  {G.}~\bibnamefont {Duvert}}, \bibinfo {author} {\bibfnamefont
  {M.}~\bibnamefont {Ebert}}, \emph {et~al.},\ }\bibfield  {title} {\bibinfo
  {title} {Test of the einstein equivalence principle near the galactic center
  supermassive black hole},\ }\href@noop {} {\bibfield  {journal} {\bibinfo
  {journal} {Physical Review Letters}\ }\textbf {\bibinfo {volume} {122}},\
  \bibinfo {pages} {101102} (\bibinfo {year} {2019})}\BibitemShut {NoStop}%
\bibitem [{\citenamefont {Abuter}\ \emph {et~al.}(2021)\citenamefont {Abuter},
  \citenamefont {Amorim}, \citenamefont {Baub{\"o}ck}, \citenamefont
  {Baganoff}, \citenamefont {Berger}, \citenamefont {Boyce}, \citenamefont
  {Bonnet}, \citenamefont {Brandner}, \citenamefont {Cl{\'e}net}, \citenamefont
  {Davies} \emph {et~al.}}]{abuter2021constraining}%
  \BibitemOpen
  \bibfield  {author} {\bibinfo {author} {\bibfnamefont {R.}~\bibnamefont
  {Abuter}}, \bibinfo {author} {\bibfnamefont {A.}~\bibnamefont {Amorim}},
  \bibinfo {author} {\bibfnamefont {M.}~\bibnamefont {Baub{\"o}ck}}, \bibinfo
  {author} {\bibfnamefont {F.}~\bibnamefont {Baganoff}}, \bibinfo {author}
  {\bibfnamefont {J.}~\bibnamefont {Berger}}, \bibinfo {author} {\bibfnamefont
  {H.}~\bibnamefont {Boyce}}, \bibinfo {author} {\bibfnamefont
  {H.}~\bibnamefont {Bonnet}}, \bibinfo {author} {\bibfnamefont
  {W.}~\bibnamefont {Brandner}}, \bibinfo {author} {\bibfnamefont
  {Y.}~\bibnamefont {Cl{\'e}net}}, \bibinfo {author} {\bibfnamefont
  {R.}~\bibnamefont {Davies}}, \emph {et~al.},\ }\bibfield  {title} {\bibinfo
  {title} {Constraining particle acceleration in sgr a$*$ with simultaneous
  gravity, spitzer, nustar, and chandra observations},\ }\href@noop {}
  {\bibfield  {journal} {\bibinfo  {journal} {Astronomy \& Astrophysics}\
  }\textbf {\bibinfo {volume} {654}},\ \bibinfo {pages} {A22} (\bibinfo {year}
  {2021})}\BibitemShut {NoStop}%
\bibitem [{\citenamefont {{Lovell}}\ \emph {et~al.}(2013)\citenamefont
  {{Lovell}}, \citenamefont {{McCallum}}, \citenamefont {{Reid}}, \citenamefont
  {{McCulloch}}, \citenamefont {{Baynes}}, \citenamefont {{Dickey}},
  \citenamefont {{Shabala}}, \citenamefont {{Watson}}, \citenamefont {{Titov}},
  \citenamefont {{Ruddick}}, \citenamefont {{Twilley}}, \citenamefont
  {{Reynolds}}, \citenamefont {{Tingay}}, \citenamefont {{Shield}},
  \citenamefont {{Adada}}, \citenamefont {{Ellingsen}}, \citenamefont
  {{Morgan}},\ and\ \citenamefont {{Bignall}}}]{Lovell2013}%
  \BibitemOpen
  \bibfield  {author} {\bibinfo {author} {\bibfnamefont {J.~E.~J.}\
  \bibnamefont {{Lovell}}}, \bibinfo {author} {\bibfnamefont {J.~N.}\
  \bibnamefont {{McCallum}}}, \bibinfo {author} {\bibfnamefont {P.~B.}\
  \bibnamefont {{Reid}}}, \bibinfo {author} {\bibfnamefont {P.~M.}\
  \bibnamefont {{McCulloch}}}, \bibinfo {author} {\bibfnamefont {B.~E.}\
  \bibnamefont {{Baynes}}}, \bibinfo {author} {\bibfnamefont {J.~M.}\
  \bibnamefont {{Dickey}}}, \bibinfo {author} {\bibfnamefont {S.~S.}\
  \bibnamefont {{Shabala}}}, \bibinfo {author} {\bibfnamefont {C.~S.}\
  \bibnamefont {{Watson}}}, \bibinfo {author} {\bibfnamefont {O.}~\bibnamefont
  {{Titov}}}, \bibinfo {author} {\bibfnamefont {R.}~\bibnamefont {{Ruddick}}},
  \bibinfo {author} {\bibfnamefont {R.}~\bibnamefont {{Twilley}}}, \bibinfo
  {author} {\bibfnamefont {C.}~\bibnamefont {{Reynolds}}}, \bibinfo {author}
  {\bibfnamefont {S.~J.}\ \bibnamefont {{Tingay}}}, \bibinfo {author}
  {\bibfnamefont {P.}~\bibnamefont {{Shield}}}, \bibinfo {author}
  {\bibfnamefont {R.}~\bibnamefont {{Adada}}}, \bibinfo {author} {\bibfnamefont
  {S.~P.}\ \bibnamefont {{Ellingsen}}}, \bibinfo {author} {\bibfnamefont
  {J.~S.}\ \bibnamefont {{Morgan}}},\ and\ \bibinfo {author} {\bibfnamefont
  {H.~E.}\ \bibnamefont {{Bignall}}},\ }\bibfield  {title} {\bibinfo {title}
  {{The AuScope geodetic VLBI array}},\ }\href
  {https://doi.org/10.1007/s00190-013-0626-3} {\bibfield  {journal} {\bibinfo
  {journal} {Journal of Geodesy}\ }\textbf {\bibinfo {volume} {87}},\ \bibinfo
  {pages} {527} (\bibinfo {year} {2013})},\ \Eprint
  {https://arxiv.org/abs/1304.3213} {arXiv:1304.3213 [astro-ph.IM]}
  \BibitemShut {NoStop}%
\bibitem [{\citenamefont {Hai-Tao}\ \emph {et~al.}(2013)\citenamefont
  {Hai-Tao}, \citenamefont {Wei-Jun}, \citenamefont {Bei}, \citenamefont
  {Gen-Ru}, \citenamefont {Jie},\ and\ \citenamefont {Cheng-Lin}}]{Yin2013}%
  \BibitemOpen
  \bibfield  {author} {\bibinfo {author} {\bibfnamefont {Y.}~\bibnamefont
  {Hai-Tao}}, \bibinfo {author} {\bibfnamefont {G.}~\bibnamefont {Wei-Jun}},
  \bibinfo {author} {\bibfnamefont {H.}~\bibnamefont {Bei}}, \bibinfo {author}
  {\bibfnamefont {X.}~\bibnamefont {Gen-Ru}}, \bibinfo {author} {\bibfnamefont
  {L.}~\bibnamefont {Jie}},\ and\ \bibinfo {author} {\bibfnamefont
  {Z.}~\bibnamefont {Cheng-Lin}},\ }\bibfield  {title} {\bibinfo {title}
  {Effects of the 2011 tohoku-oki m9.0 earthquake on the crustal movement in
  the shandong area derived from gps data},\ }\href
  {https://doi.org/https://doi.org/10.1002/cjg2.20025} {\bibfield  {journal}
  {\bibinfo  {journal} {Chinese Journal of Geophysics}\ }\textbf {\bibinfo
  {volume} {56}},\ \bibinfo {pages} {243} (\bibinfo {year} {2013})}\BibitemShut
  {NoStop}%
\bibitem [{\citenamefont {Malkin}(2024)}]{malkin2024should}%
  \BibitemOpen
  \bibfield  {author} {\bibinfo {author} {\bibfnamefont {Z.}~\bibnamefont
  {Malkin}},\ }\bibfield  {title} {\bibinfo {title} {Should we expect further
  acceleration of the earth’s rotation in the coming years?},\ }\href@noop {}
  {\bibfield  {journal} {\bibinfo  {journal} {Astronomy Reports}\ }\textbf
  {\bibinfo {volume} {68}},\ \bibinfo {pages} {1022} (\bibinfo {year}
  {2024})}\BibitemShut {NoStop}%
\bibitem [{\citenamefont {{Nelson}}\ \emph {et~al.}(2001)\citenamefont
  {{Nelson}}, \citenamefont {{McCarthy}}, \citenamefont {{Malys}},
  \citenamefont {{Levine}}, \citenamefont {{Guinot}}, \citenamefont
  {{Fliegel}}, \citenamefont {{Beard}},\ and\ \citenamefont
  {{Bartholomew}}}]{Nelson2001}%
  \BibitemOpen
  \bibfield  {author} {\bibinfo {author} {\bibfnamefont {R.~A.}\ \bibnamefont
  {{Nelson}}}, \bibinfo {author} {\bibfnamefont {D.~D.}\ \bibnamefont
  {{McCarthy}}}, \bibinfo {author} {\bibfnamefont {S.}~\bibnamefont {{Malys}}},
  \bibinfo {author} {\bibfnamefont {J.}~\bibnamefont {{Levine}}}, \bibinfo
  {author} {\bibfnamefont {B.}~\bibnamefont {{Guinot}}}, \bibinfo {author}
  {\bibfnamefont {H.~F.}\ \bibnamefont {{Fliegel}}}, \bibinfo {author}
  {\bibfnamefont {R.~L.}\ \bibnamefont {{Beard}}},\ and\ \bibinfo {author}
  {\bibfnamefont {T.~R.}\ \bibnamefont {{Bartholomew}}},\ }\bibfield  {title}
  {\bibinfo {title} {{The leap second: its history and possible future}},\
  }\href {https://doi.org/10.1088/0026-1394/38/6/6} {\bibfield  {journal}
  {\bibinfo  {journal} {Metrologia}\ }\textbf {\bibinfo {volume} {38}},\
  \bibinfo {pages} {509} (\bibinfo {year} {2001})}\BibitemShut {NoStop}%
\bibitem [{\citenamefont {Agnew}(2024)}]{Agnew2024}%
  \BibitemOpen
  \bibfield  {author} {\bibinfo {author} {\bibfnamefont {D.}~\bibnamefont
  {Agnew}},\ }\bibfield  {title} {\bibinfo {title} {A global timekeeping
  problem postponed by global warming},\ }\href
  {https://doi.org/10.1038/s41586-024-07170-0} {\bibfield  {journal} {\bibinfo
  {journal} {Nature}\ }\textbf {\bibinfo {volume} {628}},\ \bibinfo {pages} {1}
  (\bibinfo {year} {2024})}\BibitemShut {NoStop}%
\bibitem [{\citenamefont {Bizouard}\ and\ \citenamefont
  {Zotov}(2013)}]{bizouard2013}%
  \BibitemOpen
  \bibfield  {author} {\bibinfo {author} {\bibfnamefont {C.}~\bibnamefont
  {Bizouard}}\ and\ \bibinfo {author} {\bibfnamefont {L.}~\bibnamefont
  {Zotov}},\ }\bibfield  {title} {\bibinfo {title} {Asymmetric effects on
  earth’s polar motion},\ }\href
  {https://api.semanticscholar.org/CorpusID:254384568} {\bibfield  {journal}
  {\bibinfo  {journal} {Celestial Mechanics and Dynamical Astronomy}\ }\textbf
  {\bibinfo {volume} {116}},\ \bibinfo {pages} {195 } (\bibinfo {year}
  {2013})}\BibitemShut {NoStop}%
\bibitem [{Kar(2017)}]{Karbon2017}%
  \BibitemOpen
  \bibfield  {title} {\bibinfo {title} {Earth orientation parameters from vlbi
  determined with a kalman filter},\ }\href
  {https://doi.org/https://doi.org/10.1016/j.geog.2017.05.006} {\bibfield
  {journal} {\bibinfo  {journal} {Geodesy and Geodynamics}\ }\textbf {\bibinfo
  {volume} {8}},\ \bibinfo {pages} {396} (\bibinfo {year} {2017})},\ \bibinfo
  {note} {geodesy, Astronomy and Geophysics in Earth Rotation}\BibitemShut
  {NoStop}%
\bibitem [{\citenamefont {Shirai}\ and\ \citenamefont
  {Fukushima}(2001)}]{Shirai2001}%
  \BibitemOpen
  \bibfield  {author} {\bibinfo {author} {\bibfnamefont {T.}~\bibnamefont
  {Shirai}}\ and\ \bibinfo {author} {\bibfnamefont {T.}~\bibnamefont
  {Fukushima}},\ }\bibfield  {title} {\bibinfo {title} {Construction of a new
  forced nutation theory of the nonrigidearth},\ }\href@noop {} {\bibfield
  {journal} {\bibinfo  {journal} {The Astronomical Journal}\ }\textbf {\bibinfo
  {volume} {121}},\ \bibinfo {pages} {3270} (\bibinfo {year}
  {2001})}\BibitemShut {NoStop}%
\bibitem [{\citenamefont {Mathews}\ \emph {et~al.}(2002)\citenamefont
  {Mathews}, \citenamefont {Herring},\ and\ \citenamefont
  {Buffett}}]{Mathews2002}%
  \BibitemOpen
  \bibfield  {author} {\bibinfo {author} {\bibfnamefont {P.~M.}\ \bibnamefont
  {Mathews}}, \bibinfo {author} {\bibfnamefont {T.~A.}\ \bibnamefont
  {Herring}},\ and\ \bibinfo {author} {\bibfnamefont {B.~A.}\ \bibnamefont
  {Buffett}},\ }\bibfield  {title} {\bibinfo {title} {Modeling of nutation and
  precession: New nutation series for nonrigid earth and insights into the
  earth's interior},\ }\href
  {https://doi.org/https://doi.org/10.1029/2001JB000390} {\bibfield  {journal}
  {\bibinfo  {journal} {Journal of Geophysical Research: Solid Earth}\ }\textbf
  {\bibinfo {volume} {107}},\ \bibinfo {pages} {ETG 3} (\bibinfo {year}
  {2002})}\BibitemShut {NoStop}%
\bibitem [{\citenamefont {Lambert}\ and\ \citenamefont
  {Dehant}(2007)}]{Lambert2007}%
  \BibitemOpen
  \bibfield  {author} {\bibinfo {author} {\bibfnamefont {S.~B.}\ \bibnamefont
  {Lambert}}\ and\ \bibinfo {author} {\bibfnamefont {V.}~\bibnamefont
  {Dehant}},\ }\bibfield  {title} {\bibinfo {title} {The earth's core
  parameters as seen by the vlbi},\ }\href@noop {} {\bibfield  {journal}
  {\bibinfo  {journal} {Astronomy \& Astrophysics}\ }\textbf {\bibinfo {volume}
  {469}},\ \bibinfo {pages} {777} (\bibinfo {year} {2007})}\BibitemShut
  {NoStop}%
\bibitem [{\citenamefont {Charlot}\ \emph {et~al.}(2020)\citenamefont
  {Charlot}, \citenamefont {Jacobs}, \citenamefont {Gordon}, \citenamefont
  {Lambert}, \citenamefont {De~Witt}, \citenamefont {B{\"o}hm}, \citenamefont
  {Fey}, \citenamefont {Heinkelmann}, \citenamefont {Skurikhina}, \citenamefont
  {Titov} \emph {et~al.}}]{charlot2020third}%
  \BibitemOpen
  \bibfield  {author} {\bibinfo {author} {\bibfnamefont {P.}~\bibnamefont
  {Charlot}}, \bibinfo {author} {\bibfnamefont {C.}~\bibnamefont {Jacobs}},
  \bibinfo {author} {\bibfnamefont {D.}~\bibnamefont {Gordon}}, \bibinfo
  {author} {\bibfnamefont {S.}~\bibnamefont {Lambert}}, \bibinfo {author}
  {\bibfnamefont {A.}~\bibnamefont {De~Witt}}, \bibinfo {author} {\bibfnamefont
  {J.}~\bibnamefont {B{\"o}hm}}, \bibinfo {author} {\bibfnamefont
  {A.}~\bibnamefont {Fey}}, \bibinfo {author} {\bibfnamefont {R.}~\bibnamefont
  {Heinkelmann}}, \bibinfo {author} {\bibfnamefont {E.}~\bibnamefont
  {Skurikhina}}, \bibinfo {author} {\bibfnamefont {O.}~\bibnamefont {Titov}},
  \emph {et~al.},\ }\bibfield  {title} {\bibinfo {title} {The third realization
  of the international celestial reference frame by very long baseline
  interferometry},\ }\href@noop {} {\bibfield  {journal} {\bibinfo  {journal}
  {Astronomy \& Astrophysics}\ }\textbf {\bibinfo {volume} {644}},\ \bibinfo
  {pages} {A159} (\bibinfo {year} {2020})}\BibitemShut {NoStop}%
\bibitem [{Kra()}]{Krasna2023}%
  \BibitemOpen
  \href@noop {} {\ }\BibitemShut {NoStop}%
\bibitem [{\citenamefont {{Jacobs}}(2009)}]{Jacobs2009}%
  \BibitemOpen
  \bibfield  {author} {\bibinfo {author} {\bibfnamefont {C.~S.}\ \bibnamefont
  {{Jacobs}}},\ }\bibfield  {title} {\bibinfo {title} {{The Celestial Frame at
  Four Radio Frequncies}},\ }in\ \href@noop {} {\emph {\bibinfo {booktitle}
  {Journ{\'e}es Syst{\`e}mes de R{\'e}f{\'e}rence Spatio-temporels 2008}}},\
  \bibinfo {editor} {edited by\ \bibinfo {editor} {\bibfnamefont
  {M.}~\bibnamefont {{Soffel}}}\ and\ \bibinfo {editor} {\bibfnamefont
  {N.}~\bibnamefont {{Capitaine}}}}\ (\bibinfo {year} {2009})\ p.\ \bibinfo
  {pages} {251}\BibitemShut {NoStop}%
\bibitem [{\citenamefont {Lupo}\ \emph {et~al.}(2020)\citenamefont {Lupo},
  \citenamefont {Huang},\ and\ \citenamefont {Kok}}]{PhysRevLett.124.080503}%
  \BibitemOpen
  \bibfield  {author} {\bibinfo {author} {\bibfnamefont {C.}~\bibnamefont
  {Lupo}}, \bibinfo {author} {\bibfnamefont {Z.}~\bibnamefont {Huang}},\ and\
  \bibinfo {author} {\bibfnamefont {P.}~\bibnamefont {Kok}},\ }\bibfield
  {title} {\bibinfo {title} {Quantum limits to incoherent imaging are achieved
  by linear interferometry},\ }\href
  {https://doi.org/10.1103/PhysRevLett.124.080503} {\bibfield  {journal}
  {\bibinfo  {journal} {Phys. Rev. Lett.}\ }\textbf {\bibinfo {volume} {124}},\
  \bibinfo {pages} {080503} (\bibinfo {year} {2020})}\BibitemShut {NoStop}%
\bibitem [{\citenamefont {Mandel}\ and\ \citenamefont
  {Wolf}(1995)}]{mandel1995optical}%
  \BibitemOpen
  \bibfield  {author} {\bibinfo {author} {\bibfnamefont {L.}~\bibnamefont
  {Mandel}}\ and\ \bibinfo {author} {\bibfnamefont {E.}~\bibnamefont {Wolf}},\
  }\href {https://doi.org/10.1017/CBO9781139644105} {\emph {\bibinfo {title}
  {Optical Coherence and Quantum Optics}}}\ (\bibinfo  {publisher} {Cambridge
  University Press},\ \bibinfo {year} {1995})\BibitemShut {NoStop}%
\bibitem [{\citenamefont {Pearce}\ \emph {et~al.}(2017)\citenamefont {Pearce},
  \citenamefont {Campbell},\ and\ \citenamefont {Kok}}]{Pearce2017optimal}%
  \BibitemOpen
  \bibfield  {author} {\bibinfo {author} {\bibfnamefont {M.~E.}\ \bibnamefont
  {Pearce}}, \bibinfo {author} {\bibfnamefont {E.~T.}\ \bibnamefont
  {Campbell}},\ and\ \bibinfo {author} {\bibfnamefont {P.}~\bibnamefont
  {Kok}},\ }\bibfield  {title} {\bibinfo {title} {Optimal quantum metrology of
  distant black bodies},\ }\href {https://doi.org/10.22331/q-2017-07-26-21}
  {\bibfield  {journal} {\bibinfo  {journal} {Quantum}\ }\textbf {\bibinfo
  {volume} {1}},\ \bibinfo {pages} {21} (\bibinfo {year} {2017})}\BibitemShut
  {NoStop}%
\bibitem [{\citenamefont {Braunstein}\ and\ \citenamefont
  {Caves}(1994)}]{caves}%
  \BibitemOpen
  \bibfield  {author} {\bibinfo {author} {\bibfnamefont {S.~L.}\ \bibnamefont
  {Braunstein}}\ and\ \bibinfo {author} {\bibfnamefont {C.~M.}\ \bibnamefont
  {Caves}},\ }\bibfield  {title} {\bibinfo {title} {Statistical distance and
  the geometry of quantum states},\ }\href
  {https://doi.org/10.1103/PhysRevLett.72.3439} {\bibfield  {journal} {\bibinfo
   {journal} {Physical Review Letters}\ }\textbf {\bibinfo {volume} {72}},\
  \bibinfo {pages} {3439} (\bibinfo {year} {1994})}\BibitemShut {NoStop}%
\bibitem [{\citenamefont {Braunstein}\ \emph {et~al.}(1996)\citenamefont
  {Braunstein}, \citenamefont {Caves},\ and\ \citenamefont {Milburn}}]{caves1}%
  \BibitemOpen
  \bibfield  {author} {\bibinfo {author} {\bibfnamefont {S.~L.}\ \bibnamefont
  {Braunstein}}, \bibinfo {author} {\bibfnamefont {C.~M.}\ \bibnamefont
  {Caves}},\ and\ \bibinfo {author} {\bibfnamefont {G.~J.}\ \bibnamefont
  {Milburn}},\ }\bibfield  {title} {\bibinfo {title} {Generalized uncertainty
  relations: Theory, examples, and {L}orentz invariance},\ }\href@noop {}
  {\bibfield  {journal} {\bibinfo  {journal} {Annals of Physics}\ }\textbf
  {\bibinfo {volume} {247}},\ \bibinfo {pages} {135} (\bibinfo {year}
  {1996})}\BibitemShut {NoStop}%
\bibitem [{\citenamefont {Giovannetti}\ \emph {et~al.}(2011)\citenamefont
  {Giovannetti}, \citenamefont {Lloyd},\ and\ \citenamefont
  {Maccone}}]{giovannetti2011advances}%
  \BibitemOpen
  \bibfield  {author} {\bibinfo {author} {\bibfnamefont {V.}~\bibnamefont
  {Giovannetti}}, \bibinfo {author} {\bibfnamefont {S.}~\bibnamefont {Lloyd}},\
  and\ \bibinfo {author} {\bibfnamefont {L.}~\bibnamefont {Maccone}},\
  }\bibfield  {title} {\bibinfo {title} {Advances in quantum metrology},\
  }\href {https://doi.org/10.1038/nphoton.2011.35} {\bibfield  {journal}
  {\bibinfo  {journal} {Nature Photonics}\ }\textbf {\bibinfo {volume} {5}},\
  \bibinfo {pages} {222} (\bibinfo {year} {2011})}\BibitemShut {NoStop}%
\bibitem [{\citenamefont {Giovannetti}\ \emph {et~al.}(2006)\citenamefont
  {Giovannetti}, \citenamefont {Lloyd},\ and\ \citenamefont
  {Maccone}}]{giovannetti2006quantum}%
  \BibitemOpen
  \bibfield  {author} {\bibinfo {author} {\bibfnamefont {V.}~\bibnamefont
  {Giovannetti}}, \bibinfo {author} {\bibfnamefont {S.}~\bibnamefont {Lloyd}},\
  and\ \bibinfo {author} {\bibfnamefont {L.}~\bibnamefont {Maccone}},\
  }\bibfield  {title} {\bibinfo {title} {Quantum metrology},\ }\href
  {https://doi.org/10.1103/PhysRevLett.96.010401} {\bibfield  {journal}
  {\bibinfo  {journal} {Physical Review Letters}\ }\textbf {\bibinfo {volume}
  {96}},\ \bibinfo {pages} {010401} (\bibinfo {year} {2006})}\BibitemShut
  {NoStop}%
\bibitem [{\citenamefont {Paris}(2009)}]{paris2009quantum}%
  \BibitemOpen
  \bibfield  {author} {\bibinfo {author} {\bibfnamefont {M.~G.}\ \bibnamefont
  {Paris}},\ }\bibfield  {title} {\bibinfo {title} {Quantum estimation for
  quantum technology},\ }\href@noop {} {\bibfield  {journal} {\bibinfo
  {journal} {International Journal of Quantum Information}\ }\textbf {\bibinfo
  {volume} {7}},\ \bibinfo {pages} {125} (\bibinfo {year} {2009})}\BibitemShut
  {NoStop}%
\bibitem [{\citenamefont {Huang}\ \emph
  {et~al.}(2024{\natexlab{b}})\citenamefont {Huang}, \citenamefont {Baragiola},
  \citenamefont {Menicucci},\ and\ \citenamefont {Wilde}}]{huang2024limited}%
  \BibitemOpen
  \bibfield  {author} {\bibinfo {author} {\bibfnamefont {Z.}~\bibnamefont
  {Huang}}, \bibinfo {author} {\bibfnamefont {B.~Q.}\ \bibnamefont
  {Baragiola}}, \bibinfo {author} {\bibfnamefont {N.~C.}\ \bibnamefont
  {Menicucci}},\ and\ \bibinfo {author} {\bibfnamefont {M.~M.}\ \bibnamefont
  {Wilde}},\ }\bibfield  {title} {\bibinfo {title} {Limited quantum advantage
  for stellar interferometry via continuous-variable teleportation},\
  }\href@noop {} {\bibfield  {journal} {\bibinfo  {journal} {Physical Review
  A}\ }\textbf {\bibinfo {volume} {109}},\ \bibinfo {pages} {052434} (\bibinfo
  {year} {2024}{\natexlab{b}})}\BibitemShut {NoStop}%
\bibitem [{\citenamefont {Hardy}(1994)}]{hardy1994nonlocality}%
  \BibitemOpen
  \bibfield  {author} {\bibinfo {author} {\bibfnamefont {L.}~\bibnamefont
  {Hardy}},\ }\bibfield  {title} {\bibinfo {title} {Nonlocality of a single
  photon revisited},\ }\href@noop {} {\bibfield  {journal} {\bibinfo  {journal}
  {Physical review letters}\ }\textbf {\bibinfo {volume} {73}},\ \bibinfo
  {pages} {2279} (\bibinfo {year} {1994})}\BibitemShut {NoStop}%
\bibitem [{\citenamefont {Hardy}(1995)}]{hardy1995hardy}%
  \BibitemOpen
  \bibfield  {author} {\bibinfo {author} {\bibfnamefont {L.}~\bibnamefont
  {Hardy}},\ }\bibfield  {title} {\bibinfo {title} {Hardy replies},\
  }\href@noop {} {\bibfield  {journal} {\bibinfo  {journal} {Physical review
  letters}\ }\textbf {\bibinfo {volume} {75}},\ \bibinfo {pages} {2065}
  (\bibinfo {year} {1995})}\BibitemShut {NoStop}%
\bibitem [{\citenamefont {Peres}(1996)}]{PhysRevLett.76.2205}%
  \BibitemOpen
  \bibfield  {author} {\bibinfo {author} {\bibfnamefont {A.}~\bibnamefont
  {Peres}},\ }\bibfield  {title} {\bibinfo {title} {Nonlocal effects in fock
  space},\ }\href {https://doi.org/10.1103/PhysRevLett.76.2205} {\bibfield
  {journal} {\bibinfo  {journal} {Phys. Rev. Lett.}\ }\textbf {\bibinfo
  {volume} {76}},\ \bibinfo {pages} {2205} (\bibinfo {year}
  {1996})}\BibitemShut {NoStop}%
\bibitem [{\citenamefont {Wang}\ \emph {et~al.}(2019)\citenamefont {Wang},
  \citenamefont {He}, \citenamefont {Yin}, \citenamefont {Lu}, \citenamefont
  {Cui}, \citenamefont {Chen}, \citenamefont {Zhou}, \citenamefont {Guo},\ and\
  \citenamefont {Han}}]{PhysRevX.9.021046}%
  \BibitemOpen
  \bibfield  {author} {\bibinfo {author} {\bibfnamefont {S.}~\bibnamefont
  {Wang}}, \bibinfo {author} {\bibfnamefont {D.-Y.}\ \bibnamefont {He}},
  \bibinfo {author} {\bibfnamefont {Z.-Q.}\ \bibnamefont {Yin}}, \bibinfo
  {author} {\bibfnamefont {F.-Y.}\ \bibnamefont {Lu}}, \bibinfo {author}
  {\bibfnamefont {C.-H.}\ \bibnamefont {Cui}}, \bibinfo {author} {\bibfnamefont
  {W.}~\bibnamefont {Chen}}, \bibinfo {author} {\bibfnamefont {Z.}~\bibnamefont
  {Zhou}}, \bibinfo {author} {\bibfnamefont {G.-C.}\ \bibnamefont {Guo}},\ and\
  \bibinfo {author} {\bibfnamefont {Z.-F.}\ \bibnamefont {Han}},\ }\bibfield
  {title} {\bibinfo {title} {Beating the fundamental rate-distance limit in a
  proof-of-principle quantum key distribution system},\ }\href
  {https://doi.org/10.1103/PhysRevX.9.021046} {\bibfield  {journal} {\bibinfo
  {journal} {Phys. Rev. X}\ }\textbf {\bibinfo {volume} {9}},\ \bibinfo {pages}
  {021046} (\bibinfo {year} {2019})}\BibitemShut {NoStop}%
\bibitem [{\citenamefont {Zhang}\ and\ \citenamefont
  {Jennewein}(2025)}]{zhang2025criteria}%
  \BibitemOpen
  \bibfield  {author} {\bibinfo {author} {\bibfnamefont {Y.}~\bibnamefont
  {Zhang}}\ and\ \bibinfo {author} {\bibfnamefont {T.}~\bibnamefont
  {Jennewein}},\ }\bibfield  {title} {\bibinfo {title} {Criteria for optimal
  entanglement-assisted long baseline imaging protocols},\ }\href@noop {}
  {\bibfield  {journal} {\bibinfo  {journal} {arXiv preprint arXiv:2501.16670}\
  } (\bibinfo {year} {2025})}\BibitemShut {NoStop}%
\bibitem [{\citenamefont {M{\"u}nzberg}\ \emph {et~al.}(2022)\citenamefont
  {M{\"u}nzberg}, \citenamefont {Draxl}, \citenamefont {Covre~da Silva},
  \citenamefont {Karli}, \citenamefont {Manna}, \citenamefont {Rastelli},
  \citenamefont {Weihs},\ and\ \citenamefont {Keil}}]{munzberg2022fast}%
  \BibitemOpen
  \bibfield  {author} {\bibinfo {author} {\bibfnamefont {J.}~\bibnamefont
  {M{\"u}nzberg}}, \bibinfo {author} {\bibfnamefont {F.}~\bibnamefont {Draxl}},
  \bibinfo {author} {\bibfnamefont {S.~F.}\ \bibnamefont {Covre~da Silva}},
  \bibinfo {author} {\bibfnamefont {Y.}~\bibnamefont {Karli}}, \bibinfo
  {author} {\bibfnamefont {S.}~\bibnamefont {Manna}}, \bibinfo {author}
  {\bibfnamefont {A.}~\bibnamefont {Rastelli}}, \bibinfo {author}
  {\bibfnamefont {G.}~\bibnamefont {Weihs}},\ and\ \bibinfo {author}
  {\bibfnamefont {R.}~\bibnamefont {Keil}},\ }\bibfield  {title} {\bibinfo
  {title} {Fast and efficient demultiplexing of single photons from a quantum
  dot with resonantly enhanced electro-optic modulators},\ }\href@noop {}
  {\bibfield  {journal} {\bibinfo  {journal} {APL Photonics}\ }\textbf
  {\bibinfo {volume} {7}} (\bibinfo {year} {2022})}\BibitemShut {NoStop}%
\bibitem [{\citenamefont {Gao}\ \emph {et~al.}(2019)\citenamefont {Gao},
  \citenamefont {Xu}, \citenamefont {He}, \citenamefont {Chen}, \citenamefont
  {Zhang}, \citenamefont {Li}, \citenamefont {Chen}, \citenamefont {Luo},
  \citenamefont {Liu}, \citenamefont {Yu},\ and\ \citenamefont
  {Cai}}]{Gao2019}%
  \BibitemOpen
  \bibfield  {author} {\bibinfo {author} {\bibfnamefont {S.}~\bibnamefont
  {Gao}}, \bibinfo {author} {\bibfnamefont {M.}~\bibnamefont {Xu}}, \bibinfo
  {author} {\bibfnamefont {M.}~\bibnamefont {He}}, \bibinfo {author}
  {\bibfnamefont {B.}~\bibnamefont {Chen}}, \bibinfo {author} {\bibfnamefont
  {X.}~\bibnamefont {Zhang}}, \bibinfo {author} {\bibfnamefont
  {Z.}~\bibnamefont {Li}}, \bibinfo {author} {\bibfnamefont {L.}~\bibnamefont
  {Chen}}, \bibinfo {author} {\bibfnamefont {Y.}~\bibnamefont {Luo}}, \bibinfo
  {author} {\bibfnamefont {L.}~\bibnamefont {Liu}}, \bibinfo {author}
  {\bibfnamefont {S.}~\bibnamefont {Yu}},\ and\ \bibinfo {author}
  {\bibfnamefont {X.}~\bibnamefont {Cai}},\ }\bibfield  {title} {\bibinfo
  {title} {Fast polarization-insensitive optical switch based on hybrid silicon
  and lithium niobate platform},\ }\href
  {https://doi.org/10.1109/LPT.2019.2949090} {\bibfield  {journal} {\bibinfo
  {journal} {IEEE Photonics Technology Letters}\ }\textbf {\bibinfo {volume}
  {31}},\ \bibinfo {pages} {1838} (\bibinfo {year} {2019})}\BibitemShut
  {NoStop}%
\bibitem [{psi(2025)}]{psiquantum2025manufacturable}%
  \BibitemOpen
  \bibfield  {title} {\bibinfo {title} {A manufacturable platform for photonic
  quantum computing},\ }\href@noop {} {\bibfield  {journal} {\bibinfo
  {journal} {Nature}\ }\textbf {\bibinfo {volume} {641}},\ \bibinfo {pages}
  {876} (\bibinfo {year} {2025})}\BibitemShut {NoStop}%
\bibitem [{\citenamefont {Networking}(2023)}]{fs_dwdm_2023}%
  \BibitemOpen
  \bibfield  {author} {\bibinfo {author} {\bibfnamefont {F.~O.}\ \bibnamefont
  {Networking}},\ }\href
  {https://www.fs.com/blog/dwdmcwdm-wavelength-itu-channels-guide-3149.html}
  {\bibinfo {title} {Dwdm/cwdm wavelength itu channels guide}} (\bibinfo {year}
  {2023}),\ \bibinfo {note} {blog post}\BibitemShut {NoStop}%
\bibitem [{\citenamefont {{ESO}}(2025)}]{eso_harps_overview}%
  \BibitemOpen
  \bibfield  {author} {\bibinfo {author} {\bibnamefont {{ESO}}},\ }\href
  {https://www.eso.org/sci/facilities/lasilla/instruments/harps/overview.html}
  {\bibinfo {title} {Harps --- high accuracy radial velocity planet searcher}}
  (\bibinfo {year} {2025}),\ \bibinfo {note} {european Southern Observatory
  instrument overview page}\BibitemShut {NoStop}%
\bibitem [{\citenamefont {Stone}\ \emph {et~al.}(2021)\citenamefont {Stone},
  \citenamefont {Suleymanzade}, \citenamefont {Taneja}, \citenamefont
  {Schuster},\ and\ \citenamefont {Simon}}]{Stone:21}%
  \BibitemOpen
  \bibfield  {author} {\bibinfo {author} {\bibfnamefont {M.}~\bibnamefont
  {Stone}}, \bibinfo {author} {\bibfnamefont {A.}~\bibnamefont {Suleymanzade}},
  \bibinfo {author} {\bibfnamefont {L.}~\bibnamefont {Taneja}}, \bibinfo
  {author} {\bibfnamefont {D.~I.}\ \bibnamefont {Schuster}},\ and\ \bibinfo
  {author} {\bibfnamefont {J.}~\bibnamefont {Simon}},\ }\bibfield  {title}
  {\bibinfo {title} {Optical mode conversion in coupled fabry--perot
  resonators},\ }\href {https://doi.org/10.1364/OL.400998} {\bibfield
  {journal} {\bibinfo  {journal} {Opt. Lett.}\ }\textbf {\bibinfo {volume}
  {46}},\ \bibinfo {pages} {21} (\bibinfo {year} {2021})}\BibitemShut {NoStop}%
\bibitem [{\citenamefont {Reiserer}(2022)}]{RevModPhys.94.041003}%
  \BibitemOpen
  \bibfield  {author} {\bibinfo {author} {\bibfnamefont {A.}~\bibnamefont
  {Reiserer}},\ }\bibfield  {title} {\bibinfo {title} {Colloquium:
  Cavity-enhanced quantum network nodes},\ }\href
  {https://doi.org/10.1103/RevModPhys.94.041003} {\bibfield  {journal}
  {\bibinfo  {journal} {Rev. Mod. Phys.}\ }\textbf {\bibinfo {volume} {94}},\
  \bibinfo {pages} {041003} (\bibinfo {year} {2022})}\BibitemShut {NoStop}%
\bibitem [{\citenamefont {Chang}\ \emph {et~al.}(2018)\citenamefont {Chang},
  \citenamefont {Douglas}, \citenamefont {Gonz\'alez-Tudela}, \citenamefont
  {Hung},\ and\ \citenamefont {Kimble}}]{RevModPhys.90.031002}%
  \BibitemOpen
  \bibfield  {author} {\bibinfo {author} {\bibfnamefont {D.~E.}\ \bibnamefont
  {Chang}}, \bibinfo {author} {\bibfnamefont {J.~S.}\ \bibnamefont {Douglas}},
  \bibinfo {author} {\bibfnamefont {A.}~\bibnamefont {Gonz\'alez-Tudela}},
  \bibinfo {author} {\bibfnamefont {C.-L.}\ \bibnamefont {Hung}},\ and\
  \bibinfo {author} {\bibfnamefont {H.~J.}\ \bibnamefont {Kimble}},\ }\bibfield
   {title} {\bibinfo {title} {Colloquium: Quantum matter built from nanoscopic
  lattices of atoms and photons},\ }\href
  {https://doi.org/10.1103/RevModPhys.90.031002} {\bibfield  {journal}
  {\bibinfo  {journal} {Rev. Mod. Phys.}\ }\textbf {\bibinfo {volume} {90}},\
  \bibinfo {pages} {031002} (\bibinfo {year} {2018})}\BibitemShut {NoStop}%
\bibitem [{\citenamefont {Pompili}\ \emph {et~al.}(2021)\citenamefont
  {Pompili}, \citenamefont {Hermans}, \citenamefont {Baier}, \citenamefont
  {Beukers}, \citenamefont {Humphreys}, \citenamefont {Schouten}, \citenamefont
  {Vermeulen}, \citenamefont {Tiggelman}, \citenamefont {dos Santos~Martins},
  \citenamefont {Dirkse}, \citenamefont {Wehner},\ and\ \citenamefont
  {Hanson}}]{pompili21}%
  \BibitemOpen
  \bibfield  {author} {\bibinfo {author} {\bibfnamefont {M.}~\bibnamefont
  {Pompili}}, \bibinfo {author} {\bibfnamefont {S.~L.~N.}\ \bibnamefont
  {Hermans}}, \bibinfo {author} {\bibfnamefont {S.}~\bibnamefont {Baier}},
  \bibinfo {author} {\bibfnamefont {H.~K.~C.}\ \bibnamefont {Beukers}},
  \bibinfo {author} {\bibfnamefont {P.~C.}\ \bibnamefont {Humphreys}}, \bibinfo
  {author} {\bibfnamefont {R.~N.}\ \bibnamefont {Schouten}}, \bibinfo {author}
  {\bibfnamefont {R.~F.~L.}\ \bibnamefont {Vermeulen}}, \bibinfo {author}
  {\bibfnamefont {M.~J.}\ \bibnamefont {Tiggelman}}, \bibinfo {author}
  {\bibfnamefont {L.}~\bibnamefont {dos Santos~Martins}}, \bibinfo {author}
  {\bibfnamefont {B.}~\bibnamefont {Dirkse}}, \bibinfo {author} {\bibfnamefont
  {S.}~\bibnamefont {Wehner}},\ and\ \bibinfo {author} {\bibfnamefont
  {R.}~\bibnamefont {Hanson}},\ }\bibfield  {title} {\bibinfo {title}
  {Realization of a multinode quantum network of remote solid-state qubits},\
  }\href {https://doi.org/10.1126/science.abg1919} {\bibfield  {journal}
  {\bibinfo  {journal} {Science}\ }\textbf {\bibinfo {volume} {372}},\ \bibinfo
  {pages} {259} (\bibinfo {year} {2021})}\BibitemShut {NoStop}%
\bibitem [{\citenamefont {Reiserer}\ \emph {et~al.}(2016)\citenamefont
  {Reiserer}, \citenamefont {Kalb}, \citenamefont {Blok}, \citenamefont {van
  Bemmelen}, \citenamefont {Taminiau}, \citenamefont {Hanson}, \citenamefont
  {Twitchen},\ and\ \citenamefont {Markham}}]{PhysRevX.6.021040}%
  \BibitemOpen
  \bibfield  {author} {\bibinfo {author} {\bibfnamefont {A.}~\bibnamefont
  {Reiserer}}, \bibinfo {author} {\bibfnamefont {N.}~\bibnamefont {Kalb}},
  \bibinfo {author} {\bibfnamefont {M.~S.}\ \bibnamefont {Blok}}, \bibinfo
  {author} {\bibfnamefont {K.~J.~M.}\ \bibnamefont {van Bemmelen}}, \bibinfo
  {author} {\bibfnamefont {T.~H.}\ \bibnamefont {Taminiau}}, \bibinfo {author}
  {\bibfnamefont {R.}~\bibnamefont {Hanson}}, \bibinfo {author} {\bibfnamefont
  {D.~J.}\ \bibnamefont {Twitchen}},\ and\ \bibinfo {author} {\bibfnamefont
  {M.}~\bibnamefont {Markham}},\ }\bibfield  {title} {\bibinfo {title} {Robust
  quantum-network memory using decoherence-protected subspaces of nuclear
  spins},\ }\href {https://doi.org/10.1103/PhysRevX.6.021040} {\bibfield
  {journal} {\bibinfo  {journal} {Phys. Rev. X}\ }\textbf {\bibinfo {volume}
  {6}},\ \bibinfo {pages} {021040} (\bibinfo {year} {2016})}\BibitemShut
  {NoStop}%
\bibitem [{\citenamefont {Knaut}\ \emph {et~al.}(2024)\citenamefont {Knaut},
  \citenamefont {Suleymanzade}, \citenamefont {Wei}, \citenamefont {Assumpcao},
  \citenamefont {Stas}, \citenamefont {Huan}, \citenamefont {Machielse},
  \citenamefont {Knall}, \citenamefont {Sutula}, \citenamefont {Baranes} \emph
  {et~al.}}]{knaut2024entanglement}%
  \BibitemOpen
  \bibfield  {author} {\bibinfo {author} {\bibfnamefont {C.~M.}\ \bibnamefont
  {Knaut}}, \bibinfo {author} {\bibfnamefont {A.}~\bibnamefont {Suleymanzade}},
  \bibinfo {author} {\bibfnamefont {Y.-C.}\ \bibnamefont {Wei}}, \bibinfo
  {author} {\bibfnamefont {D.~R.}\ \bibnamefont {Assumpcao}}, \bibinfo {author}
  {\bibfnamefont {P.-J.}\ \bibnamefont {Stas}}, \bibinfo {author}
  {\bibfnamefont {Y.~Q.}\ \bibnamefont {Huan}}, \bibinfo {author}
  {\bibfnamefont {B.}~\bibnamefont {Machielse}}, \bibinfo {author}
  {\bibfnamefont {E.~N.}\ \bibnamefont {Knall}}, \bibinfo {author}
  {\bibfnamefont {M.}~\bibnamefont {Sutula}}, \bibinfo {author} {\bibfnamefont
  {G.}~\bibnamefont {Baranes}}, \emph {et~al.},\ }\bibfield  {title} {\bibinfo
  {title} {Entanglement of nanophotonic quantum memory nodes in a telecom
  network},\ }\href@noop {} {\bibfield  {journal} {\bibinfo  {journal}
  {Nature}\ }\textbf {\bibinfo {volume} {629}},\ \bibinfo {pages} {573}
  (\bibinfo {year} {2024})}\BibitemShut {NoStop}%
\bibitem [{\citenamefont {Reiserer}\ and\ \citenamefont
  {Rempe}(2015)}]{RevModPhys.87.1379}%
  \BibitemOpen
  \bibfield  {author} {\bibinfo {author} {\bibfnamefont {A.}~\bibnamefont
  {Reiserer}}\ and\ \bibinfo {author} {\bibfnamefont {G.}~\bibnamefont
  {Rempe}},\ }\bibfield  {title} {\bibinfo {title} {Cavity-based quantum
  networks with single atoms and optical photons},\ }\href
  {https://doi.org/10.1103/RevModPhys.87.1379} {\bibfield  {journal} {\bibinfo
  {journal} {Rev. Mod. Phys.}\ }\textbf {\bibinfo {volume} {87}},\ \bibinfo
  {pages} {1379} (\bibinfo {year} {2015})}\BibitemShut {NoStop}%
\bibitem [{kor()}]{korber_decoherence-protected_2018}%
  \BibitemOpen
  \bibfield  {title} {\bibinfo {title} {Decoherence-protected memory for a
  single-photon qubit},\ }\href@noop {} {\ \textbf {\bibinfo {volume}
  {12}}}\BibitemShut {NoStop}%
\bibitem [{\citenamefont {Wang}\ \emph {et~al.}(2021)\citenamefont {Wang},
  \citenamefont {Luan}, \citenamefont {Qiao}, \citenamefont {Um}, \citenamefont
  {Zhang}, \citenamefont {Wang}, \citenamefont {Yuan}, \citenamefont {Gu},
  \citenamefont {Zhang},\ and\ \citenamefont {Kim}}]{wang_single_2021}%
  \BibitemOpen
  \bibfield  {author} {\bibinfo {author} {\bibfnamefont {P.}~\bibnamefont
  {Wang}}, \bibinfo {author} {\bibfnamefont {C.-Y.}\ \bibnamefont {Luan}},
  \bibinfo {author} {\bibfnamefont {M.}~\bibnamefont {Qiao}}, \bibinfo {author}
  {\bibfnamefont {M.}~\bibnamefont {Um}}, \bibinfo {author} {\bibfnamefont
  {J.}~\bibnamefont {Zhang}}, \bibinfo {author} {\bibfnamefont
  {Y.}~\bibnamefont {Wang}}, \bibinfo {author} {\bibfnamefont {X.}~\bibnamefont
  {Yuan}}, \bibinfo {author} {\bibfnamefont {M.}~\bibnamefont {Gu}}, \bibinfo
  {author} {\bibfnamefont {J.}~\bibnamefont {Zhang}},\ and\ \bibinfo {author}
  {\bibfnamefont {K.}~\bibnamefont {Kim}},\ }\bibfield  {title} {\bibinfo
  {title} {Single ion qubit with estimated coherence time exceeding one hour},\
  }\href {https://doi.org/10.1038/s41467-020-20330-w} {\bibfield  {journal}
  {\bibinfo  {journal} {Nature Communications}\ }\textbf {\bibinfo {volume}
  {12}},\ \bibinfo {pages} {233} (\bibinfo {year} {2021})}\BibitemShut
  {NoStop}%
\bibitem [{\citenamefont {Ritter}\ \emph {et~al.}(2012)\citenamefont {Ritter},
  \citenamefont {N{\"o}lleke}, \citenamefont {Hahn}, \citenamefont {Reiserer},
  \citenamefont {Neuzner}, \citenamefont {Uphoff}, \citenamefont {M{\"u}cke},
  \citenamefont {Figueroa}, \citenamefont {Bochmann},\ and\ \citenamefont
  {Rempe}}]{ritter_elementary_2012}%
  \BibitemOpen
  \bibfield  {author} {\bibinfo {author} {\bibfnamefont {S.}~\bibnamefont
  {Ritter}}, \bibinfo {author} {\bibfnamefont {C.}~\bibnamefont {N{\"o}lleke}},
  \bibinfo {author} {\bibfnamefont {C.}~\bibnamefont {Hahn}}, \bibinfo {author}
  {\bibfnamefont {A.}~\bibnamefont {Reiserer}}, \bibinfo {author}
  {\bibfnamefont {A.}~\bibnamefont {Neuzner}}, \bibinfo {author} {\bibfnamefont
  {M.}~\bibnamefont {Uphoff}}, \bibinfo {author} {\bibfnamefont
  {M.}~\bibnamefont {M{\"u}cke}}, \bibinfo {author} {\bibfnamefont
  {E.}~\bibnamefont {Figueroa}}, \bibinfo {author} {\bibfnamefont
  {J.}~\bibnamefont {Bochmann}},\ and\ \bibinfo {author} {\bibfnamefont
  {G.}~\bibnamefont {Rempe}},\ }\bibfield  {title} {\bibinfo {title} {An
  elementary quantum network of single atoms in optical cavities},\ }\href@noop
  {} {\bibfield  {journal} {\bibinfo  {journal} {Nature}\ }\textbf {\bibinfo
  {volume} {484}},\ \bibinfo {pages} {195} (\bibinfo {year}
  {2012})}\BibitemShut {NoStop}%
\bibitem [{\citenamefont {Stephenson}\ \emph {et~al.}(2020)\citenamefont
  {Stephenson}, \citenamefont {Nadlinger}, \citenamefont {Nichol},
  \citenamefont {An}, \citenamefont {Drmota}, \citenamefont {Ballance},
  \citenamefont {Thirumalai}, \citenamefont {Goodwin}, \citenamefont {Lucas},\
  and\ \citenamefont {Ballance}}]{PhysRevLett.124.110501}%
  \BibitemOpen
  \bibfield  {author} {\bibinfo {author} {\bibfnamefont {L.~J.}\ \bibnamefont
  {Stephenson}}, \bibinfo {author} {\bibfnamefont {D.~P.}\ \bibnamefont
  {Nadlinger}}, \bibinfo {author} {\bibfnamefont {B.~C.}\ \bibnamefont
  {Nichol}}, \bibinfo {author} {\bibfnamefont {S.}~\bibnamefont {An}}, \bibinfo
  {author} {\bibfnamefont {P.}~\bibnamefont {Drmota}}, \bibinfo {author}
  {\bibfnamefont {T.~G.}\ \bibnamefont {Ballance}}, \bibinfo {author}
  {\bibfnamefont {K.}~\bibnamefont {Thirumalai}}, \bibinfo {author}
  {\bibfnamefont {J.~F.}\ \bibnamefont {Goodwin}}, \bibinfo {author}
  {\bibfnamefont {D.~M.}\ \bibnamefont {Lucas}},\ and\ \bibinfo {author}
  {\bibfnamefont {C.~J.}\ \bibnamefont {Ballance}},\ }\bibfield  {title}
  {\bibinfo {title} {High-rate, high-fidelity entanglement of qubits across an
  elementary quantum network},\ }\href
  {https://doi.org/10.1103/PhysRevLett.124.110501} {\bibfield  {journal}
  {\bibinfo  {journal} {Phys. Rev. Lett.}\ }\textbf {\bibinfo {volume} {124}},\
  \bibinfo {pages} {110501} (\bibinfo {year} {2020})}\BibitemShut {NoStop}%
\bibitem [{\citenamefont {Langenfeld}\ \emph {et~al.}(2021)\citenamefont
  {Langenfeld}, \citenamefont {Thomas}, \citenamefont {Morin},\ and\
  \citenamefont {Rempe}}]{Langenfeld}%
  \BibitemOpen
  \bibfield  {author} {\bibinfo {author} {\bibfnamefont {S.}~\bibnamefont
  {Langenfeld}}, \bibinfo {author} {\bibfnamefont {P.}~\bibnamefont {Thomas}},
  \bibinfo {author} {\bibfnamefont {O.}~\bibnamefont {Morin}},\ and\ \bibinfo
  {author} {\bibfnamefont {G.}~\bibnamefont {Rempe}},\ }\bibfield  {title}
  {\bibinfo {title} {Quantum repeater node demonstrating unconditionally secure
  key distribution},\ }\href {https://doi.org/10.1103/PhysRevLett.126.230506}
  {\bibfield  {journal} {\bibinfo  {journal} {Phys. Rev. Lett.}\ }\textbf
  {\bibinfo {volume} {126}},\ \bibinfo {pages} {230506} (\bibinfo {year}
  {2021})}\BibitemShut {NoStop}%
\bibitem [{\citenamefont {Leu}\ \emph {et~al.}(2023)\citenamefont {Leu},
  \citenamefont {Gely}, \citenamefont {Weber}, \citenamefont {Smith},
  \citenamefont {Nadlinger},\ and\ \citenamefont
  {Lucas}}]{PhysRevLett.131.120601}%
  \BibitemOpen
  \bibfield  {author} {\bibinfo {author} {\bibfnamefont {A.~D.}\ \bibnamefont
  {Leu}}, \bibinfo {author} {\bibfnamefont {M.~F.}\ \bibnamefont {Gely}},
  \bibinfo {author} {\bibfnamefont {M.~A.}\ \bibnamefont {Weber}}, \bibinfo
  {author} {\bibfnamefont {M.~C.}\ \bibnamefont {Smith}}, \bibinfo {author}
  {\bibfnamefont {D.~P.}\ \bibnamefont {Nadlinger}},\ and\ \bibinfo {author}
  {\bibfnamefont {D.~M.}\ \bibnamefont {Lucas}},\ }\bibfield  {title} {\bibinfo
  {title} {Fast, high-fidelity addressed single-qubit gates using efficient
  composite pulse sequences},\ }\href
  {https://doi.org/10.1103/PhysRevLett.131.120601} {\bibfield  {journal}
  {\bibinfo  {journal} {Phys. Rev. Lett.}\ }\textbf {\bibinfo {volume} {131}},\
  \bibinfo {pages} {120601} (\bibinfo {year} {2023})}\BibitemShut {NoStop}%
\bibitem [{\citenamefont {Postler}\ \emph {et~al.}(2022)\citenamefont
  {Postler}, \citenamefont {Heu$\beta$en}, \citenamefont {Pogorelov},
  \citenamefont {Rispler}, \citenamefont {Feldker}, \citenamefont {Meth},
  \citenamefont {Marciniak}, \citenamefont {Stricker}, \citenamefont
  {Ringbauer}, \citenamefont {Blatt} \emph
  {et~al.}}]{postler2022demonstration}%
  \BibitemOpen
  \bibfield  {author} {\bibinfo {author} {\bibfnamefont {L.}~\bibnamefont
  {Postler}}, \bibinfo {author} {\bibfnamefont {S.}~\bibnamefont
  {Heu$\beta$en}}, \bibinfo {author} {\bibfnamefont {I.}~\bibnamefont
  {Pogorelov}}, \bibinfo {author} {\bibfnamefont {M.}~\bibnamefont {Rispler}},
  \bibinfo {author} {\bibfnamefont {T.}~\bibnamefont {Feldker}}, \bibinfo
  {author} {\bibfnamefont {M.}~\bibnamefont {Meth}}, \bibinfo {author}
  {\bibfnamefont {C.~D.}\ \bibnamefont {Marciniak}}, \bibinfo {author}
  {\bibfnamefont {R.}~\bibnamefont {Stricker}}, \bibinfo {author}
  {\bibfnamefont {M.}~\bibnamefont {Ringbauer}}, \bibinfo {author}
  {\bibfnamefont {R.}~\bibnamefont {Blatt}}, \emph {et~al.},\ }\bibfield
  {title} {\bibinfo {title} {Demonstration of fault-tolerant universal quantum
  gate operations},\ }\href@noop {} {\bibfield  {journal} {\bibinfo  {journal}
  {Nature}\ }\textbf {\bibinfo {volume} {605}},\ \bibinfo {pages} {675}
  (\bibinfo {year} {2022})}\BibitemShut {NoStop}%
\bibitem [{\citenamefont {Evered}\ \emph {et~al.}(2023)\citenamefont {Evered},
  \citenamefont {Bluvstein}, \citenamefont {Kalinowski}, \citenamefont {Ebadi},
  \citenamefont {Manovitz}, \citenamefont {Zhou}, \citenamefont {Li},
  \citenamefont {Geim}, \citenamefont {Wang}, \citenamefont {Maskara} \emph
  {et~al.}}]{evered2023high}%
  \BibitemOpen
  \bibfield  {author} {\bibinfo {author} {\bibfnamefont {S.~J.}\ \bibnamefont
  {Evered}}, \bibinfo {author} {\bibfnamefont {D.}~\bibnamefont {Bluvstein}},
  \bibinfo {author} {\bibfnamefont {M.}~\bibnamefont {Kalinowski}}, \bibinfo
  {author} {\bibfnamefont {S.}~\bibnamefont {Ebadi}}, \bibinfo {author}
  {\bibfnamefont {T.}~\bibnamefont {Manovitz}}, \bibinfo {author}
  {\bibfnamefont {H.}~\bibnamefont {Zhou}}, \bibinfo {author} {\bibfnamefont
  {S.~H.}\ \bibnamefont {Li}}, \bibinfo {author} {\bibfnamefont {A.~A.}\
  \bibnamefont {Geim}}, \bibinfo {author} {\bibfnamefont {T.~T.}\ \bibnamefont
  {Wang}}, \bibinfo {author} {\bibfnamefont {N.}~\bibnamefont {Maskara}}, \emph
  {et~al.},\ }\bibfield  {title} {\bibinfo {title} {High-fidelity parallel
  entangling gates on a neutral-atom quantum computer},\ }\href@noop {}
  {\bibfield  {journal} {\bibinfo  {journal} {Nature}\ }\textbf {\bibinfo
  {volume} {622}},\ \bibinfo {pages} {268} (\bibinfo {year}
  {2023})}\BibitemShut {NoStop}%
\bibitem [{\citenamefont {Chew}\ \emph {et~al.}(2022)\citenamefont {Chew},
  \citenamefont {Tomita}, \citenamefont {Mahesh}, \citenamefont {Sugawa},
  \citenamefont {de~L{\'e}s{\'e}leuc},\ and\ \citenamefont
  {Ohmori}}]{chew2022ultrafast}%
  \BibitemOpen
  \bibfield  {author} {\bibinfo {author} {\bibfnamefont {Y.}~\bibnamefont
  {Chew}}, \bibinfo {author} {\bibfnamefont {T.}~\bibnamefont {Tomita}},
  \bibinfo {author} {\bibfnamefont {T.~P.}\ \bibnamefont {Mahesh}}, \bibinfo
  {author} {\bibfnamefont {S.}~\bibnamefont {Sugawa}}, \bibinfo {author}
  {\bibfnamefont {S.}~\bibnamefont {de~L{\'e}s{\'e}leuc}},\ and\ \bibinfo
  {author} {\bibfnamefont {K.}~\bibnamefont {Ohmori}},\ }\bibfield  {title}
  {\bibinfo {title} {Ultrafast energy exchange between two single rydberg atoms
  on a nanosecond timescale},\ }\href@noop {} {\bibfield  {journal} {\bibinfo
  {journal} {Nature Photonics}\ }\textbf {\bibinfo {volume} {16}},\ \bibinfo
  {pages} {724} (\bibinfo {year} {2022})}\BibitemShut {NoStop}%
\bibitem [{\citenamefont {Phuttitarn}\ \emph {et~al.}(2024)\citenamefont
  {Phuttitarn}, \citenamefont {Becker}, \citenamefont {Chinnarasu},
  \citenamefont {Graham},\ and\ \citenamefont {Saffman}}]{Phuttitarn:24}%
  \BibitemOpen
  \bibfield  {author} {\bibinfo {author} {\bibfnamefont {L.}~\bibnamefont
  {Phuttitarn}}, \bibinfo {author} {\bibfnamefont {B.~M.}\ \bibnamefont
  {Becker}}, \bibinfo {author} {\bibfnamefont {R.}~\bibnamefont {Chinnarasu}},
  \bibinfo {author} {\bibfnamefont {T.~M.}\ \bibnamefont {Graham}},\ and\
  \bibinfo {author} {\bibfnamefont {M.}~\bibnamefont {Saffman}},\ }\bibfield
  {title} {\bibinfo {title} {Enhanced measurement of neutral-atom qubits with
  machine learning},\ }\href {https://doi.org/10.1103/PhysRevApplied.22.024011}
  {\bibfield  {journal} {\bibinfo  {journal} {Phys. Rev. Appl.}\ }\textbf
  {\bibinfo {volume} {22}},\ \bibinfo {pages} {024011} (\bibinfo {year}
  {2024})}\BibitemShut {NoStop}%
\bibitem [{\citenamefont {Arute}\ \emph {et~al.}(2019)\citenamefont {Arute},
  \citenamefont {Arya}, \citenamefont {Babbush}, \citenamefont {Bacon},
  \citenamefont {Bardin}, \citenamefont {Barends}, \citenamefont {Biswas},
  \citenamefont {Boixo}, \citenamefont {Brandao}, \citenamefont {Buell} \emph
  {et~al.}}]{arute2019quantum}%
  \BibitemOpen
  \bibfield  {author} {\bibinfo {author} {\bibfnamefont {F.}~\bibnamefont
  {Arute}}, \bibinfo {author} {\bibfnamefont {K.}~\bibnamefont {Arya}},
  \bibinfo {author} {\bibfnamefont {R.}~\bibnamefont {Babbush}}, \bibinfo
  {author} {\bibfnamefont {D.}~\bibnamefont {Bacon}}, \bibinfo {author}
  {\bibfnamefont {J.~C.}\ \bibnamefont {Bardin}}, \bibinfo {author}
  {\bibfnamefont {R.}~\bibnamefont {Barends}}, \bibinfo {author} {\bibfnamefont
  {R.}~\bibnamefont {Biswas}}, \bibinfo {author} {\bibfnamefont
  {S.}~\bibnamefont {Boixo}}, \bibinfo {author} {\bibfnamefont {F.~G.}\
  \bibnamefont {Brandao}}, \bibinfo {author} {\bibfnamefont {D.~A.}\
  \bibnamefont {Buell}}, \emph {et~al.},\ }\bibfield  {title} {\bibinfo {title}
  {Quantum supremacy using a programmable superconducting processor},\
  }\href@noop {} {\bibfield  {journal} {\bibinfo  {journal} {Nature}\ }\textbf
  {\bibinfo {volume} {574}},\ \bibinfo {pages} {505} (\bibinfo {year}
  {2019})}\BibitemShut {NoStop}%
\bibitem [{\citenamefont {Neumann}\ \emph {et~al.}(2022)\citenamefont
  {Neumann}, \citenamefont {Selimovic}, \citenamefont {Bohmann},\ and\
  \citenamefont {Ursin}}]{neumann2022experimental}%
  \BibitemOpen
  \bibfield  {author} {\bibinfo {author} {\bibfnamefont {S.~P.}\ \bibnamefont
  {Neumann}}, \bibinfo {author} {\bibfnamefont {M.}~\bibnamefont {Selimovic}},
  \bibinfo {author} {\bibfnamefont {M.}~\bibnamefont {Bohmann}},\ and\ \bibinfo
  {author} {\bibfnamefont {R.}~\bibnamefont {Ursin}},\ }\bibfield  {title}
  {\bibinfo {title} {Experimental entanglement generation for quantum key
  distribution beyond 1 gbit/s},\ }\href@noop {} {\bibfield  {journal}
  {\bibinfo  {journal} {Quantum}\ }\textbf {\bibinfo {volume} {6}},\ \bibinfo
  {pages} {822} (\bibinfo {year} {2022})}\BibitemShut {NoStop}%
\bibitem [{\citenamefont {Rahmouni}\ \emph {et~al.}(2024)\citenamefont
  {Rahmouni}, \citenamefont {Wang}, \citenamefont {Li}, \citenamefont {Tang},
  \citenamefont {Gerrits}, \citenamefont {Slattery}, \citenamefont {Li},\ and\
  \citenamefont {Ma}}]{rahmouni2024entangled}%
  \BibitemOpen
  \bibfield  {author} {\bibinfo {author} {\bibfnamefont {A.}~\bibnamefont
  {Rahmouni}}, \bibinfo {author} {\bibfnamefont {R.}~\bibnamefont {Wang}},
  \bibinfo {author} {\bibfnamefont {J.}~\bibnamefont {Li}}, \bibinfo {author}
  {\bibfnamefont {X.}~\bibnamefont {Tang}}, \bibinfo {author} {\bibfnamefont
  {T.}~\bibnamefont {Gerrits}}, \bibinfo {author} {\bibfnamefont
  {O.}~\bibnamefont {Slattery}}, \bibinfo {author} {\bibfnamefont
  {Q.}~\bibnamefont {Li}},\ and\ \bibinfo {author} {\bibfnamefont
  {L.}~\bibnamefont {Ma}},\ }\bibfield  {title} {\bibinfo {title} {Entangled
  photon pair generation in an integrated sic platform},\ }\href@noop {}
  {\bibfield  {journal} {\bibinfo  {journal} {Light: Science \& Applications}\
  }\textbf {\bibinfo {volume} {13}},\ \bibinfo {pages} {110} (\bibinfo {year}
  {2024})}\BibitemShut {NoStop}%
\bibitem [{\citenamefont {Chopin}\ \emph {et~al.}(2023)\citenamefont {Chopin},
  \citenamefont {Barone}, \citenamefont {Ghorbel}, \citenamefont {Combri{\'e}},
  \citenamefont {Bajoni}, \citenamefont {Raineri}, \citenamefont {Galli},\ and\
  \citenamefont {De~Rossi}}]{chopin2023ultra}%
  \BibitemOpen
  \bibfield  {author} {\bibinfo {author} {\bibfnamefont {A.}~\bibnamefont
  {Chopin}}, \bibinfo {author} {\bibfnamefont {A.}~\bibnamefont {Barone}},
  \bibinfo {author} {\bibfnamefont {I.}~\bibnamefont {Ghorbel}}, \bibinfo
  {author} {\bibfnamefont {S.}~\bibnamefont {Combri{\'e}}}, \bibinfo {author}
  {\bibfnamefont {D.}~\bibnamefont {Bajoni}}, \bibinfo {author} {\bibfnamefont
  {F.}~\bibnamefont {Raineri}}, \bibinfo {author} {\bibfnamefont
  {M.}~\bibnamefont {Galli}},\ and\ \bibinfo {author} {\bibfnamefont
  {A.}~\bibnamefont {De~Rossi}},\ }\bibfield  {title} {\bibinfo {title}
  {Ultra-efficient generation of time-energy entangled photon pairs in an ingap
  photonic crystal cavity},\ }\href@noop {} {\bibfield  {journal} {\bibinfo
  {journal} {Communications Physics}\ }\textbf {\bibinfo {volume} {6}},\
  \bibinfo {pages} {77} (\bibinfo {year} {2023})}\BibitemShut {NoStop}%
\bibitem [{\citenamefont {Bhaskar}\ \emph {et~al.}(2020)\citenamefont
  {Bhaskar}, \citenamefont {Riedinger}, \citenamefont {Machielse},
  \citenamefont {Levonian}, \citenamefont {Nguyen}, \citenamefont {Knall},
  \citenamefont {Park}, \citenamefont {Englund}, \citenamefont {Lon{\v{c}}ar},
  \citenamefont {Sukachev} \emph {et~al.}}]{bhaskar2020experimental}%
  \BibitemOpen
  \bibfield  {author} {\bibinfo {author} {\bibfnamefont {M.~K.}\ \bibnamefont
  {Bhaskar}}, \bibinfo {author} {\bibfnamefont {R.}~\bibnamefont {Riedinger}},
  \bibinfo {author} {\bibfnamefont {B.}~\bibnamefont {Machielse}}, \bibinfo
  {author} {\bibfnamefont {D.~S.}\ \bibnamefont {Levonian}}, \bibinfo {author}
  {\bibfnamefont {C.~T.}\ \bibnamefont {Nguyen}}, \bibinfo {author}
  {\bibfnamefont {E.~N.}\ \bibnamefont {Knall}}, \bibinfo {author}
  {\bibfnamefont {H.}~\bibnamefont {Park}}, \bibinfo {author} {\bibfnamefont
  {D.}~\bibnamefont {Englund}}, \bibinfo {author} {\bibfnamefont
  {M.}~\bibnamefont {Lon{\v{c}}ar}}, \bibinfo {author} {\bibfnamefont {D.~D.}\
  \bibnamefont {Sukachev}}, \emph {et~al.},\ }\bibfield  {title} {\bibinfo
  {title} {Experimental demonstration of memory-enhanced quantum
  communication},\ }\href@noop {} {\bibfield  {journal} {\bibinfo  {journal}
  {Nature}\ }\textbf {\bibinfo {volume} {580}},\ \bibinfo {pages} {60}
  (\bibinfo {year} {2020})}\BibitemShut {NoStop}%
\bibitem [{\citenamefont {Liu}\ \emph {et~al.}(2021)\citenamefont {Liu},
  \citenamefont {Hu}, \citenamefont {Li}, \citenamefont {Li}, \citenamefont
  {Li}, \citenamefont {Liang}, \citenamefont {Zhou}, \citenamefont {Li},\ and\
  \citenamefont {Guo}}]{liu2021heralded}%
  \BibitemOpen
  \bibfield  {author} {\bibinfo {author} {\bibfnamefont {X.}~\bibnamefont
  {Liu}}, \bibinfo {author} {\bibfnamefont {J.}~\bibnamefont {Hu}}, \bibinfo
  {author} {\bibfnamefont {Z.-F.}\ \bibnamefont {Li}}, \bibinfo {author}
  {\bibfnamefont {X.}~\bibnamefont {Li}}, \bibinfo {author} {\bibfnamefont
  {P.-Y.}\ \bibnamefont {Li}}, \bibinfo {author} {\bibfnamefont {P.-J.}\
  \bibnamefont {Liang}}, \bibinfo {author} {\bibfnamefont {Z.-Q.}\ \bibnamefont
  {Zhou}}, \bibinfo {author} {\bibfnamefont {C.-F.}\ \bibnamefont {Li}},\ and\
  \bibinfo {author} {\bibfnamefont {G.-C.}\ \bibnamefont {Guo}},\ }\bibfield
  {title} {\bibinfo {title} {Heralded entanglement distribution between two
  absorptive quantum memories},\ }\href@noop {} {\bibfield  {journal} {\bibinfo
   {journal} {Nature}\ }\textbf {\bibinfo {volume} {594}},\ \bibinfo {pages}
  {41} (\bibinfo {year} {2021})}\BibitemShut {NoStop}%
\bibitem [{\citenamefont {Liu}\ \emph {et~al.}(2024{\natexlab{a}})\citenamefont
  {Liu}, \citenamefont {Luo}, \citenamefont {Yu}, \citenamefont {Wang},
  \citenamefont {Wang}, \citenamefont {Hu}, \citenamefont {Li}, \citenamefont
  {Zheng}, \citenamefont {Yao}, \citenamefont {Yan} \emph
  {et~al.}}]{liu2024creation}%
  \BibitemOpen
  \bibfield  {author} {\bibinfo {author} {\bibfnamefont {J.-L.}\ \bibnamefont
  {Liu}}, \bibinfo {author} {\bibfnamefont {X.-Y.}\ \bibnamefont {Luo}},
  \bibinfo {author} {\bibfnamefont {Y.}~\bibnamefont {Yu}}, \bibinfo {author}
  {\bibfnamefont {C.-Y.}\ \bibnamefont {Wang}}, \bibinfo {author}
  {\bibfnamefont {B.}~\bibnamefont {Wang}}, \bibinfo {author} {\bibfnamefont
  {Y.}~\bibnamefont {Hu}}, \bibinfo {author} {\bibfnamefont {J.}~\bibnamefont
  {Li}}, \bibinfo {author} {\bibfnamefont {M.-Y.}\ \bibnamefont {Zheng}},
  \bibinfo {author} {\bibfnamefont {B.}~\bibnamefont {Yao}}, \bibinfo {author}
  {\bibfnamefont {Z.}~\bibnamefont {Yan}}, \emph {et~al.},\ }\bibfield  {title}
  {\bibinfo {title} {Creation of memory--memory entanglement in a metropolitan
  quantum network},\ }\href@noop {} {\bibfield  {journal} {\bibinfo  {journal}
  {Nature}\ }\textbf {\bibinfo {volume} {629}},\ \bibinfo {pages} {579}
  (\bibinfo {year} {2024}{\natexlab{a}})}\BibitemShut {NoStop}%
\bibitem [{\citenamefont {Chakraborty}\ \emph {et~al.}(2025)\citenamefont
  {Chakraborty}, \citenamefont {Das}, \citenamefont {van Brug}, \citenamefont
  {Pietx-Casas}, \citenamefont {Wang}, \citenamefont {Amaral}, \citenamefont
  {Tchebotareva},\ and\ \citenamefont {Tittel}}]{chakraborty2025towards}%
  \BibitemOpen
  \bibfield  {author} {\bibinfo {author} {\bibfnamefont {T.}~\bibnamefont
  {Chakraborty}}, \bibinfo {author} {\bibfnamefont {A.}~\bibnamefont {Das}},
  \bibinfo {author} {\bibfnamefont {H.}~\bibnamefont {van Brug}}, \bibinfo
  {author} {\bibfnamefont {O.}~\bibnamefont {Pietx-Casas}}, \bibinfo {author}
  {\bibfnamefont {P.-C.}\ \bibnamefont {Wang}}, \bibinfo {author}
  {\bibfnamefont {G.~C.~d.}\ \bibnamefont {Amaral}}, \bibinfo {author}
  {\bibfnamefont {A.~L.}\ \bibnamefont {Tchebotareva}},\ and\ \bibinfo {author}
  {\bibfnamefont {W.}~\bibnamefont {Tittel}},\ }\bibfield  {title} {\bibinfo
  {title} {Towards a spectrally multiplexed quantum repeater},\ }\href@noop {}
  {\bibfield  {journal} {\bibinfo  {journal} {npj Quantum Information}\
  }\textbf {\bibinfo {volume} {11}},\ \bibinfo {pages} {3} (\bibinfo {year}
  {2025})}\BibitemShut {NoStop}%
\bibitem [{\citenamefont {Liu}\ \emph {et~al.}(2024{\natexlab{b}})\citenamefont
  {Liu}, \citenamefont {Hu}, \citenamefont {Zhu}, \citenamefont {Zhang},
  \citenamefont {Xiao}, \citenamefont {Miao}, \citenamefont {Ou}, \citenamefont
  {Li}, \citenamefont {Liu}, \citenamefont {Zhou} \emph
  {et~al.}}]{liu2024nonlocal}%
  \BibitemOpen
  \bibfield  {author} {\bibinfo {author} {\bibfnamefont {X.}~\bibnamefont
  {Liu}}, \bibinfo {author} {\bibfnamefont {X.-M.}\ \bibnamefont {Hu}},
  \bibinfo {author} {\bibfnamefont {T.-X.}\ \bibnamefont {Zhu}}, \bibinfo
  {author} {\bibfnamefont {C.}~\bibnamefont {Zhang}}, \bibinfo {author}
  {\bibfnamefont {Y.-X.}\ \bibnamefont {Xiao}}, \bibinfo {author}
  {\bibfnamefont {J.-L.}\ \bibnamefont {Miao}}, \bibinfo {author}
  {\bibfnamefont {Z.-W.}\ \bibnamefont {Ou}}, \bibinfo {author} {\bibfnamefont
  {P.-Y.}\ \bibnamefont {Li}}, \bibinfo {author} {\bibfnamefont {B.-H.}\
  \bibnamefont {Liu}}, \bibinfo {author} {\bibfnamefont {Z.-Q.}\ \bibnamefont
  {Zhou}}, \emph {et~al.},\ }\bibfield  {title} {\bibinfo {title} {Nonlocal
  photonic quantum gates over 7.0 km},\ }\href@noop {} {\bibfield  {journal}
  {\bibinfo  {journal} {Nature Communications}\ }\textbf {\bibinfo {volume}
  {15}},\ \bibinfo {pages} {8529} (\bibinfo {year}
  {2024}{\natexlab{b}})}\BibitemShut {NoStop}%
\bibitem [{\citenamefont {Goswami}\ and\ \citenamefont
  {Dhara}(2023)}]{PhysRevApplied.20.024048}%
  \BibitemOpen
  \bibfield  {author} {\bibinfo {author} {\bibfnamefont {S.}~\bibnamefont
  {Goswami}}\ and\ \bibinfo {author} {\bibfnamefont {S.}~\bibnamefont
  {Dhara}},\ }\bibfield  {title} {\bibinfo {title} {Satellite-relayed global
  quantum communication without quantum memory},\ }\href
  {https://doi.org/10.1103/PhysRevApplied.20.024048} {\bibfield  {journal}
  {\bibinfo  {journal} {Phys. Rev. Appl.}\ }\textbf {\bibinfo {volume} {20}},\
  \bibinfo {pages} {024048} (\bibinfo {year} {2023})}\BibitemShut {NoStop}%
\bibitem [{\citenamefont {Carpenter}\ \emph {et~al.}(2025)\citenamefont
  {Carpenter}, \citenamefont {Boyajian}, \citenamefont {Buzasi}, \citenamefont
  {Clark}, \citenamefont {Creech-Eakman}, \citenamefont {Dean}, \citenamefont
  {Elliott}, \citenamefont {Foster}, \citenamefont {Gong}, \citenamefont
  {Karovska} \emph {et~al.}}]{carpenter2025nasa}%
  \BibitemOpen
  \bibfield  {author} {\bibinfo {author} {\bibfnamefont {K.~G.}\ \bibnamefont
  {Carpenter}}, \bibinfo {author} {\bibfnamefont {T.}~\bibnamefont {Boyajian}},
  \bibinfo {author} {\bibfnamefont {D.}~\bibnamefont {Buzasi}}, \bibinfo
  {author} {\bibfnamefont {J.}~\bibnamefont {Clark}}, \bibinfo {author}
  {\bibfnamefont {M.}~\bibnamefont {Creech-Eakman}}, \bibinfo {author}
  {\bibfnamefont {B.}~\bibnamefont {Dean}}, \bibinfo {author} {\bibfnamefont
  {A.}~\bibnamefont {Elliott}}, \bibinfo {author} {\bibfnamefont
  {J.}~\bibnamefont {Foster}}, \bibinfo {author} {\bibfnamefont
  {Q.}~\bibnamefont {Gong}}, \bibinfo {author} {\bibfnamefont {M.}~\bibnamefont
  {Karovska}}, \emph {et~al.},\ }\bibfield  {title} {\bibinfo {title} {Nasa
  innovative advanced concepts phase i final report--a lunar long-baseline
  uv/optical imaging interferometer: Artemis-enabled stellar imager (aesi)},\
  }\href@noop {} {\bibfield  {journal} {\bibinfo  {journal} {arXiv preprint
  arXiv:2503.02105}\ } (\bibinfo {year} {2025})}\BibitemShut {NoStop}%
\bibitem [{\citenamefont {Brown}\ \emph {et~al.}(2023)\citenamefont {Brown},
  \citenamefont {Allgaier}, \citenamefont {Thiel}, \citenamefont {Monnier},
  \citenamefont {Raymer},\ and\ \citenamefont
  {Smith}}]{brown2023interferometric}%
  \BibitemOpen
  \bibfield  {author} {\bibinfo {author} {\bibfnamefont {M.~R.}\ \bibnamefont
  {Brown}}, \bibinfo {author} {\bibfnamefont {M.}~\bibnamefont {Allgaier}},
  \bibinfo {author} {\bibfnamefont {V.}~\bibnamefont {Thiel}}, \bibinfo
  {author} {\bibfnamefont {J.~D.}\ \bibnamefont {Monnier}}, \bibinfo {author}
  {\bibfnamefont {M.~G.}\ \bibnamefont {Raymer}},\ and\ \bibinfo {author}
  {\bibfnamefont {B.~J.}\ \bibnamefont {Smith}},\ }\bibfield  {title} {\bibinfo
  {title} {Interferometric imaging using shared quantum entanglement},\
  }\href@noop {} {\bibfield  {journal} {\bibinfo  {journal} {Physical Review
  Letters}\ }\textbf {\bibinfo {volume} {131}},\ \bibinfo {pages} {210801}
  (\bibinfo {year} {2023})}\BibitemShut {NoStop}%
\bibitem [{\citenamefont {Bland-Hawthorn}\ \emph {et~al.}(2021)\citenamefont
  {Bland-Hawthorn}, \citenamefont {Sellars},\ and\ \citenamefont
  {Bartholomew}}]{bland2021quantum}%
  \BibitemOpen
  \bibfield  {author} {\bibinfo {author} {\bibfnamefont {J.}~\bibnamefont
  {Bland-Hawthorn}}, \bibinfo {author} {\bibfnamefont {M.~J.}\ \bibnamefont
  {Sellars}},\ and\ \bibinfo {author} {\bibfnamefont {J.~G.}\ \bibnamefont
  {Bartholomew}},\ }\bibfield  {title} {\bibinfo {title} {Quantum memories and
  the double-slit experiment: implications for astronomical interferometry},\
  }\href@noop {} {\bibfield  {journal} {\bibinfo  {journal} {JOSA B}\ }\textbf
  {\bibinfo {volume} {38}},\ \bibinfo {pages} {A86} (\bibinfo {year}
  {2021})}\BibitemShut {NoStop}%
\bibitem [{\citenamefont {Czupryniak}\ \emph {et~al.}(2023)\citenamefont
  {Czupryniak}, \citenamefont {Steinmetz}, \citenamefont {Kwiat},\ and\
  \citenamefont {Jordan}}]{czupryniak2023optimal}%
  \BibitemOpen
  \bibfield  {author} {\bibinfo {author} {\bibfnamefont {R.}~\bibnamefont
  {Czupryniak}}, \bibinfo {author} {\bibfnamefont {J.}~\bibnamefont
  {Steinmetz}}, \bibinfo {author} {\bibfnamefont {P.~G.}\ \bibnamefont
  {Kwiat}},\ and\ \bibinfo {author} {\bibfnamefont {A.~N.}\ \bibnamefont
  {Jordan}},\ }\bibfield  {title} {\bibinfo {title} {Optimal qubit circuits for
  quantum-enhanced telescopes},\ }\href@noop {} {\bibfield  {journal} {\bibinfo
   {journal} {Physical Review A}\ }\textbf {\bibinfo {volume} {108}},\ \bibinfo
  {pages} {052408} (\bibinfo {year} {2023})}\BibitemShut {NoStop}%
\bibitem [{\citenamefont {Wang}\ \emph {et~al.}(2023)\citenamefont {Wang},
  \citenamefont {Zhang},\ and\ \citenamefont {Lorenz}}]{wang2023astronomical}%
  \BibitemOpen
  \bibfield  {author} {\bibinfo {author} {\bibfnamefont {Y.}~\bibnamefont
  {Wang}}, \bibinfo {author} {\bibfnamefont {Y.}~\bibnamefont {Zhang}},\ and\
  \bibinfo {author} {\bibfnamefont {V.~O.}\ \bibnamefont {Lorenz}},\ }\bibfield
   {title} {\bibinfo {title} {Astronomical interferometry using continuous
  variable quantum teleportation},\ }\href@noop {} {\bibfield  {journal}
  {\bibinfo  {journal} {arXiv preprint arXiv:2308.12851}\ } (\bibinfo {year}
  {2023})}\BibitemShut {NoStop}%
\bibitem [{\citenamefont {Wang}\ and\ \citenamefont
  {Zhou}(2025)}]{wang2025limitations}%
  \BibitemOpen
  \bibfield  {author} {\bibinfo {author} {\bibfnamefont {Y.}~\bibnamefont
  {Wang}}\ and\ \bibinfo {author} {\bibfnamefont {S.}~\bibnamefont {Zhou}},\
  }\bibfield  {title} {\bibinfo {title} {Limitations of gaussian measurements
  in quantum imaging},\ }\href@noop {} {\bibfield  {journal} {\bibinfo
  {journal} {arXiv preprint arXiv:2503.06363}\ } (\bibinfo {year}
  {2025})}\BibitemShut {NoStop}%
\bibitem [{\citenamefont {Tang}\ \emph {et~al.}(2025)\citenamefont {Tang},
  \citenamefont {Zhang}, \citenamefont {Guo}, \citenamefont {Cui},
  \citenamefont {Li},\ and\ \citenamefont {Ou}}]{tang2025phase}%
  \BibitemOpen
  \bibfield  {author} {\bibinfo {author} {\bibfnamefont {X.}~\bibnamefont
  {Tang}}, \bibinfo {author} {\bibfnamefont {Y.}~\bibnamefont {Zhang}},
  \bibinfo {author} {\bibfnamefont {X.}~\bibnamefont {Guo}}, \bibinfo {author}
  {\bibfnamefont {L.}~\bibnamefont {Cui}}, \bibinfo {author} {\bibfnamefont
  {X.}~\bibnamefont {Li}},\ and\ \bibinfo {author} {\bibfnamefont
  {Z.}~\bibnamefont {Ou}},\ }\bibfield  {title} {\bibinfo {title}
  {Phase-dependent hanbury-brown and twiss effect for the complete measurement
  of the complex coherence function},\ }\href@noop {} {\bibfield  {journal}
  {\bibinfo  {journal} {Light: Science \& Applications}\ }\textbf {\bibinfo
  {volume} {14}},\ \bibinfo {pages} {46} (\bibinfo {year} {2025})}\BibitemShut
  {NoStop}%
\bibitem [{\citenamefont {Liu}\ \emph {et~al.}(2024{\natexlab{c}})\citenamefont
  {Liu}, \citenamefont {Wu}, \citenamefont {Li}, \citenamefont {Chen},
  \citenamefont {Wilczek}, \citenamefont {Shao}, \citenamefont {Xu},
  \citenamefont {Zhang},\ and\ \citenamefont {Pan}}]{liu2024super}%
  \BibitemOpen
  \bibfield  {author} {\bibinfo {author} {\bibfnamefont {L.-C.}\ \bibnamefont
  {Liu}}, \bibinfo {author} {\bibfnamefont {C.}~\bibnamefont {Wu}}, \bibinfo
  {author} {\bibfnamefont {W.}~\bibnamefont {Li}}, \bibinfo {author}
  {\bibfnamefont {Y.-A.}\ \bibnamefont {Chen}}, \bibinfo {author}
  {\bibfnamefont {F.}~\bibnamefont {Wilczek}}, \bibinfo {author} {\bibfnamefont
  {X.-P.}\ \bibnamefont {Shao}}, \bibinfo {author} {\bibfnamefont
  {F.}~\bibnamefont {Xu}}, \bibinfo {author} {\bibfnamefont {Q.}~\bibnamefont
  {Zhang}},\ and\ \bibinfo {author} {\bibfnamefont {J.-W.}\ \bibnamefont
  {Pan}},\ }\bibfield  {title} {\bibinfo {title} {Super-resolution imaging
  based on active optical intensity interferometry},\ }\href@noop {} {\bibfield
   {journal} {\bibinfo  {journal} {arXiv:2404.15685}\ } (\bibinfo {year}
  {2024}{\natexlab{c}})}\BibitemShut {NoStop}%
\bibitem [{\citenamefont {Czupryniak}\ \emph {et~al.}(2022)\citenamefont
  {Czupryniak}, \citenamefont {Chitambar}, \citenamefont {Steinmetz},\ and\
  \citenamefont {Jordan}}]{PhysRevA.106.032424}%
  \BibitemOpen
  \bibfield  {author} {\bibinfo {author} {\bibfnamefont {R.}~\bibnamefont
  {Czupryniak}}, \bibinfo {author} {\bibfnamefont {E.}~\bibnamefont
  {Chitambar}}, \bibinfo {author} {\bibfnamefont {J.}~\bibnamefont
  {Steinmetz}},\ and\ \bibinfo {author} {\bibfnamefont {A.~N.}\ \bibnamefont
  {Jordan}},\ }\bibfield  {title} {\bibinfo {title} {Quantum telescopy clock
  games},\ }\href {https://doi.org/10.1103/PhysRevA.106.032424} {\bibfield
  {journal} {\bibinfo  {journal} {Phys. Rev. A}\ }\textbf {\bibinfo {volume}
  {106}},\ \bibinfo {pages} {032424} (\bibinfo {year} {2022})}\BibitemShut
  {NoStop}%
\bibitem [{\citenamefont {Marchese}\ and\ \citenamefont
  {Kok}(2023)}]{marchese2023large}%
  \BibitemOpen
  \bibfield  {author} {\bibinfo {author} {\bibfnamefont {M.~M.}\ \bibnamefont
  {Marchese}}\ and\ \bibinfo {author} {\bibfnamefont {P.}~\bibnamefont {Kok}},\
  }\bibfield  {title} {\bibinfo {title} {Large baseline optical imaging
  assisted by single photons and linear quantum optics},\ }\href@noop {}
  {\bibfield  {journal} {\bibinfo  {journal} {Physical Review Letters}\
  }\textbf {\bibinfo {volume} {130}},\ \bibinfo {pages} {160801} (\bibinfo
  {year} {2023})}\BibitemShut {NoStop}%
\bibitem [{\citenamefont {Modak}\ and\ \citenamefont
  {Kok}(2024)}]{modak2024large}%
  \BibitemOpen
  \bibfield  {author} {\bibinfo {author} {\bibfnamefont {S.}~\bibnamefont
  {Modak}}\ and\ \bibinfo {author} {\bibfnamefont {P.}~\bibnamefont {Kok}},\
  }\bibfield  {title} {\bibinfo {title} {Large baseline quantum telescopes
  assisted by partially distinguishable photons},\ }\href@noop {} {\bibfield
  {journal} {\bibinfo  {journal} {arXiv preprint arXiv:2412.16571}\ } (\bibinfo
  {year} {2024})}\BibitemShut {NoStop}%
\bibitem [{\citenamefont {Vasilev}\ \emph {et~al.}(2009)\citenamefont
  {Vasilev}, \citenamefont {Kuhn},\ and\ \citenamefont
  {Vitanov}}]{PhysRevA.80.013417}%
  \BibitemOpen
  \bibfield  {author} {\bibinfo {author} {\bibfnamefont {G.~S.}\ \bibnamefont
  {Vasilev}}, \bibinfo {author} {\bibfnamefont {A.}~\bibnamefont {Kuhn}},\ and\
  \bibinfo {author} {\bibfnamefont {N.~V.}\ \bibnamefont {Vitanov}},\
  }\bibfield  {title} {\bibinfo {title} {Optimum pulse shapes for stimulated
  raman adiabatic passage},\ }\href
  {https://doi.org/10.1103/PhysRevA.80.013417} {\bibfield  {journal} {\bibinfo
  {journal} {Phys. Rev. A}\ }\textbf {\bibinfo {volume} {80}},\ \bibinfo
  {pages} {013417} (\bibinfo {year} {2009})}\BibitemShut {NoStop}%
\bibitem [{\citenamefont {Grinkemeyer}\ \emph {et~al.}(2025)\citenamefont
  {Grinkemeyer}, \citenamefont {Guardado-Sanchez}, \citenamefont {Dimitrova},
  \citenamefont {Shchepanovich}, \citenamefont {Mandopoulou}, \citenamefont
  {Borregaard}, \citenamefont {Vuletić},\ and\ \citenamefont
  {Lukin}}]{Lukin:2025}%
  \BibitemOpen
  \bibfield  {author} {\bibinfo {author} {\bibfnamefont {B.}~\bibnamefont
  {Grinkemeyer}}, \bibinfo {author} {\bibfnamefont {E.}~\bibnamefont
  {Guardado-Sanchez}}, \bibinfo {author} {\bibfnamefont {I.}~\bibnamefont
  {Dimitrova}}, \bibinfo {author} {\bibfnamefont {D.}~\bibnamefont
  {Shchepanovich}}, \bibinfo {author} {\bibfnamefont {G.~E.}\ \bibnamefont
  {Mandopoulou}}, \bibinfo {author} {\bibfnamefont {J.}~\bibnamefont
  {Borregaard}}, \bibinfo {author} {\bibfnamefont {V.}~\bibnamefont
  {Vuletić}},\ and\ \bibinfo {author} {\bibfnamefont {M.~D.}\ \bibnamefont
  {Lukin}},\ }\bibfield  {title} {\bibinfo {title} {Error-detected quantum
  operations with neutral atoms mediated by an optical cavity},\ }\href
  {https://doi.org/10.1126/science.adr7075} {\bibfield  {journal} {\bibinfo
  {journal} {Science}\ }\textbf {\bibinfo {volume} {387}},\ \bibinfo {pages}
  {1301} (\bibinfo {year} {2025})},\ \Eprint
  {https://arxiv.org/abs/https://www.science.org/doi/pdf/10.1126/science.adr7075}
  {https://www.science.org/doi/pdf/10.1126/science.adr7075} \BibitemShut
  {NoStop}%
\bibitem [{\citenamefont {Srivastava}\ \emph {et~al.}(2024)\citenamefont
  {Srivastava}, \citenamefont {Jandura}, \citenamefont {Brennen},\ and\
  \citenamefont
  {Pupillo}}]{srivastava2024entanglementenhancedquantumsensingoptimal}%
  \BibitemOpen
  \bibfield  {author} {\bibinfo {author} {\bibfnamefont {V.}~\bibnamefont
  {Srivastava}}, \bibinfo {author} {\bibfnamefont {S.}~\bibnamefont {Jandura}},
  \bibinfo {author} {\bibfnamefont {G.~K.}\ \bibnamefont {Brennen}},\ and\
  \bibinfo {author} {\bibfnamefont {G.}~\bibnamefont {Pupillo}},\ }\href
  {https://arxiv.org/abs/2409.12932} {\bibinfo {title} {Entanglement-enhanced
  quantum sensing via optimal global control}} (\bibinfo {year} {2024}),\
  \Eprint {https://arxiv.org/abs/2409.12932} {arXiv:2409.12932 [quant-ph]}
  \BibitemShut {NoStop}%
\bibitem [{\citenamefont {Hunger}\ \emph {et~al.}(2010)\citenamefont {Hunger},
  \citenamefont {Steinmetz}, \citenamefont {Colombe}, \citenamefont {Deutsch},
  \citenamefont {Hänsch},\ and\ \citenamefont {Reichel}}]{Hunger_2010}%
  \BibitemOpen
  \bibfield  {author} {\bibinfo {author} {\bibfnamefont {D.}~\bibnamefont
  {Hunger}}, \bibinfo {author} {\bibfnamefont {T.}~\bibnamefont {Steinmetz}},
  \bibinfo {author} {\bibfnamefont {Y.}~\bibnamefont {Colombe}}, \bibinfo
  {author} {\bibfnamefont {C.}~\bibnamefont {Deutsch}}, \bibinfo {author}
  {\bibfnamefont {T.~W.}\ \bibnamefont {Hänsch}},\ and\ \bibinfo {author}
  {\bibfnamefont {J.}~\bibnamefont {Reichel}},\ }\bibfield  {title} {\bibinfo
  {title} {A fiber fabry–perot cavity with high finesse},\ }\href
  {https://doi.org/10.1088/1367-2630/12/6/065038} {\bibfield  {journal}
  {\bibinfo  {journal} {New Journal of Physics}\ }\textbf {\bibinfo {volume}
  {12}},\ \bibinfo {pages} {065038} (\bibinfo {year} {2010})}\BibitemShut
  {NoStop}%
\bibitem [{\citenamefont {Manetsch}\ \emph {et~al.}(2024)\citenamefont
  {Manetsch}, \citenamefont {Nomura}, \citenamefont {Bataille}, \citenamefont
  {Leung}, \citenamefont {Lv},\ and\ \citenamefont
  {Endres}}]{manetsch2024tweezer}%
  \BibitemOpen
  \bibfield  {author} {\bibinfo {author} {\bibfnamefont {H.~J.}\ \bibnamefont
  {Manetsch}}, \bibinfo {author} {\bibfnamefont {G.}~\bibnamefont {Nomura}},
  \bibinfo {author} {\bibfnamefont {E.}~\bibnamefont {Bataille}}, \bibinfo
  {author} {\bibfnamefont {K.~H.}\ \bibnamefont {Leung}}, \bibinfo {author}
  {\bibfnamefont {X.}~\bibnamefont {Lv}},\ and\ \bibinfo {author}
  {\bibfnamefont {M.}~\bibnamefont {Endres}},\ }\bibfield  {title} {\bibinfo
  {title} {A tweezer array with 6100 highly coherent atomic qubits},\
  }\href@noop {} {\bibfield  {journal} {\bibinfo  {journal} {arXiv preprint
  arXiv:2403.12021}\ } (\bibinfo {year} {2024})}\BibitemShut {NoStop}%
\bibitem [{\citenamefont {Niemietz}\ \emph {et~al.}(2021)\citenamefont
  {Niemietz}, \citenamefont {Farrera}, \citenamefont {Langenfeld},\ and\
  \citenamefont {Rempe}}]{Niemietz2021-nf}%
  \BibitemOpen
  \bibfield  {author} {\bibinfo {author} {\bibfnamefont {D.}~\bibnamefont
  {Niemietz}}, \bibinfo {author} {\bibfnamefont {P.}~\bibnamefont {Farrera}},
  \bibinfo {author} {\bibfnamefont {S.}~\bibnamefont {Langenfeld}},\ and\
  \bibinfo {author} {\bibfnamefont {G.}~\bibnamefont {Rempe}},\ }\bibfield
  {title} {\bibinfo {title} {Nondestructive detection of photonic qubits},\
  }\href@noop {} {\bibfield  {journal} {\bibinfo  {journal} {Nature}\ }\textbf
  {\bibinfo {volume} {591}},\ \bibinfo {pages} {570} (\bibinfo {year}
  {2021})}\BibitemShut {NoStop}%
\bibitem [{\citenamefont {Serafini}(2017)}]{serafini2017quantum}%
  \BibitemOpen
  \bibfield  {author} {\bibinfo {author} {\bibfnamefont {A.}~\bibnamefont
  {Serafini}},\ }\href {https://doi.org/10.1201/9781315118727} {\emph {\bibinfo
  {title} {Quantum Continuous Variables: A Primer of Theoretical Methods}}}\
  (\bibinfo  {publisher} {CRC press},\ \bibinfo {year} {2017})\BibitemShut
  {NoStop}%
\bibitem [{\citenamefont {Weedbrook}\ \emph {et~al.}(2012)\citenamefont
  {Weedbrook}, \citenamefont {Pirandola}, \citenamefont {Garc\'{\i}a-Patr\'on},
  \citenamefont {Cerf}, \citenamefont {Ralph}, \citenamefont {Shapiro},\ and\
  \citenamefont {Lloyd}}]{RevModPhys.84.621}%
  \BibitemOpen
  \bibfield  {author} {\bibinfo {author} {\bibfnamefont {C.}~\bibnamefont
  {Weedbrook}}, \bibinfo {author} {\bibfnamefont {S.}~\bibnamefont
  {Pirandola}}, \bibinfo {author} {\bibfnamefont {R.}~\bibnamefont
  {Garc\'{\i}a-Patr\'on}}, \bibinfo {author} {\bibfnamefont {N.~J.}\
  \bibnamefont {Cerf}}, \bibinfo {author} {\bibfnamefont {T.~C.}\ \bibnamefont
  {Ralph}}, \bibinfo {author} {\bibfnamefont {J.~H.}\ \bibnamefont {Shapiro}},\
  and\ \bibinfo {author} {\bibfnamefont {S.}~\bibnamefont {Lloyd}},\ }\bibfield
   {title} {\bibinfo {title} {Gaussian quantum information},\ }\href
  {https://doi.org/10.1103/RevModPhys.84.621} {\bibfield  {journal} {\bibinfo
  {journal} {Reviews of Modern Physics}\ }\textbf {\bibinfo {volume} {84}},\
  \bibinfo {pages} {621} (\bibinfo {year} {2012})}\BibitemShut {NoStop}%
\bibitem [{\citenamefont {Braunstein}\ and\ \citenamefont
  {Kimble}(1998)}]{PhysRevLett.80.869}%
  \BibitemOpen
  \bibfield  {author} {\bibinfo {author} {\bibfnamefont {S.~L.}\ \bibnamefont
  {Braunstein}}\ and\ \bibinfo {author} {\bibfnamefont {H.~J.}\ \bibnamefont
  {Kimble}},\ }\bibfield  {title} {\bibinfo {title} {Teleportation of
  continuous quantum variables},\ }\href
  {https://doi.org/10.1103/PhysRevLett.80.869} {\bibfield  {journal} {\bibinfo
  {journal} {Physical Review Letters}\ }\textbf {\bibinfo {volume} {80}},\
  \bibinfo {pages} {869} (\bibinfo {year} {1998})}\BibitemShut {NoStop}%
\bibitem [{\citenamefont {Wang}\ \emph {et~al.}(2025)\citenamefont {Wang},
  \citenamefont {Oh}, \citenamefont {Liu}, \citenamefont {Jiang},\ and\
  \citenamefont {Zhou}}]{wang2025advancing}%
  \BibitemOpen
  \bibfield  {author} {\bibinfo {author} {\bibfnamefont {Y.}~\bibnamefont
  {Wang}}, \bibinfo {author} {\bibfnamefont {C.}~\bibnamefont {Oh}}, \bibinfo
  {author} {\bibfnamefont {J.}~\bibnamefont {Liu}}, \bibinfo {author}
  {\bibfnamefont {L.}~\bibnamefont {Jiang}},\ and\ \bibinfo {author}
  {\bibfnamefont {S.}~\bibnamefont {Zhou}},\ }\bibfield  {title} {\bibinfo
  {title} {Advancing quantum imaging through learning theory},\ }\href@noop {}
  {\bibfield  {journal} {\bibinfo  {journal} {arXiv preprint arXiv:2501.15685}\
  } (\bibinfo {year} {2025})}\BibitemShut {NoStop}%
\end{thebibliography}
%

\appendix
\begin{widetext}

\section{Calculating the mean photon numbers of nearby stars and exoplanets}\label{app:a}


\begin{table}[h!]
\centering
\begin{tabular}{|l|c|c|c| c| c|}
\hline
Planet & mass ($M_\oplus$) & temperature (K) & distance (light year) &host apparant brightness ($m_{AB}$)\\
\hline
Earth              & 1            & 255 & 0   &  N/A \\ \hline
Proxima Centauri b & $\geq 1.07$  & 228 & 4.2 & 11.05 \\
Ross 128 b         & $\geq 1.4$   & 280 & 11  & 11.13\\
GJ 1061 d          & $\geq 1.64$  & 218 & 12  & 13.03 \\
Luyten b           & $\geq 2.89$  & 258 & 12.4 & 9.85\\
Gliese 1002 b      & $\geq 1.08$  & 231 & 15.8& 13.837\\
\hline
\end{tabular}
\caption{\label{tab:list_of_exoplanets} A list of nearby exoplanets with potentially habitable temperature, their masses relative to Earth, distance, and magnitudes of their host star}

\end{table}
In this section, we include details of how we arrive at the mean photon numbers of nearby stars and exoplanets.
Here the monochromatic absolute magnitude $m_{AB}$ is defined as the logarithm of a spectral flux density with the usual scaling of astronomical magnitudes. 
The zero-point is approximately about 3631 Jy, where
\begin{align}
1   \text{ Jy} &= \text{$10^{-26} $ W Hz$^{-1 }$ m$^{-2}$}\nn
   &= \text{ $10^{-23}$ erg s$^{-1}$ Hz$^{-1}$ cm $^{-2}$} .
\end{align}
Denote $F_\nu$ as the spectral flux density.
\begin{align}
m_\text{AB} &\approx -2.5 \log_{10}\left(  \frac{F_\nu}{3631 ~\text{Jy}} \right) 
\end{align}
The relationship between $m_{AB}$ and flux is
\begin{align}
F_\nu = 10^{-0.4 (m_{AB} + 48.6)} \text{erg} ~ s^{-1} \text{cm}^{-2} \text{Hz}^{-1}
\end{align}
Convert this to SI units
\begin{align}
1~ \text{erg}  &= 10^{-7} \text{joules}\\
1~ \text{cm}^2 &= 10^{-4} \text{m}^2
\end{align}
Therefore in SI units, $F_\nu$ has units of $10^{-3}$ W m$^{-2}$ Hz$^{-1}$:

\begin{align}
F_\nu = 10^{-0.4 (m_{AB} + 48.6)} \times 10^{-3} \text{W m}^{-2} \text{Hz}^{-1}
\end{align}

Converting to flux per unit wavelength $F_\lambda$
\begin{align}
F_\lambda = \frac{F_\nu \times c}{ \lambda^2}
\end{align}
where $c$ is the speed of light and $\lambda$ is the wavelength. Calculate the photon flux density
\begin{align}
\Phi_\lambda = \frac{F_\lambda}{ h c/\lambda} = \frac{F_\nu}{h \times\lambda}
\end{align}

Now, to calculate the total photon-specific bandwidth, we assume the density is approximately constant over the bandwidth (this holds for narrow bandwidths):
\begin{align}
\Phi_\text{tot} &= \Phi_\lambda \times \Delta \lambda \\
&=\frac{1}{ h \times \lambda } 10^{-0.4 (m_{AB} + 48.6)}  \times 10^{-3} \times
\Delta \lambda
\end{align}

Now, to relate that to $\epsilon$, we need to figure out 'how long' the photon is, i.e. the coherence time. Then we'll just call it the size of the time bin, here $\Delta \nu$ is the frequency bandwidth
\begin{align}
\tau_c = \frac{1}{\Delta \nu}
\end{align}
For a 10 GHz bandwidth signal, $\Delta \tau \sim 10^{-10}$.

To convert between a frequency bandwidth to a wavelength bandwidth,
\begin{align}
\Delta\lambda = \Delta \nu \times \lambda^2/c \\
\Delta \nu  = \Delta \lambda \times c/ \lambda^2
\end{align}

Repeating the calculation of [PRL 123, 070504 (2019)]: for a $\Delta \nu = 10$ GHz, a total collection area of $10^2$,
\begin{align}
\epsilon = \text{ No of photons per second} \times \text{coherence time} \times \text{ collection area}
\end{align}

For the CHARA array, visible interferometry has a bandwidth between 10-50 nm.

Putting in the numbers, $\Delta \nu = 10\times 10^{9}$, $\lambda = 555\times 10^{-9}, m =10$
\begin{align}
\epsilon &= \Phi_\text{tot}  \times \tau_c \times 10 \\
         &\approx 10^{-7}
\end{align}

In comparison,~\cite{khabiboulline2019quantum} obtains $ \epsilon \approx 7\times 10^{-7}$.

Now, let us examine at the bands of interest, corresponding to bio-signature molecular absorption bands:
\begin{table}[h!]
\centering
\begin{tabular}{|c| c| c|  c| c| c| c| c| c| }
\hline
Host  
($m_\text{AB}$) & $\lambda$   & $\Delta \lambda$ & $\Delta \nu$  & Photons/s ($\Phi_\text{tot} \times  $ area)  & $\tau_c$ &  $\epsilon$ (star) & $\epsilon$ (planet)\\ \hline
 9             & 760 nm      & 1 nm      &  500 GHz    &  181106 &  2$\times 10^{-12}$ & $3.5\times 10^{-7}$  & $3.5\times 10^{-16}$ \\
 9             & 1.65 $\mu$m  & 1 nm     &  100 GHz    & 83418   &    $ 10^{-12}$      & $7.5 \times 10^{-7}$ & $7.5 \times 10^{-16}$  \\ \hline
 11             & 760 nm      & 10 nm    &  5 THz     &  287035 &  2$\times 10^{-13}$ &  $5.5\times 10^{-8}$  &  $5.5\times 10^{-17}$\\
 11             & 1.65 $\mu$m & 10 nm    &  1.1 THz   & 132210  &    $ 10^{-12}$      &  $1.2 \times 10^{-7}$ &  $1.2 \times 10^{-16}$ \\\hline
 11             & 760 nm      & 1 nm     &  500 GHz    &  28703 &  2$\times 10^{-12}$ &  $5.5\times 10^{-8}$   &  $5.5\times 10^{-17}$\\
 11             & 1.65 $\mu$m & 1 nm     &  100 GHz    &  13221  &    $ 10^{-11}$      &  $1.2 \times 10^{-7}$  &  $1.2 \times 10^{-16}$\\\hline
 13             & 760 nm      & 1 nm     &  500 GHz    &  4549 &  2$\times 10^{-12}$ & $9\times 10^{-9}$      &  $9\times 10^{-18}$\\
 13             & 1.65 $\mu$m & 1 nm     &  100 GHz    & 2095  &    $ 10^{-12}$      & $1.9 \times 10^{-8} $  &  $1.9 \times 10^{-17}$  \\
 %
 %
%
\hline
\end{tabular}
\caption{Expected number of photons to collect at a given wavelength, bandwidth for the host star. The wavelengths of interest are: 760nm is the absorption line for molecular oxygen; for Earth-like conditions the pressure-broadening linewidth is typically 
\mbox{$ \sim$ 2 pm} at FWHM;  1.65 $\mu$m is a methane absorption line, and for earth-like conditions the FWHM 
\mbox{$ \times 10^{-4} \mu$m}. Here we have assumed a collection area of 10 m$^2$. Spectral demultiplexing doesn't decrease $\epsilon$ due to the fact that the effective time bins get larger. We assume a conservative $10^{-9}$ factor between the star and the planet's brightness. These numbers might differ depending on what assumption is made about the temperature of the star.
\label{tab:conversion}}
\end{table}
\end{widetext}

\end{document}